\documentclass[a4paper,12pt,titlepage,twoside,headsepline,openany,BCOR=13mm,DIV=13]{scrbook}

\makeatletter
\DeclareOldFontCommand{\rm}{\normalfont\rmfamily}{\mathrm}
\DeclareOldFontCommand{\sf}{\normalfont\sffamily}{\mathsf}
\DeclareOldFontCommand{\tt}{\normalfont\ttfamily}{\mathtt}
\DeclareOldFontCommand{\bf}{\normalfont\bfseries}{\mathbf}
\DeclareOldFontCommand{\it}{\normalfont\itshape}{\mathit}
\DeclareOldFontCommand{\sl}{\normalfont\slshape}{\@nomath\sl}
\DeclareOldFontCommand{\sc}{\normalfont\scshape}{\@nomath\sc}
\makeatother

\usepackage[Sonny]{fncychap}

\usepackage{amsmath}
\usepackage[usenames,dvipsnames]{xcolor}
\usepackage{graphicx}
\usepackage{caption}
\usepackage{float}
\usepackage{subfloat}
\usepackage{rotating}
\usepackage{nameref}
\usepackage{mathptmx} 
\usepackage{tabularx}
\usepackage{subfigure}
\usepackage{makeidx}
\usepackage{titlesec}
\usepackage[T1]{fontenc}
\usepackage{chapterbib}
\usepackage[intoc]{nomencl}
\usepackage{amssymb}
\usepackage{tabularx}
\usepackage{graphicx}
\usepackage{hyperref}

\usepackage{framed}

\usepackage{fancyhdr}
\usepackage[english]{babel}
\usepackage{afterpage}
\usepackage{titlesec, blindtext, color}
\definecolor{gray75}{gray}{0.75}
\usepackage[english]{babel}
\usepackage{nomencl}
\usepackage{siunitx}
\usepackage{tikz}
\usetikzlibrary{arrows,positioning}

\makeatletter
\newcounter {subsubsubsection}[subsubsection]
\renewcommand\thesubsubsubsection{\thesubsubsection .\@alph\c@subsubsubsection}
\newcommand\subsubsubsection{\@startsection{subsubsubsection}{4}{\z@}%
                                     {-3.25ex\@plus -1ex \@minus -.2ex}%
                                     {1.5ex \@plus .2ex}%
                                     {\normalfont\normalsize\bfseries}}
\renewcommand\paragraph{\@startsection{paragraph}{5}{\z@}%
                                    {3.25ex \@plus1ex \@minus.2ex}%
                                    {-1em}%
                                    {\normalfont\normalsize\bfseries}}
\renewcommand\subparagraph{\@startsection{subparagraph}{6}{\parindent}%
                                       {3.25ex \@plus1ex \@minus .2ex}%
                                       {-1em}%
                                      {\normalfont\normalsize\bfseries}}
\newcommand*\l@subsubsubsection{\@dottedtocline{4}{10.0em}{4.1em}}
\renewcommand*\l@paragraph{\@dottedtocline{5}{10em}{5em}}
\renewcommand*\l@subparagraph{\@dottedtocline{6}{12em}{6em}}
\newcommand*{\subsubsubsectionmark}[1]{}
\makeatother

\usepackage{hyperref}
\usepackage{lineno}
\makeatletter
\def\toclevel@subsubsubsection{4}
\def\toclevel@paragraph{5}
\def\toclevel@subparagraph{6}
\makeatother

\graphicspath{{./Figures/}}

\begin{document}
\thispagestyle{fancy}
\fancyhead[C]{Name}
\fancyfoot[C]{\footnotesize Page \thepage\ of \pageref{LastPage}}

\vspace{6cm}

\begin{titlepage}
\begin{figure}[H]
\centering
\includegraphics[scale=0.33]{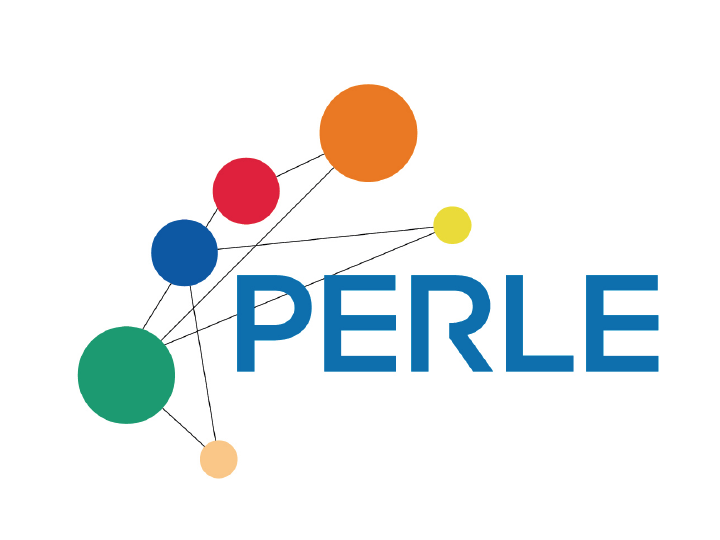}

\end{figure}

\vspace{.1cm}
\begin{center}
{\LARGE{\bf{PERLE}}\\
{\Large{Powerful Energy Recovery Linac for Experiments\\ }}}
  \vspace{0.6cm}

        {\LARGE{Conceptual Design Report\\
        \vspace{0.5cm}}}
        $to~be~published~in~J.Phys.G$

        \vspace{3cm}
        {\Large{CELIA Bordeaux, MIT Boston, CERN, Cockcroft and ASTeC Daresbury, TU Darmstadt,  U Liverpool,  Jefferson 
        Lab Newport News, BINP Novosibirsk, IPN and LAL Orsay}}

  \vspace{2cm}
        
        {\Large{May 13$^{th}$, 2017 \\}}


\vspace{1.8cm}
  \end{center}
\end{titlepage}

\newpage
\section*{List of Authors}

\noindent
D. Angal-Kalinin$^3$,
G. Arduini$^1$,
B. Auchmann$^{1}$,
J. Bernauer$^{10}$,
A. Bogacz$^{4}$,
F.~Bordry$^1$,
S.~Bousson$^{9}$, 
C.~Bracco$^{1}$, 
O.~Br\"{u}ning$^{1}$,
R. Calaga$^{1}$,
K. Cassou$^{2}$,
V. Chetvertkova$^{1}$,
E.~Cormier$^{6}$,
E. Daly$^{4}$,
D. Douglas$^{4}$,
K. Dupraz$^{2}$,
B. Goddard$^{1}$, 
J. Henry$^{4}$,
A. Hutton$^4$,
E.~Jensen$^{1}$,
W.~Kaabi$^{2}$,
M.~Klein$^{5}$, 
P. Kostka$^{5}$, 
F.~Marhauser$^{4}$,
A. Martens,$^{2}$,
A. Milanese$^{1}$,
B. Militsyn$^{3}$,
Y. Peinaud$^{2}$,
D. Pellegrini$^{1}$,
N. Pietralla$^{8}$,
Y.A.~Pupkov$^{7}$,
R. A. Rimmer$^{4}$,
K. Schirm$^{1}$,
D.~Schulte$^{1}$,
S. Smith$^3$,
A.~Stocchi$^{2}$,
A.~Valloni$^{1}$,
C.~Welsch$^{5}$,
G. Willering$^{1}$,
D.~Wollmann$^{1}$,
F.~Zimmermann$^{1}$,
F.~Zomer$^{2}$

\bigskip{\it\noindent
$^{1}$ CERN, Geneva, Switzerland\\
$^{2}$ LAL, CNRS-IN2P3, Universit\'e Paris-Sud, Centre Scientifique d'Orsay, France\\
$^{3}$ ASTeC, STFC, Daresbury, UK\\
$^{4}$ Jefferson Lab, Newport News, VA, USA\\
$^{5}$ University of Liverpool, UK\\
$^{6}$ CELIA, University of Bordeaux 1, CNRS UMR 5107, Talence, France\\
$^{7}$ BINP, Novosibirsk\\
$^{8}$ Institut f\"{u}r Kernphysik Technische Universit\"{a}t Darmstadt\\
$^{9}$ Institute de Physique Nucleaire Orsay, France \\
$^{10}$ Massachusetts Institute of Technology, Cambridge, MA, USA
}

\newpage

\section*{Abstract}
A conceptual design is presented of a novel ERL facility for the development and application of the energy recovery
technique to linear electron accelerators in the multi-turn, 
large current 
and large energy regime. 
The main characteristics of 
the powerful energy recovery linac experiment facility (PERLE) are derived from
the design of the Large Hadron electron Collider, an electron beam upgrade under study for the LHC,
 for which it would be the key demonstrator. PERLE is thus projected as a 
facility to investigate efficient, high current ($>$\,$10$\,mA) ERL operation with three
re-circulation passages through newly designed SCRF cavities, at $801.58$\,MHz frequency,
and following deceleration
over another three re-circulations. In its fully equipped configuration, PERLE provides an electron beam of
approximately $1$\,GeV energy. A physics programme possibly associated with PERLE is sketched,
 consisting of high precision elastic electron-proton
scattering experiments, as well as photo-nuclear reactions of unprecedented intensities with
 up to $30$\,MeV photon beam energy as may be obtained using Fabry-Perot cavities. The facility
has further applications as a general technology test bed that can investigate 
and validate novel superconducting magnets
(beam induced quench tests) and superconducting RF structures (structure tests with high current beams, beam
loading and transients).
Besides a chapter on operation aspects,
the report contains detailed considerations on the choices for the SCRF structure,
 optics and lattice design, solutions for arc magnets, source and injector 
and on further essential components. A suitable configuration derived from the here
presented design concept may next be moved forward to a technical design and possibly be built
by an international collaboration which is being established. 

\tableofcontents

\markboth{Introduction} {right head}\markright{Introduction}
\chapter{Introduction}

\vspace{-1cm}
The development of the Large Hadron Electron Collider (LHeC)~\cite{Abelleira2012}
opens the horizon for
turning the LHC facility, with accurate $pp \rightarrow HX$ and $ep \rightarrow  \nu HX$
measurements, into a precision Higgs ($H$) physics factory. It  also represented the
world's cleanest, high resolution microscope for exploring the substructure of 
hadronic matter and  parton dynamics at smallest  
dimensions which also complements the LHC proton-proton ($pp$) and 
heavy-ion ($AA$ and $pA$) physics.
The genuine deep inelastic electron-hadron scattering  programme of the  
LHeC~\cite{Klein:2016uwv} 
is of unprecedented richness. It may lead to beyond the Standard Model
through discoveries in $ep$ interactions in the new energy regime
and as well through the clarification of proton structure effects
in the region of very high mass, corresponding
to large Bjorken $x$, in $pp$ interactions.

As demonstrated in the  conceptual design report~\cite{Abelleira2012}, 
the LHeC may be realised by the addition
of an intense electron beam to the LHC proton (and ion) beams
following the now commencing upgrade of the LHC
for increased luminosity.  It uses two electron linear accelerators arranged 
in a racetrack configuration, tangential to the LHC tunnel. 
In three-turn operation mode one is able to generate an electron beam 
of $60~(50)$\,GeV energy for a 
circumference of U(LHeC)=U(LHC)/n   of
approximately $9~(5)$\,km length, for $n=3~(5)$.
This configuration would be of immediate use and immense value
if the LHC proton energy was doubled, 
and it has also been considered as the default option for a future electron-hadron 
operation of the FCC~\cite{fccnote}. 
The value of the Higgs production cross section at the LHeC of O($100$)\,fb sets
a luminosity goal of O($10^{34}$)\,cm$^{-2}$s$^{-1}$ which in the linac-ring configuration
of the LHeC, at a total power limit of $100$\,MW, 
can only be achieved~\cite{fccnote,Zimmermann2013lhec,Bruening:2013bga}
by application of the energy-recovery technique recently reviewed in~\cite{icfanl,tenerl}.
This sets a target for the electron current of PERLE to be of order $10$\,mA. 

The demonstration and optimisation of the LHeC principles and parameters
require building a high current, multi-turn ERL facility. Its main parameters
shall correspond to the LHeC design, and experience with PERLE's operation would be transferable to the LHeC.
 The LHeC frequency was chosen to be 
$801.58$\,MHz, which is compliant with the LHC, keeps beam-beam interactions low
and further corresponds well to general optimisation considerations
including power, surface resistance and cost. That frequency is also a base
frequency for the FCC development such that there is a multiple use envisaged of
the here described SCRF developments.
The electron beam current should be in the range of $10-20$\,mA, leading to
a 6-fold load in the cavity operation.  
Three passages through two oppositely positioned linear SCRF accelerator structures of $1$\,km length each
are required for reaching a $60$ GeV beam energy for the LHeC
as well as for FCC-eh.
PERLE will enable developing main accelerator components, such as the SCRF 
cavity-cryomodule which 
comprises four 5-cell cavities with a $15-20$\,MV/m gradient and operated in CW mode. 

The facility offers a range of unique technical and physics applications
through powerful energy recovery linac experiments from which its name, PERLE, is derived. The input electron current of about $15$\,mA leads to  high power tests of the SCRF with currents as large as
$100$\,mA following from three-turn acceleration and deceleration
in the energy recovery mode. 
The choice of  electron beam energy  depends on 
its main goals. An LHeC demonstrator, with the here mentioned parameters, may
be laid out as a machine with one (or two) cryomodule and deliver a beam of about $220~(440)$\,MeV 
energy. Physics applications, as are discussed below, may suggest to choose a higher energy.
In the here presented design a maximum size racetrack
configuration is considered using two
opposite linacs, each comprising  two cryomodules. This leads to a  nearly $1$\,GeV energy 
electron beam  suitable for
$ep$ scattering physics, possibly using polarised electrons
in weak interaction measurements.
Backscattering may generate  a photon beam of $30$\,MeV energy
which is of interest to reach beyond the so-called giant dipole resonance. Physics, site, cost and time schedule considerations make a step-wise development of such a facility attractive and likely.


The design parameters of the facility, its purpose and range of applications distinguish it
from a number of further new ERL developments, such as MESA at 
Mainz~\cite{Aulenbacher:2013xla}, BERLinPRO~\cite{Abo-Bakr:2015qow}, 
C$\beta$~\cite{Bazarov:2015xia,cbeta} at Cornell, and the recent ER$@$CEBAF~\cite{ercebaf} 
proposal for a new experiment at the Thomas Jefferson Laboratory. 
The frequencies of MESA, 
BERLinPRO and C$\beta$ are 1.3 GHz, while CEBAF operates 
at 1.5 GHz. MESA is directed primarily to weak interaction measurements.
BERLinPRO and C$\beta$ push for  very high
current developments. The ER$@$CEBAF intention is for a test at small currents but high energies, 
of about $6$\,GeV, in order to study 
synchrotron radiation effects on the ERL performance~\cite{Douglas:2015vsm}.

The present paper describes a conceptual design of an LHeC demonstrator and
some of its possible applications. 
PERLE would be of use for the beam based development of SCRF technology, regarding
for example the determination of current load limits and the control of higher order
modes. It would provide the necessary infrastructure for testing the 3-turn behaviour, 
stability and reproducibility of the ERL, beam quality measurements in (de)acceleration etc.
As is described, the facility would be of use for testing equipment, such as SC magnets and their quench behaviour, under beam conditions. It may also provide a low energy electron test beam  for developments
of detector technology such as thin Silicon trackers.
Various selected and particularly attractive
 physics applications of PERLE are sketched, comprising, with electron beams, 
searches for dark photons, weak interaction or proton radius measurements, and,
with photon beams,  the physics of photo-nuclear
reactions, nuclear structure, particle physics metrology and astrophysics,
 at photon intensities hugely exceeding that of the ELI facility~\cite{eliro}  currently
under construction in Southern Europe.

 This paper is organised as follows: Section\,2 describes the multiple purpose of PERLE, including a
possible later application as an injector to the LHeC. Section\,3 presents the conceptual design of the facility,
its system architecture, optics layout etc. Section\,4 characterises the main components, the electron source, injector,
SC cavity, cryomodule, magnets, transfers, beam dumps and also the generation of a photon beam through backscattered
laser light. Section\,5 describes aspects of monitoring and operating such a facility, largely based
on experience from CEBAF at the Thomas Jefferson Laboratory. Section\,6 provides initial considerations of site requirements, followed
by a brief summary in Section\,7.
 \clearpage 

\chapter{Purpose}

\section{SCRF and ERL  Tests  with PERLE}
PERLE is designed to be a multi-purpose and flexible machine that will be able to provide unique test beams in either ERL mode or as a multi-pass re-circulated linac (like CEBAF). It can also be constructed in a phased approach enabling early operation and logical, minimally invasive upgrades. The high intensity, low emittance beams will be invaluable for many hardware and instrumentation test programs as well as offering the potential for low energy physics experiments, dark matter searches, unique light sources etc.
Besides these many advantages, PERLE is also a ground breaking accelerator and SRF demonstration and development facility. The principles of multi-pass acceleration and energy recovery using SRF recirculating linacs have already been demonstrated, however this has usually been with SRF cavities and cryomodules developed for, or adapted from, other purposes such as CEBAF or TESLA. Even dedicated ERL demo machines such as the KEK compact ERL and the Cornell ERL injector/ ERL demo project derive their frequency and much of their DNA from the TESLA collaboration technology. JLab's ERL based FEL was also based closely on the CEBAF technology, although a new high current upgrade design was proposed but never funded. PERLE has the opportunity to be a clean-sheet globally optimised design for a new generation of high average power efficient ERL based machines. It will be an ideal facility for testing advanced concepts in cavity design, surface treatments, HOM damping, couplers, tuners, microphonics, etc., as well as emittance preserving optics, multi-pass and high dynamic range diagnostics, instability suppression and feedback, advanced LLRF techniques, etc.
\subsection{High quality SCRF cavity - status and tests}
There has been much progress in SRF cavity design and processing in recent years, stimulated by projects like ILC, XFELs, factory-type colliders, light sources and ADS. This has triggered a diversification of designs, materials, techniques, and applications and no longer does any project have to depend on a set frequency or cell design just because of history or convenience. There now exist in many places around the world the knowledge, experience and tool sets to design, build, test and integrate fully customized and optimized SRF designs for new and exacting requirements. Recent examples include crab cavities for short pulse X-ray sources and colliders, HOM-damped cavities for $e^+e^-$ colliders, high power proton linacs for ADS, etc.
The cavity shape optimisation for ERLs is somewhat different than for high-gradient pulsed linacs. The CW operation and potential for high circulating currents require careful attention to heat load (both from RF losses and field emission) and beam break up.  In this regard a balance needs to be found between peak electric and peak magnetic fields while maintaining good efficiency and, very importantly, keeping HOMs well away from strong harmonics of the beam current. Because the ERL beam current spectrum depends strongly on the filling pattern and recirculation time, some assumptions must be made about machine operation when examining the HOM spectrum. This is discussed further in section 3.3.2. An important parameter in maintaining good HOM damping is to have strong cell-to-cell coupling. This allows HOMs to propagate easily to the end cells, where the dampers are typically located, and makes the cavity less sensitive to tuning and fabrication errors. In particular it minimises the possibility for HOMs to become trapped in the cavity center or tilted away from HOM couplers. Stronger cell-to-cell coupling implies a larger iris between cells, whereas efficiency is favoured by a smaller iris, so a compromise must be reached. Dangerous HOMs can be detuned if necessary by altering the profile of the cell. The gradient and impact energy of the cell multipacting barrier can be calculated and it is prudent to avoid operating close to this gradient. The impact energy can be minimised by flattening the cell profile in the equator region to make the barrier softer and easier to transition or process away. 
\subsection{Cavity module - principle and tests}
The cryostat is the less glamorous cousin of the cavity and is often something of an afterthought, despite being the major share of the cost of the cryomodule. Previous SRF ERLs have used or adapted cryostats from other projects, in some cases converting them from pulsed to CW operation. Some important considerations are pressure code compliance, static heat load, maintainability and operability and cost. The number of magnetic and thermal shields and intercepts, the mechanical support and alignment scheme and whether the linac is continuous (like ILC) or segmented (like CEBAF and SNS) are all variables. For a large machine like LHeC it is worth performing a careful evaluation or even a new, clean sheet design optimised for this purpose, however, for a test machine like PERLE it is advantageous to use an existing well proven design. For this study we have used the SNS style cryostat as it can easily accommodate the 805 MHz 5-cell beta=1 cavities with very minimal modifications, has plenty of heat load capacity, is a segmented design allowing phased construction of the facility and ease of maintenance, and has existing tooling and operational experience. More details are presented in section 3.3
\subsection{Goals of the ERL design and operation}
The purposes of the PERLE ERL demonstrator are to provide flexible test beams for component development, low energy physics experiments, and also to demonstrate and gain operational experience with low-frequency high-current SRF cavities and cryomodules of a type suitable for scale up to a high-energy machine. Since the cavity design, HOM couplers, FPC's etc. will be all new or at least heavily modified, PERLE will serve as a technology test bed that will explore all the parameters needed for a larger machine. There is no other high current ERL test bed in the world that can do this. PERLE will also feature emittance preserving recirculation optics and this will also be an important demonstration that these can be constructed and operated in a flexible user-facility environment. The machine must run with high reliability to provide test beams for experimenters or ultimately provide Compton or FEL radiation to light source users. This demonstration of stability and high reliability will be essential for any future large facility.

\section{Technical Applications}
An intense beam facility will offer new opportunities for auxiliary
applications. In view of a possible placement of PERLE 
in the vicinity of or even at CERN
various test options have been studied and results are described subsequently of simulations dedicated to the possibility for
beam based investigations of quench levels of superconducting magnets and cables. 
As is also sketched below, PERLE may offer versatile possibilities for
tests of cavities with different frequencies with a suitably chosen injector frequency.
With, for example, a 12.146 MHz injector,  one may test cavities for frequencies including
values of 352, 401, 704, 802 and 1300 MHz, which are of direct
interest for CERN's Linac4 and ESS, FCC, ESS, LHeC and FCC, and the ILC, respectively. 

\subsection{Magnets, cables, quench tests}
Understanding the quench levels of superconducting cables and magnets is important for an efficient design and the safe and optimal operation of an accelerator using superconducting magnets. Quench levels are used as an input to define requirements for controlling beam losses, therefore influencing e.g. beam cleaning and collimation, beam loss monitor positions and thresholds, interlock delays etc..

The quench level defines the maximum amount of energy that can be deposited locally in a superconducting magnet or cable to cause the phase change from superconducting to normal-conducting state. The quench level is a function of the energy deposition distribution and the duration of the impact, the local temperature before the impact, the cooling capacity, and the local magnetic field.

State of the art electro-thermal solvers, which are used to predict the quench levels of superconducting cables and magnets, are mainly based on lab experiments without beam.
To verify their predictions in case of beam impact, quench levels have been extensively studied with beam in the LHC at the end of Run 1 in February 2013. The results for short duration ($< 50 \mu s$) and steady state ($> 5 s$) energy deposition are in good agreement with predictions based on electro-thermal simulation codes like QP3~\cite{QP3} and THEA~\cite{THEA}. For intermediate duration energy depositions the electro-thermal models predict a factor 4 lower quench levels than found during the experiment~\cite{Auchmann+2015}, which still needs to be understood.

Currently the LHC is the only accelerator at CERN, where quench tests with beam can be performed for all relevant time scales. Nevertheless, the LHC is not an adequate test bed to perform quench tests as:
		i) only magnets installed in the LHC can be tested, ii)
	 non-trivial beam dynamic studies are required to interpret experimental results and iii)
	 the LHC is a sophisticated accelerator which is ultimately optimized to deliver luminosity to the particle physics experiments.
The other facilities at CERN either lack the availability of cryogenics (PS, HiRadMat) or the particle beams (SM18). Furthermore, using the fast extraction from the SPS the HiRadMat facility could only cover the regime of short duration energy deposition. Therefore, a dedicated facility equipped with cryogenics to perform quench tests is required.

\subsubsection*{Energy deposition studies}
\begin{figure}\center
\includegraphics[width=0.49\columnwidth]{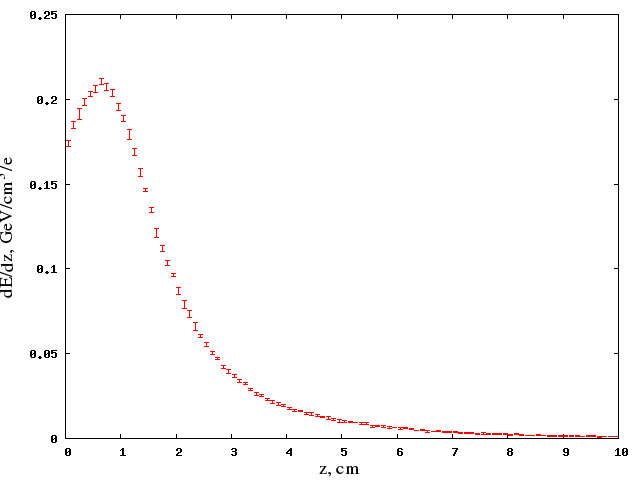}
\includegraphics[width=0.49\columnwidth]{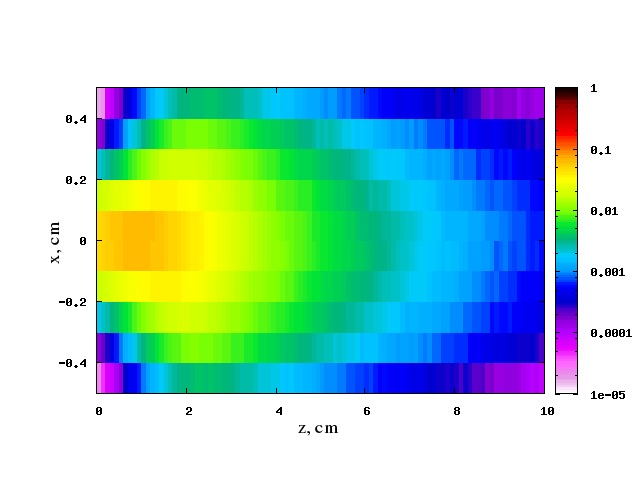}
\caption{Maximum values of energy deposition (left) and projection of energy deposition (right) for $150$\,MeV electrons impacting in a solid copper block as calculated by FLUKA~\cite{FLUKA, FLUKA2}. An emittance of $50$\,$\mu$m and a beta-function at extraction of $5$\,m was used. The bin size was 1\,mm$^3$.}
	\label{fig:EnDep150MeV}
\end{figure}

\begin{figure}\center
\includegraphics[width=0.49\columnwidth]{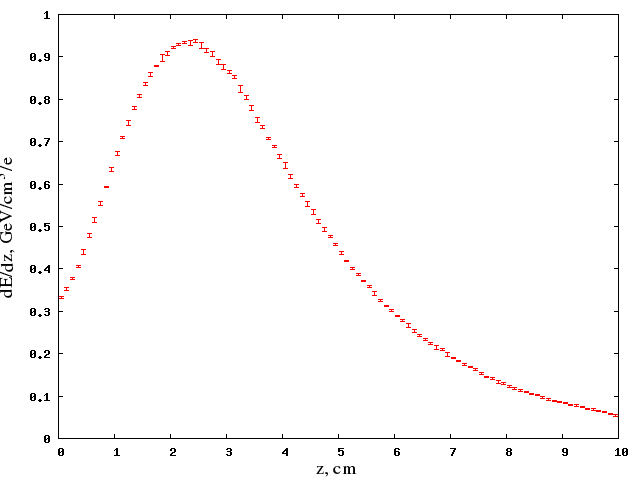}
\includegraphics[width=0.49\columnwidth]{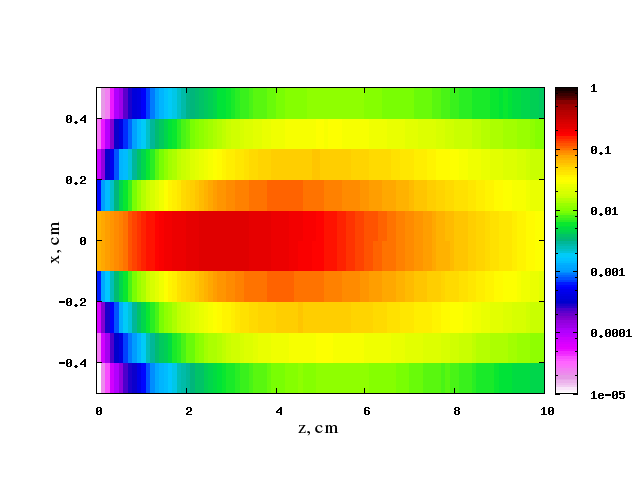}
\caption{Maximum values of energy deposition (left) and projection of energy deposition (right) for $1$\,GeV electrons impacting in a solid copper block as calculated by FLUKA~\cite{FLUKA, FLUKA2}. An emittance of $50$\,$\mu$m and a beta-function at extraction of $5$\,m was used. The bin size was 1\,mm$^3$.}
	\label{fig:EnDep1GeV}
\end{figure}

\begin{figure}\center
\includegraphics[width=0.85\columnwidth]{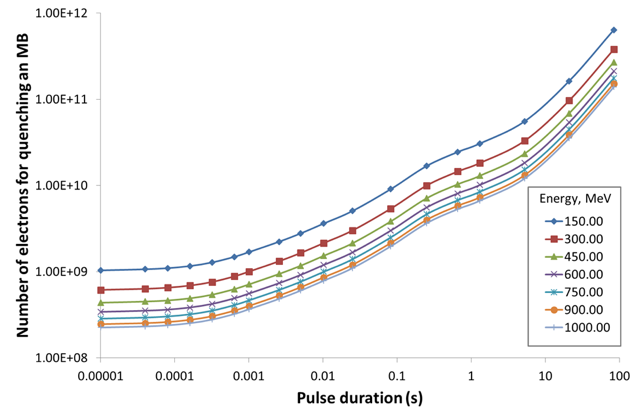}
\caption{Amount of impacting particles versus pulse length to reach the quench level of a LHC main dipole. The energy density distribution is taken from the FLUKA simulations shown in Fig. \ref{fig:EnDep150MeV} and \ref{fig:EnDep1GeV}.}
 \label{fig:PartNumber}
\end{figure}

Figures~\ref{fig:EnDep150MeV}, \ref{fig:EnDep1GeV} show the energy deposition per primary electron in a solid copper target for 150~MeV and 1~GeV electrons, respectively. For the simulations an emittance of 50~$\mu$m and a beta-function at extraction of 5~m was considered. The bin size was 1~mm$^3$. Combining the peak energy deposition with the quench levels for the LHC main dipoles, as calculated by QP3, the number of primary particles required to reach quench levels for different durations of the energy deposition can be derived. Figure \ref{fig:PartNumber} summarises the required number of primary particles in case of different particle energies and pulse length durations.

Comparing these numbers to the baseline beam parameters shows that PERLE can provide sufficient beam to perform quench tests during all stages of its construction. It is important to assure in a subsequent detailed design process that the facility can provide fast and slow extracted beams to the quench test experiments, to allow for experiments in all energy deposition duration regimes.

\subsubsection*{Quench test facility}
Besides  a high energy electron beam, the quench experiments require a dedicated facility. The detailed design and space requirements of such a facility change strongly depending whether it should allow for testing full size magnets like the LHC dipoles or if testing of cable and short magnet samples would be sufficient.
In both cases such a facility requires power converters, which deliver currents up to $\sim 25$~kA to power the samples and possible solenoid magnets providing external magnetic fields. Furthermore instrumentation racks for quench protection, measurement of voltages, temperatures and other parameters are required. Most importantly it requires a dedicated cryogenic installation, to avoid impacting the operation of PERLE.

One may start with a facility for testing cable samples and short sample coils, which, at a later stage, can be extended with a test bench to perform quench tests with full size magnets, as it is e.g. done at CERN in the  SM18 test area. The space and power requirements of the final facility have to be taken into account from the beginning.


\subsection{Cavity tests at different frequencies}
PERLE is described below in a default configuration including cavities at $801.58$\,MHz in up to $4$ cryomodules and a bunch spacing of 
$25$\,ns. To gain flexibility and widen its potential as a development facility for testing cavities and cryomodules with beam, PERLE may, however, also be configured to a number of different frequencies, especially those which are
commonly used in accelerator facilities world-wide, i.e. 
$352$\,MHz (Linac4, ESS), $401$\,MHz (LHC, FCC), $704$\,MHz (ESS), the PERLE default $802$\,MHz (LHC, FCC and LHeC) 
and $1300$\,MHz (ILC, XFEL, ...). To make this possible, the injector must be based on a photocathode with a laser pulser that can be operated at  $f_0 = 12.146$\,MHz with a buncher/booster system adjusted to a harmonic of $f_0$. The frequency of 
$12.146$\,MHz is chosen as a joint sub-harmonic of these commonly used frequencies. The exact harmonic frequencies accessible as PERLE's main RFs are given in Table\,\ref{taberk11}, assuming the possibility to tune the subharmonic $f_0$ by moderate variations of  $\pm~4$\, kHz. This assumption translates to certain tuning range, for example at
$801.58$\,MHz of $\pm 0.26$\,MHz, which would have to
be implemented in the buncher/booster.
\begin{table}
\begin{center}
\begin{tabular}{lccccc}
\hline
h &	29	& 33 &	58 &	66 &	107 \\
\hline
h $\cdot (f_0$  -4 kHz) &	352.118 &	400.686 &	704.236 &	801.372 &	1299.19 \\
h $\cdot (f_0$ +4 kHz)	  &    352.350  &	400.950 &	704.700 &	801.900 &	1300.05 \\
\hline
\end{tabular}
\end{center}
\caption{Main RF ranges for selected harmonics, in MHz, accessible to PERLE with an injector pulsed at $f_0=12.146$\,MHz, a configuration suitable for beam based RF developments at most commonly used frequencies with this facility.}
\label{taberk11}
\end{table}

Referring to the description of source and injector in 
Sect.\,\ref{sec:sourceandinjector} 
below, it is clear that a bunch repetition frequency of $40.1$\,MHz ($25$\,ns) is not compatible with most of the above frequencies and should be adapted to either $12.146$\,MHz, where it could be used with all mentioned frequencies, with the caveat that a larger bunch charge would have to be generated for a similar average current (challenging $1$\,nC for $12$\,mA). For tests at $401$\,MHz and $802$\,MHz, however, a bunch repetition frequency of $36.438$\,MHz can be chosen, which would be close enough to the LHeC parameters to be relevant. It would produce $12$\,mA of beam current with $329$\,pC bunch charge. The filling scheme and bunch recombination pattern, see Sect.\,\ref{bunchpat} 
(Figs.\,\ref{FIG_REC_PATTERN_1},\,\ref{FIG_REC_PATTERN_2})
would have to be adapted $mutatis~mutandis$ (the harmonic $20$ becoming harmonic $22$) with individual bunch spacings $7 \lambda$\,-\,$8 \lambda$\,-\,$7 \lambda$. The buncher/booster system described in Sect.\,\ref{sec:sourceandinjector}  remained unchanged. It is noteworthy that for a bunch repetition frequency of $12.146$\,MHz, captured and accelerated in this booster at $801$\,MHz, the frequency of the cavities in the ERL might still be $704$\,MHz and $1300$\,MHz, and even the simultaneous operation at different frequencies in the same linac would not be impossible.

\section{Injector for the LHeC} \label{injectorLHeC}
In the course of the PERLE development, it had been studied
whether a suitably modified PERLE facility could serve as an injector 
to the LHeC eventually.
From the beam dynamics point of view, many parameters are shared between the PERLE and the LHeC designs (emittance, bunch spacing, beam current...). 
When operated as an injector, PERLE would need to deliver beam without energy recovery, as the highly disrupted beam from the LHeC cannot accept a further deceleration.
In the Higgs factory configuration, the LHeC requires bunches up to \SI{640}{pC} at an energy of \SI{500}{MeV} which results in an average beam power at injection of about \SI{10}{MW}. Assuming that the cavity design can handle such power flow, this would nevertheless drive the requirements for the klystrons and power converters, requiring new sets of them.

Concerning the layout, PERLE could, for example, be reconfigured keeping only two passages and lowering the accelerating field to \SI{125}{MV/linac} in order to balance the power between the two of them. 
Further considerations have to be made:
\begin{itemize}
\item The LHeC requires continuous beam injection, therefore other applications of PERLE would be relegated to the LHeC downtime, thus disrupting its user program;
\item If PERLE would be located at ground level on the CERN site, a some-hundred-metres tunnel, with a reasonable slope, has to be dug from the location of  PERLE to the LHeC tunnel. A kilometre-scale transfer line will probably be needed to transport the beam to the LHeC injection chicane.
\end{itemize}
 
It should be noted that with the PERLE accelerating gradient of \SI{15}{MV/m}, an active length of just \SI{33}{m} is required to reach the LHeC injection energy even without recirculation. A dedicated linac, placed in a \SI{\sim 100}{m} tunnel close to the LHeC injection chicane could be a preferable option. The possibility to reuse PERLE components for this machine could be taken into account.
It so seems less preferable, though possible, to consider the genuine
PERLE facility when located at CERN as an injector to the LHeC. 

\section{Physics with Electron Beam}
Elastic $ep$ scattering has been of fundamental
importance since, now 60 years ago, 
it lead to the discovery of a finite 
radius of the proton of about $1$\,fm by 
Hofstadter~\cite{Hofstadter:1956qs}.
This process has a major revival as recent determinations of the
proton radius with electrons and muons strongly
disagree, see below. With its outstanding luminosity
and large energy range, hugely interesting opportunities
open up with PERLE measurements of unprecedented precision.
These, as sketched below, concern measurements of the scale
dependence of the electroweak mixing angle, sin$^2\theta$, of
the electric and magnetic formfactors, $G_E$ and $G_M$, of
hyperon physics and searches for physics complementing the Standard
Model. New physics may appear 
in loop corrections or in direct manifestations of new particles,
for which dark photons, leading to the reaction 
$e^-A \rightarrow e^+e^-e^-A$, are currently a prime 
example~\cite{Essig:2013lka,ArkaniHamed:2008qn}.

Following a brief recollection of the elastic scattering characteristics
and the luminosity prospects of PERLE, three 
interesting physics applications are illustrated subsequently
i) the potential for weak interaction
measurements using polarised $e^-p$ scattering; ii) a discussion of
the status and possibilities for new precision measurements of
the proton form factors, pion production and iii)
 the search for light dark matter and new physics.
\subsection{Elastic ep scattering and luminosity}
For a given electron beam of energy, $E$, scattered off a fixed proton 
target, the elastic $ep$ cross section depends only on the polar
angle $\theta$ of the scattered electron. This determines both the
negative four-momentum transfer squared, $Q^2$, and the energy $E'$ of
the scattered electron through the relations
\begin{equation}
Q^2 =\frac{2ME^2(1-\cos\theta)}{M+E(1-\cos\theta)}~~~~~~~~~~E'=\frac{E}{1+\frac{E}{M}(1-\cos\theta)},
\end{equation}
where $M$ is the proton mass. The cross section, in its Born approximation, is given as the
product of four factors, the Rutherford formula, the Mott electron spin modification,
a correction, equal to $E'/E$, for the proton recoil and finally a function $f(G_E,G_M,\theta)$, which 
characterises the spin  and the spatial extension of the proton
\begin{equation}
\label{eq:sig}
\frac{d\sigma}{d\Omega} = \frac{\alpha^2}{[E(1-\cos\theta)]^2} \cdot \cos^2\frac{\theta}{2}
 \cdot \frac{1}{1+\frac{E}{M}(1-\cos\theta)} \cdot f(G_E,G_M,\theta),
\end{equation} 
with $\alpha$ the fine-structure constant. With the convention $\tau = Q^2/4M^2$ the 
form factor term is given by
\begin{equation}
 f(G_E,G_M,\theta) = \frac{G_E^2 +\tau G_M^2}{1+\tau} + 2 \tau G_M^2 \tan^2 \frac{\theta}{2}.
\end{equation} 
To some first approximation, one has  $G_M= \mu_p G_E$ and $G_E=1/(1+Q^2/0.71GeV^2)^2$, with the 
anomalous magnetic moment $\mu_p$ of the proton. The two form factors
$G_E$ and $G_M$ can be separated through
a variation of the energy following Rosenbluth. This should be an advantage of PERLE
as with its variable energy it may cover a large 
range from a few hundreds of MeV to almost 1\,GeV.
The formulae above are sufficient for 
practical estimates of counting rates, but neglect all the physics which is contained
in corrections to Eq.\ref{eq:sig} as arise from electroweak, BSM and higher order QED effects.

The luminosity of a facility like PERLE is obtained as
$ L = \rho l N_A N_e$. For a hydrogen target of 
density $\rho = 0.07$\,g cm$^{-3}$ and length  
$l=10$\,cm one gets $L= 4.3 \cdot 10^{23}$\,cm$^{-2}$ $N_e$. 
For a source  delivering 320\,pC of charge and a
25\,ns bunch spacing one obtains a current of 12.8\,mA corresponding to 
about $8 \cdot 10^{16}$e\,s$^{-1}$, or 
a number of electrons per bunch of $N_e=2 \cdot 10^9$.
As a consequence the luminosity for elastic $ep$ scattering can be expected to be
as high as $3 \cdot 10^{40}$\,cm$^{-2}$s$^{-1}$ with a 10\,cm  proton
target. 


%
\subsection{Parity violation and the Weinberg angle}
\begin{figure}[H]
 \centering
  \includegraphics[width=0.85\textwidth]{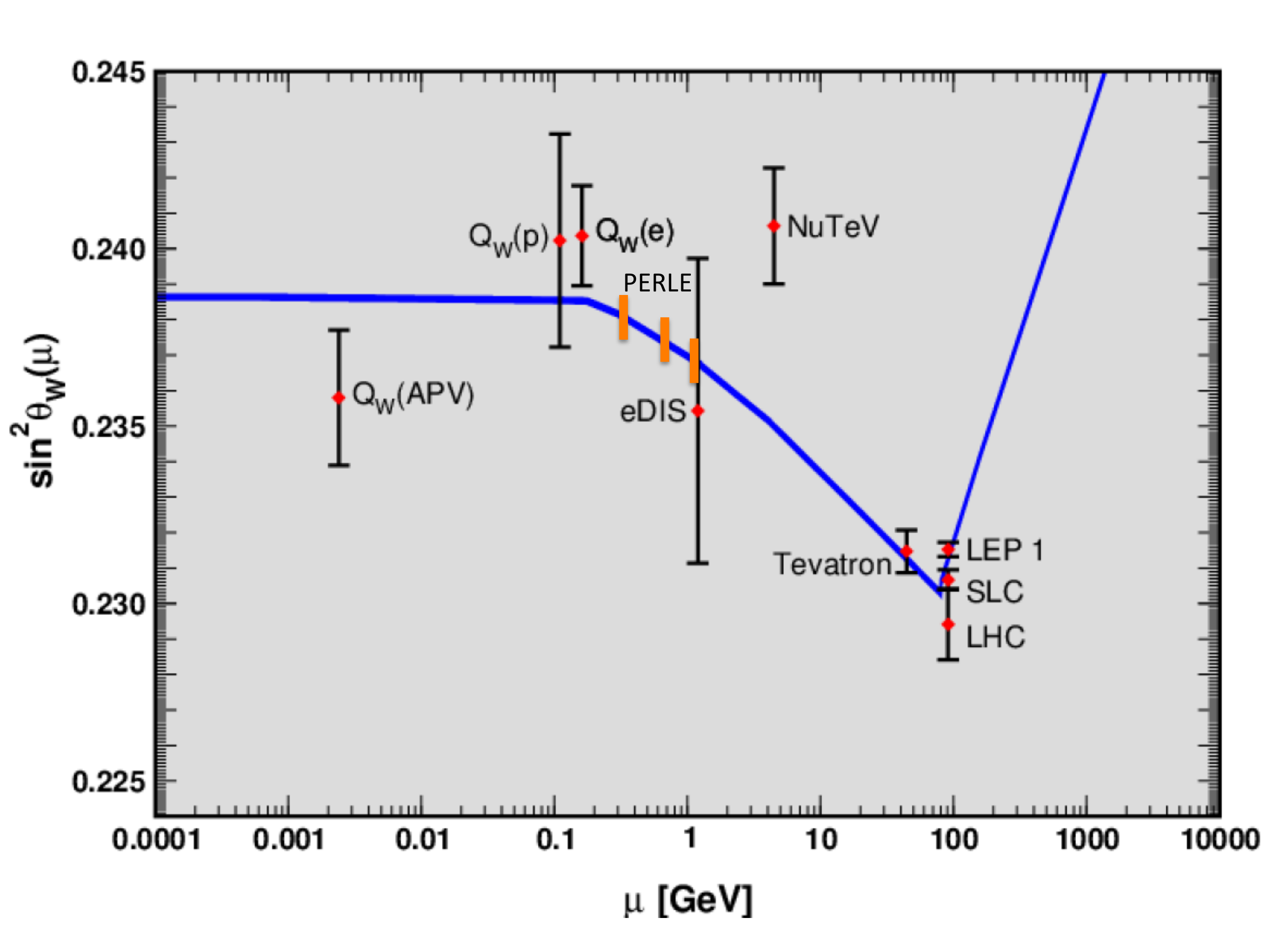}
\caption{Prospect for the measurement of the weak mixing angle
with PERLE (illustration of half a percent accuracy
measurement) based on the polarisation
asymmetry $A^-$, as compared to the current status
of sin$^2\theta_W$ measurements, from PDG2014.}\label{Fig:sinusth}
\end{figure}

The unification of the electromagnetic and weak interactions within the SU(2)$_L$xU(1)
theory is expressed by the Weinberg angle $\sin^2 \theta_W$, which has a strong characteristic
dependence on the momentum scale ($\sqrt{Q^2}$ in ep scattering) due to loop 
corrections~\cite{Erler:2004in}
to the tree-level expressions, see Fig.\,\ref{Fig:sinusth}.
 The most precise $\sin^2\theta_W$ measurements so far were  performed at the
$Z$ pole at LEP and SLC, leading to an unresolved discrepancy of about three standard deviations.
Various measurements of so far limited precision were performed at low scales, with a departure 
from theory observed by the NuTeV Collaboration in $\nu N$ scattering which caused a multitude of
subsequent considerations as on the amount of strange quarks in the nucleus and the behaviour of
nuclear corrections. Measurements of the mixing angle are very complex challenges and lead
to new insight often beyond the genuine intention to determine $\sin^2\theta_W$.
Measurements with the LHeC (FCC-he), as presented in the 
LHeC CDR~\cite{Abelleira2012},
 will be based on very large electroweak asymmetry effects 
and determine the electroweak mixing angle precisely for a range
below the $Z$ mass up to high scales of 1 (3) TeV.

With PERLE one can access effects from $Z$-boson exchange with polarised electron scattering,
as well as with charge asymmetry measurements, for $\sqrt{Q^2}$ between about
$0.1$ and $1$\,GeV. The intensity of a polarised electron source
is probably an order of magnitude  higher than that of a positron source.
This makes the measurement of a polarised electron scattering asymmetry, $A^-$ more likely than that
of a charged or combined charge and polarisation asymmetry, $B$. Both have been discussed in 
\cite{Klein:1979dy}. The polarisation asymmetry can be expressed as
\begin{equation}
A^-(P,P') = \frac{\sigma(P) - \sigma(P')}{\sigma(P)+\sigma(P')} = - \kappa \frac{P-P'}{2}
 \cdot (v_eA - a_e V) 
\end{equation} 
where $\kappa = Q^2 G/\sqrt{2} 2 \pi \alpha$ determines the size of the asymmetry to be
O($10^{-4}Q^2/$GeV$^2$).
Here $v_e$ and $a_e$ are the weak neutral current (NC) couplings of the electron and
$V$ and $A$ are new combinations of the form factors $G_E$ and $G_M$ which also depend
on the quark NC couplings as well as the charged current axial vector form factor.
Evidently, the asymmetry $A^-$ is different from zero through parity violation.
With PERLE, it allows to measure the mixing angle in a particularly interesting
range of scale, as is illustrated in Fig.\,\ref{Fig:sinusth}. 
Besides providing a measurement of $\sin^2 \theta_W$, with $ep$ scattering asymmetries, one
accesses also new combinations of quark couplings. Following \cite{Klein:1979dy} one
sees, for example, that the hadronic axial vector factor $A$ determines
a combination of $a_d+3.55a_u$ which can be compared with ep scattering at HERA and the LHeC
where $A = a_d-2a_u$. 

\begin{figure}[H]
 \centering
  \includegraphics[width=0.85\textwidth]{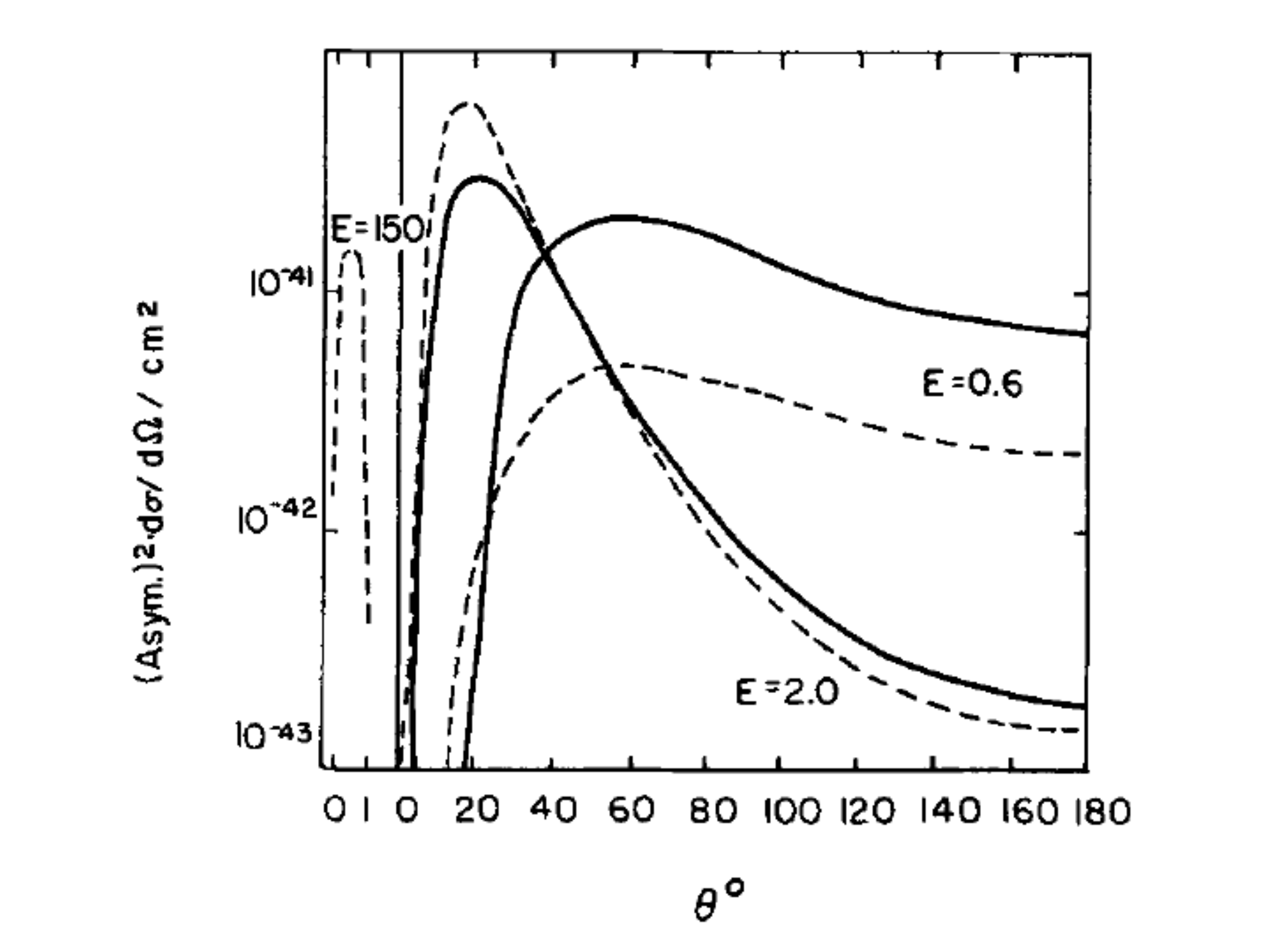}
\caption{Variation of the statistical accuracy represented
as asymmetry squared times cross section  in cm$^2$ for two
kinds of asymmetry, solid: beam charge conjugation and
dashed: polarisation, from \cite{Klein:1979dy}.}\label{Fig:statasy}
\end{figure}

The measurement accuracy depends on the beam energy and scattering kinematics. This is illustrated in
Fig.\ref{Fig:statasy}. Since the asymmetries vanish at small angles while the cross section decreases
towards larger angles, an optimum is observed, with striking variations. One finds for a beam  energy 
$E \sim 1$\,GeV that asymmetry measurements at $\theta \sim 30-90^{\circ}$ can be expected
to be especially precise. 

The measurement of the weak mixing angle at small scales
is an area of vigorous activity, because of the new level of
precision anticipated in a coming generation of tests of
its predicted scale dependence, as at Mainz and Jefferson Lab,
and because of the relation these measurements have
to new physics such as rare Higgs decays and dark $Z$ bosons,
see~\cite{Davoudiasl:2015bua} and references therein.
The salient potential of the here presented ERL facility consists in its
potential large energy coverage and particularly high luminosity which make further studies of the possibility to measure
that process with PERLE interesting indeed. 

\subsection{Proton form factors}
The proton electromagnetic form factors, $G_E$ and $G_M$, which have been studied for many decades, have become the focus of recent research mainly due to the proton radius puzzle, recognised even in the popular press \cite{Bernauer2014}. It is the more than $7\sigma$ discrepancy between the determination of the proton radius with electrons ($r_E=0.8775(51)\ \mathsf{fm}$ \cite{Mohr2012}) and using muon spectroscopy ($r_M=0.84087(39)\ \mathsf{fm}$). Since its observation in 2010, the discrepancy has sparked large work efforts on both the experimental and theoretical side, but no widely accepted explanation has yet been found.

On the electron side, both spectroscopy and scattering experiments agree. In the latter, the radius is extracted from the slope of the form factors at $Q^2=0$. Since data can only be taken at finite $Q^2$, the form factors have to be extrapolated to 0. Currently, the most precise data set from 
scattering experiments~\cite{Bernauer2010,Distler2010,Bernauer2013} has
 been measured by the A1 collaboration in Mainz at the MAMI accelerator. It contains more than 1400 measured cross sections and reaches closest to the static limit with $Q_{min.}^2\approx 0.003\ (\mathsf{GeV}/c)^2$. While there are no structures/changes of curvature expected below this point, it is not possible to rule them out. Such structures would invalidate the extrapolation and may resolve part of the puzzle.

\begin{figure}[t]
  \centering  \includegraphics{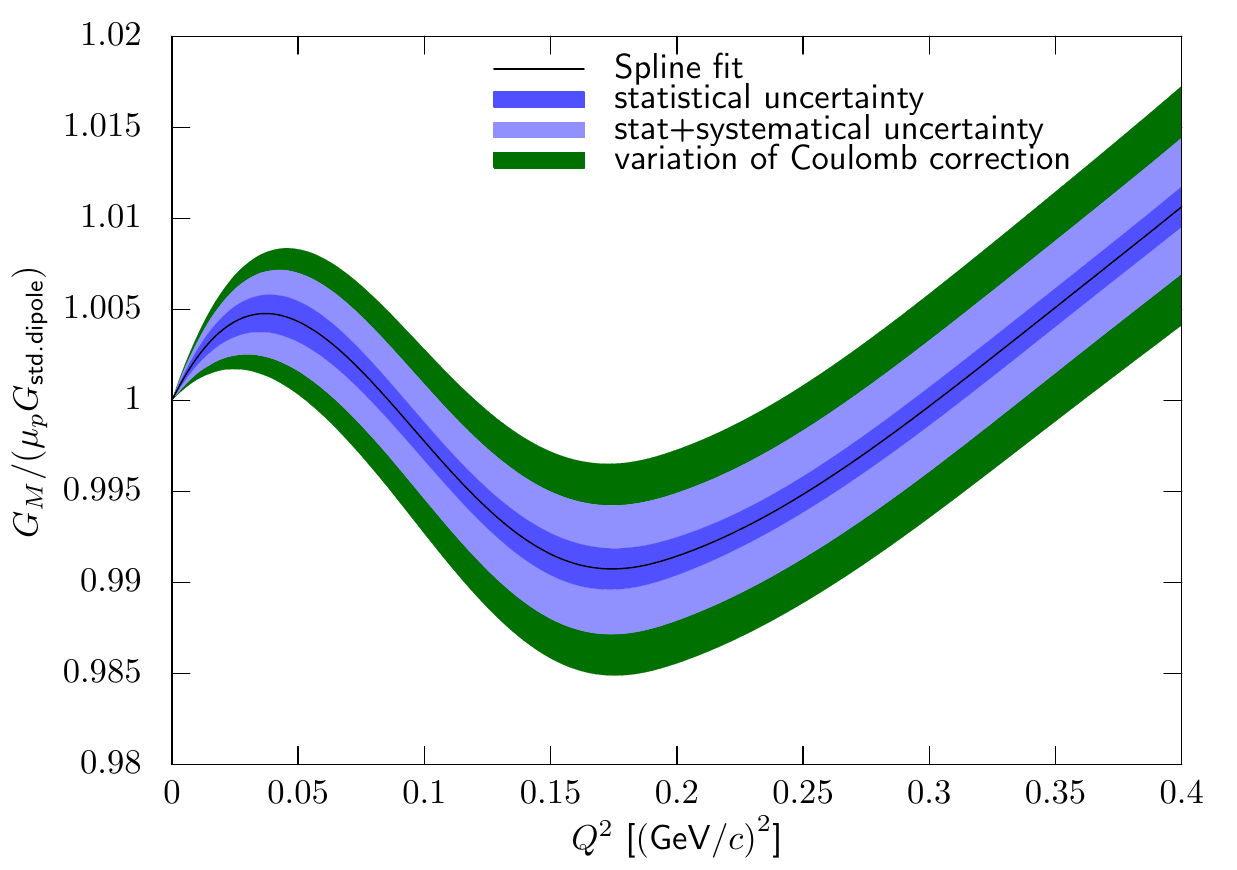}
  \caption{\label{gmstructurelowq} The fits in \cite{Bernauer2013} for the magnetic form factor $G_M$, divided by the standard dipole, exhibit a maximum-minimum structure at low $Q^2$. While the local minimum around $0.2\ (\mathsf{GeV}/c)^2$ is seen in earlier fits, the local maximum around $0.03\ (\mathsf{GeV}/c)^2$ has not been observed before.
  }
  \end{figure}
\begin{figure}[h]
  \centering  \includegraphics{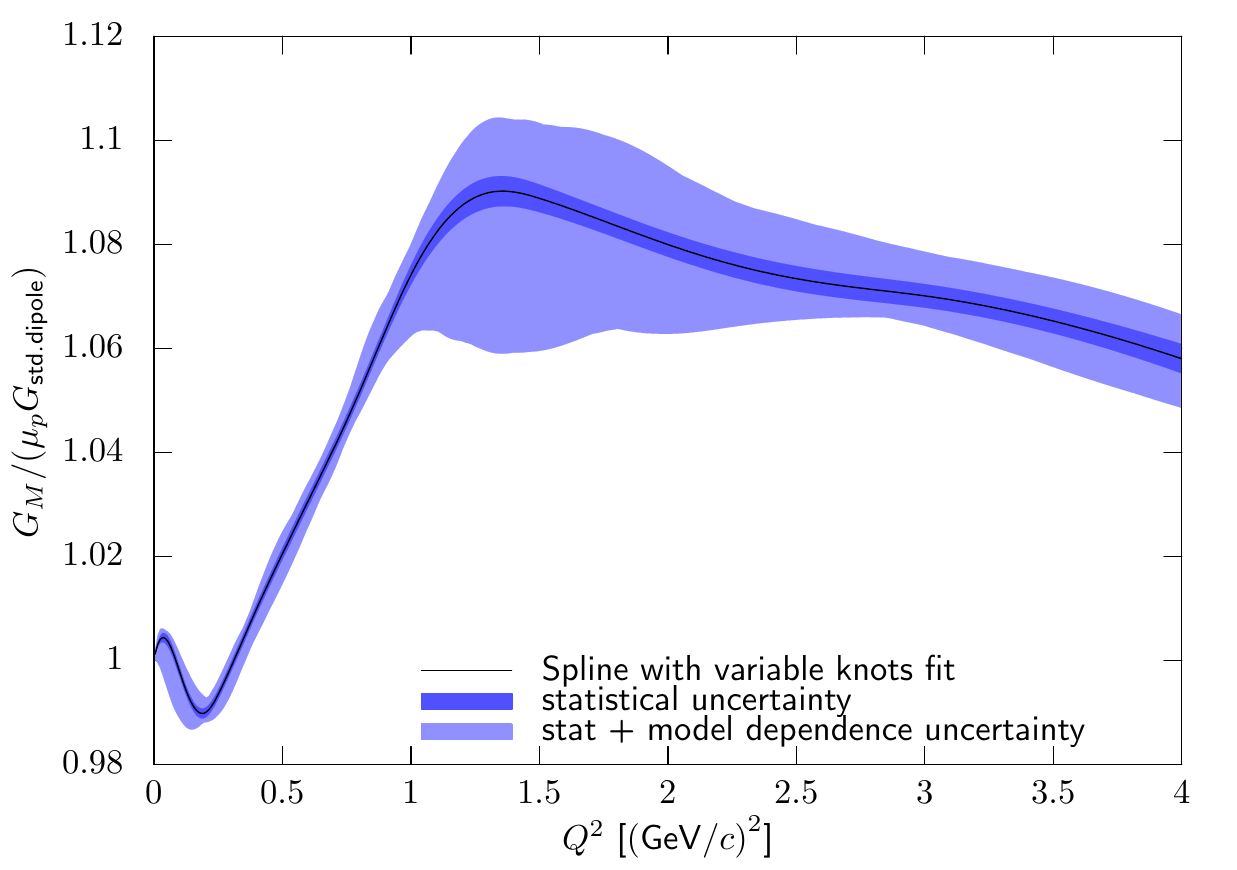}
  \caption{\label{gmstructurehighq} Fits to the world data (here from \cite{Bernauer2013}) for the magnetic form factor $G_M$, divided by the standard dipole, show a cusp or strong bend between $1$ and $1.5\ (\mathsf{GeV}/c)^2$. The exact shape strongly depends on the form factor model used to fit the data.  }
 \end{figure}

This data set also found an interesting structure in $G_M$ at low $Q^2$, shown in Fig.\ \ref{gmstructurelowq}. The magnetic form factor, divided by the standard dipole, exhibits two local extrema. While the minimum is found in earlier extractions, the maximum has not been seen and is in fact below the resolution of previous data. This leads to a significantly different magnetic radius compared to earlier findings. The strength of the maximum is strongly affected by radiative corrections and could be a statistical aberration.  An external validation is important as the existence of structures like this points to corresponding length scales in the physics inside the proton.

Fits to the world data set exhibit a cusp around $Q^2=1.5\ (\mathsf{GeV}/c)^2$ in $G_M$, shown in Fig.\ \ref{gmstructurehighq}, again pointing to underlying length scales in the internal structure of the proton. However, the cusp is only visible in the combination of multiple data sets and could be an artefact.

PERLE could provide crucial new high-precision data to study these three phenomena using different experimental approaches:
\begin{itemize}
\item Possible structures below $Q^2_{min}$ and their influence on the proton radius could be studied with a single, low beam energy and forward scattering experiment, similar to the PRad experiment \cite{Gasparian2014}. At lower energies and higher beam currents than planned for PRad, an ERL beam with a point-like target (e.g.\ a gas jet) could provide higher rates and smaller systematic uncertainties. An alternative approach is to exploit initial state radiation, measuring deep into the radiative tail to probe $Q^2$-values that are orders of magnitude smaller than directly accessible. This approach is described in more detail e.g.\ in \cite{Mihovilovic2014}.
\item
The low-$Q^2$ structure in $G_M$ could be studied in an experimental setup similar to \cite{Bernauer2010}. The interesting region in $Q^2$ would be covered by performing an angular scan of the cross section and multiple energies up to 300 MeV. Such an experiment would benefit substantially from a point-like target without target walls, which are the main background of \cite{Bernauer2010}. It would produce an electric radius with similar uncertainties, and a magnetic radius with substantially improved precision compared to current results.
Additionally, with a polarised beam and target, an asymmetry measurement, sensitive to the ratio $G_E/G_M$, could be performed. Such a measurement would help to disentangle $G_E$ and $G_M$ from the cross-section measurement and would make it possible to study whether the structure is related to imperfect radiative corrections.
\item
The high-$Q^2$ structure could be studied with high precision using beam energies of 1 GeV and up, possibly with just one angular scan of the cross section at a fixed energy around 1.3 GeV. Without a good connection to lower beam energies, the precision of the absolute normalisation is not likely to better than a few percent, however the cusp structure is large enough that a good relative normalisation of the data points, e.g.\ using a detector at forward angles as a luminosity monitor, is enough to extract a meaningful result.
\end{itemize}

\subsection{Pion electroproduction}
Using virtual photon tagging, it is possible to study confinement-scale QCD. In photo-production, the photon tagger sets the rate limit and only a small fraction of the tagged photons interact with the target, leading to low data-taking efficiency. At forward angles, the virtual photons are almost real, so that a forward scattering electron tagger can be used as a highly efficient substitute. Because of the high efficiency and high beam currents, it is possible to use pure, thin targets and detect low energy recoil particles which would not escape traditional, thick targets. It is thus possible to measure the reactions $\gamma p\rightarrow \pi^0p,\pi^+n$, $\gamma n\rightarrow \pi^0n,\pi^-p$ and $\gamma D\rightarrow \pi^0D$. Coherent $\pi^0$ production in $D$ and $^3He$ measure relative signs of the $\gamma p\rightarrow \pi^0p$,$\gamma n\rightarrow \pi^0n$ amplitudes.

Such an experiment requires beam energies of 300 MeV or more. Depending on the target, beam current and polarization capabilities, different experiments are possible:
\begin{itemize}
\item With about 1\ mA unpolarized beam, a measurement with a thin, windowless, unpolarized gas target, detecting either the $\pi^+$ or the recoiling proton, could be performed. This would allow a test of $a_{nn}=a_{pp}$ and few-body calculations via $\gamma D\rightarrow nn\pi^+$, and also check $a_{np}$ with $\gamma D\rightarrow np\pi^0$. It would further be possible to test isospin conservation by testing
  $$ A(\gamma p\rightarrow \pi^+n)+A(\gamma n\rightarrow \pi^-p)=\sqrt{2}[A(\gamma n\rightarrow \pi^0n)-A(\gamma p\rightarrow \pi^0p)].$$
  
  \item At about 100\ mA unpolarized beam with a windowless transverse polarized gas target, one could test isospin breaking through a measurement of $\gamma N\rightarrow \pi^0N$ near threshold.
\end{itemize}
For more information, see e.g.\ \cite{Bernstein2013}.

\subsection{Light dark matter}
The search for new physics beyond the Standard Model is a major focus of the nuclear and particle physics community. A simple extension of the SM Lagrangian \cite{Holdom1985,ArkaniHamed2008} leads to new ``dark'' Abelian forces with a new dark gauge field $A^\prime$.
Among many others, a possible production mechanism is $e^-p\rightarrow e^-pA^\prime(\rightarrow e^-pe^+e^-)$, i.e.\ the elastic scattering with a radiated ``dark'' photon, and the possible  subsequent decay of the radiated $A^\prime$ into a lepton pair (``visible decay'')
The DarkLight experiment \cite{Balewski2014}, planned to be run at the Jefferson Lab ERL, aims to search for these visible decays in the region preferred by the muon g-2 results, detecting all four outgoing particles. A variant also looking for invisible decays is planned \cite{Kahn2012}.
The PERLE facility could be an option for a version 2 of the experiment, with increased luminosity.

Alternatively, with high-precision, high-rate detectors measuring just the recoiling proton and electron, it should be possible to mount a competitive search sensible to both visible and invisible decays. More work is needed to study this further.

\subsection{Speculative ideas}
At $Q^2$ above $1$\,GeV$^2$, determinations of the form factor ratio from unpolarized and polarized measurement do not agree. This has been attributed to two-photon exchange, whose size is directly tested in current experiments \cite{Rachek2014,Adikaram2014,Milner2013}. At lower $Q^2$, this effect is believed to be small, but could explain part of the proton ratio discrepancy. A positron source would make it possible to measure the effect directly at small $Q^2$, validating theoretical calculations.

The experiments described so far require a fixed target. Colliding beams open additional interesting possibilities. Head-on collisions with a high-momentum proton beam can probably not help with the physics described, however, if it could be arranged that the beam collide almost colinearly, i.e.\ essentially with the same, not opposite, direction, one would access the fixed-target equivalent of backward scattering at very low $Q^2$, accessing the magnetic form factor at unprecedentedly small four-momentum transfer. Similar, a collision of a muon and electron beam in this way would test lepton universality, a further possible explanation for the radius puzzle.


\section{Physics with Photon Beam}
This section is meant to briefly sketch the potential for fundamental research with $\gamma$-ray beams that the PERLE facility will be capable of producing by laser-Compton back-scattering off the intense cw electron beam. The production mechanism and expected $\gamma$-ray beam parameters will be described below. 
Since the scope of this Conceptual Design Report does not allow a comprehensive compilation of all possible research venues, this section includes only a limited selection of research opportunities.

Photonuclear science is currently witnessing a transformation of the field which has started 
\cite{Pietralla2002} with the advent of intense, energy-tunable, completely polarized, quasi mono-chromatic $\gamma$-ray beams from laser-Compton back-scattering at the High Intensity $\gamma$-ray Source (HI$\gamma$S) \cite{Litvinenko1998} at the Duke Free Electron Laser Laboratory (DFELL) at 
Duke University, Durham, NC, U.S.A., and will continue with the 
European Extreme Light Infrastructure - Nuclear Physics (ELI-NP) which is currently under construction in Magurele, Romania~\cite{eliro}. 
ELI-NP is expected to deliver first $\gamma$-ray beams in the energy range from 0.5 - 19.5 MeV with a band width of 0.5\% 
and a peak-spectral density of $10^4$ $\gamma$s/(eV\,s\,cm$^2$) starting in 2017. Photonuclear science at ELI-NP is enjoying a strong international user community of 100 - 200 scientists who potentially could later be attracted to the PERLE $\gamma$-beam due to its expected superior performance, in particular 
with respect to intensity, band-width, and the CW time structure.  

Photonuclear reactions impact on a variety of research topics in nuclear structure physics, particle physics metrology, and nuclear astrophysics. From each of these fields, a selection of one or two examples is sketched below, in order to give a flavor of the research potential for an advanced $\gamma$-ray beam to be established at the PERLE facility, apart from additional commercial or medical applications. 

\subsection{Photonuclear reactions} 

Gamma-rays with energies up to 30 MeV can induce a variety of photonuclear reactions. 
Photoinduced nuclear excitations  below the nuclear separation energy 
will decay by subsequent re-emission of $\gamma$-radiation. 
When this reaction proceeds via a nuclear resonance it is addressed as nuclear resonance fluorescence (NRF). 
The NRF process may populate an excited low-lying nuclear isomer which may decay by 
$\beta$-decay processes addressed as internal photoactivation. 

\begin{figure}[t!]
\centering 
      \includegraphics[width=.6\textwidth]{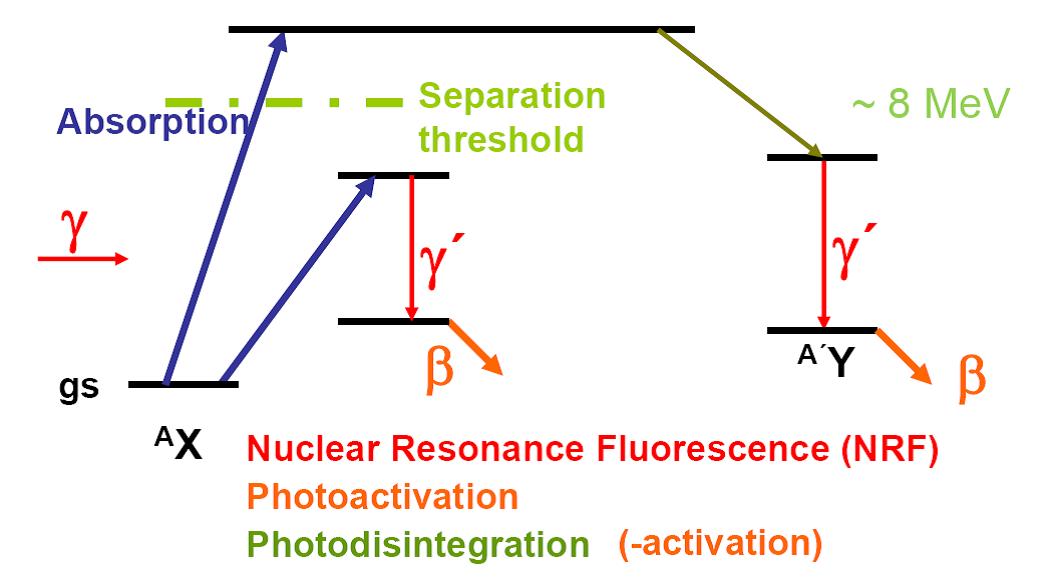}
  \caption{Photonuclear reaction modes that can be induced by photons with energies in the range of PERLE.
  } \label{fig:phnuclreact}
\end{figure}

Photodisintegration reactions become possible when a nucleus is photo-excited above the 
separation threshold. Then either neutrons or charged nuclear constituents
such as protons or even $\alpha$-particles can be emitted. Photodisintegration reactions
that result in a daughter nucleus which is radioactive are called external photoactivation. 
An extreme mode of photodisintegration is photofission where a nuclear fission process occurs 
once the nucleus has been activated by the absorption of the $\gamma$-ray. 
The various photonuclear reactions are sketched in Fig. \ref{fig:phnuclreact}.

\subsection{Nuclear structure physics} 

The field of nuclear structure physics addresses the investigation of the nuclear many-body problem and  
its understanding in terms of effective nucleon-nucleon interactions that emerge from QCD as the effective 
interaction between hadrons. Since the electromagnetic interaction is  understood quantitatively, 
photonuclear reactions enable the separation of the photonuclear reaction mechanisms from the nuclear 
properties and thus nuclearmodel-independent measurements. 
Due to the clean reaction mechanism of $\gamma$-rays with the nucleus, its iso-vector and 
one-step character, the field of nuclear structure physics has tremendously profited from photonuclear 
research since the seminal works of Bothe and Gentner in 1937 \cite{Bothe1937}. 

\subsubsection{Nuclear single-particle structure} 

The recent understanding of nuclear shell-evolution as a function of nucleon number 
and the contribution of effective three-body forces \cite{Otsuka2010} to it 
make the precise measurement of effective single-particle energies in nuclei 
a research topic of high current interest. 
Photonuclear reactions offer a unique tool to study $E1$ and $M1$ single-particle excitations from the 
ground state. 
Of particular interest is the study of the nuclear spin-orbit splitting between a nuclear level with total 
spin quantum number $j_> = l+1/2$ and its spin-orbit partner with spin quantum number $j_< = l - 1/2$. 
These single-particle orbitals are connected by a strong $M1$ matrix element of the order of 1 nuclear magneton ($\mu_N$) that can be measured precisely by photonuclear reactions, e.g., by the measurement of ground state excitation widths $\Gamma_0$ in NRF measurements. 

Also the relative assignment of various Nilsson orbitals in deformed nuclei can be clarified with photonuclear reactions. Once sufficiently intense and narrow band-width $\gamma$-ray beams will be available at the PERLE facility, it will become possible to study the electromagnetic excitation cross sections of the rotational band-head states of deformed, odd-mass isotopes in the rare-earth mass region \cite{Herzberg1997}.

\subsubsection{Collective nuclear structures} 

Of particular interest is the study of collective nuclear excitation modes with photons. Prime examples are the Isovector Giant Dipole Resonance (IV-GDR) for a collective $E1$ excitation or the Scissors Mode of deformed nuclei for a collective $M1$ excitation mode. Both are fundamental modes of the nuclear many-body system and have intensely been studied by photonuclear reactions \cite{Kneissl2006}. 
Due to the limited spectral density and abundant low-energy background at previous bremsstrahlung sources, important questions are still not resolved. What is the quadrupole deformation of the scissors mode? How does the IV-GDR emerge as a function of excitation energy and what is its fine-structure? How does the decay of the components of the IV-GDR depend on their $K$-quantum number? What is the nature of the Pygmy Dipole Resonance (PDR) that rides on the low-energy tail of the 
IV-GDR and dominates the nuclear $E1$ response near the particle separation threshold? 
PERLE could contribute to answering these questions. 
Measurement of the intrinsic $E2$ matrix element between the scissors mode and the nuclear ground state 
requires the determination of the absolute monopolar $E2$ decay width between a state of the scissor mode band and the ground state band, e.g., the $J^\pi = 1^+$ band head of the scissors mode band and the $2^+_1$ state of the ground state rotational band in a deformed even-even nucleus. The measurement of the monopolar  partial decay width of interest, 
$\Gamma_{1^+ \to 2^+_1, E2} = \delta^2/(1+\delta^2)\,\Gamma_{1^+ \to 2^+_1}$, 
requires the measurement of partial decay width $\Gamma_{1^+ \to 2^+_1}$, which is routinely done in NRF experiments on the Scissors Mode, and the $E2/M1$ multipole mixing ratio, $\delta$, of this $\gamma$-decay transition. 
This has not been done so far. 
Such a measurement will be achievable at the Compton-backscattered $\gamma$-beam of the PERLE facility by measuring the azimuthal NRF intensity distribution about the polarization plane of the $\gamma$-beam. 
The measurement will determine the quadrupole collectivity of the scissors mode and will open up a research program on how this collectivity is related to the nuclear shape (prolate, oblate or triaxial,...) and its underlying single-particle structure. 
The polarization and high intensity of the new $\gamma$-beam will open up another research field 
on the electric dipole response of nuclei below and above the nuclear separation threshold. 
Along the lines of research that have been started at the HI$\gamma$S facility at DFELL, the strength, energy distribution and decay properties of the PDR can be studied with PERLE at much higher sensitivity than before. 
In particular it will become possible to excite the nucleus at a preselected excitation energy region in the 
PDR or in the IV-GDR and then to measure the decay $\gamma$-ray transitions either to the ground state 
or to low-energy excited states of interest. 
It will become possible to search for the PDR of deformed nuclei and to thereby answer the question if 
the PDR in deformed nuclei exhibits a splitting according to its $K$-quantum number components, 
$K = 0$ or $1$. 
Until now, neither has the PDR been observed in deformed nuclei, nor has it been clarified if
 the $\gamma$-decay of the IV-GDR in deformed nuclei differs between its $K = 0$ or $K=1$ 
components. A detailed understanding of these phenomena as a function of deformation, neutron excess, or excitation energy above particle separation threshold will become possible. 

\subsubsection{Nuclear photofission} 
Nuclear fission represents an extreme case of collective nuclear behavior. 
It can be triggered by incident $\gamma$-rays in photofission processes. 
The cross section for photofission reactions is tremendously enhanced when the energy of the 
initially absorbed photon coincides with the excitation energy of a quasi-bound resonance in 
the hyperdeformed well of the nuclear fission barrier. 
Information on these photofission resonances provides valuable insight in the structure of heavy fissile 
isotopes that is very difficult to obtain otherwise. 
The geometrical type of the various fission resonances dictates the subsequent fission modes and thereby 
the distribution of resulting fission fragments. 
A technological, and even a commercial, impact of photofission resonances with respect to the 
handling of radioactive waist is conceivable. 

An intense, narrow bandwidth $\gamma$-ray beam at the PERLE facility opens up an entire 
new route of research on photofission processes of long-lived actinides. 
Its high photon flux will make photonuclear experiments on small samples in the milligram range 
possible. 
Its narrow bandwidth allows for a high energy resolution in experimental searches for new photofission 
resonances by energy-scans through the relevant excitation energy region. 
A better understanding of the fission processes, in particular of long-lived trans-uranium actinides 
is of very high interest of the society. 

\subsection{Particle physics metrology}
Due to our understanding of the unified electroweak interaction, the electromagnetic reaction processes of photons with nuclei are closely related to nuclear reactions involving the weak interaction \cite{Langanke2004}. 
Consequently, photonuclear studies can, at least partly, shed light on weak interactions in materials 
that are employed in detectors for weak-interaction processes such as detectors for searching for neutrinoless double-beta 
decay or for neutrino signals from supernovae. 

\subsubsection{Nuclear matrix elements for 0$\nu\beta\beta$-decay} 
It has recently been demonstrated \cite{Beller2013} how photonuclear investigations on the $M1$ strength distribution of initial and final nuclei in $0\nu\beta\beta$-decay reactions can help to improve the theory for 
$0\nu\beta\beta$-decay matrix elements. 
Knowledge of these matrix elements will be mandatory for the determination of the neutrino mass once the 
$0\nu\beta\beta$-decay rate would have been measured. 
The $M1$ decay branching ratio was recently found to be linked to the  $0\nu\beta\beta$-decay branching ratio to the low-energy $0^+$ states of the final nucleus. 

\subsubsection{Detector response to stellar neutrinos} 
Supernovae are bright sources for neutrinos. Detectors for the measurement of neutrinos from supernovae are operational or under construction. Due to neutrino oscillations, not all of the neutrinos reaching the detector will be electron-neutrinos $\nu_e$ but may have oscillated to other possible neutrino-flavors. Non-$\nu_e$ neutrinos with typical energies of a few MeV may react on the detector material by neutral-current scattering processes, that may be inelastic and are expected to be dominated by Gamow-Teller type matrix elements from the ground state. These are closely related to the matrix elements for $M1$ excitations. 
In order to be able to quantitatively interpret the signals from neutral-current neutrino scattering on detector material it is important to precisely know and understand the $M1$ excitation strength distributions of nuclei 
present in the detectors searching for stellar neutrinos. 

\subsection{Nuclear astrophyics}
Energetic $\gamma$-rays belong to the thermal environment in stars. 
Understanding of nuclei in the variety of stellar conditions requires a detailed knowledge of photonuclear reactions. Research opportunities for photonuclear reactions in nuclear astrophysics are numerous. 
We will mention only two examples. 

\subsubsection{Stellar capture reactions}
Stellar capture reactions, such as ($p,\gamma$), ($n,\gamma$), or ($\alpha,\gamma$) determine the 
vital "energy production" in stars. 
For stars slightly heavier than our sun the CNO-cycle dominates, by which 4 protons are converted into an 
$\alpha$-particle and released binding energy in a sequence of capture and decay reactions on 
carbon, nitrogen, and oxygen isotopes. 
Break-out of the CNO-cycle can occur, when the stable ground state of $^{16}$O 
will be populated. 
Of particular interest is the cross section for the $^{12}$C($\alpha,\gamma$)$^{16}$O reaction at energies corresponding to stellar temperatures. 
This cross section is very small, therefore difficult to measure, and despite of its importance, not known. 
By the principle of detailed balance in time-reversal invariant reactions valuable constraints could be obtained 
from the inverse reaction $^{16}$O($\gamma,\alpha$)$^{12}$C which could be studied with an intense quasi-monochromatic $\gamma$-ray beam. 
A corresponding research program has started at HI$\gamma$S but suffers from too low intensity ($10^3$ $\gamma$/(eV\,s)) and too large energy-spread (1 - 3\%). 
The superior properties of PERLEs $\gamma$-ray beam will facilitate these measurements. 

\subsubsection{Nuclear synthesis} 
One of the most outstanding physics questions is that to origin of the chemical elements in nature. Heavy nuclei beyond iron are produced in the various capture processes in stars, while latest research results indicate that supernova explosions are not capable of producing a sufficient amount of elements heavier than 
silver \cite{Arcones2014}. 
Very heavy elements, such as Thorium or Uranium, undoubtedly require a rapid-neutron capture 
process ($r$-process) in a dense and hot environment with a high neutron flux. 
In order to understand the survival rate of just synthesized heavy nuclei one needs to understand 
their reactions on the thermal radiation. 
Thermal $\gamma$-rays are capable of inducing photoactivation reactions on seed-nuclei and 
transforming them in other species. 
Stellar photonuclear reactions on stable nuclei will become possible to be studied at the 
PERLE $\gamma$-ray beam with unprecedented sensitivity.


\section{Detector Test Beam Use}
\label{sec:DetectorTest}
PERLE will accelerate electrons up to about 1GeV of energy. 
Complementary to other user test beam options
 world-wide (see \cite{Ramberg_ILCWS2008}, \cite{world_accelerators} ) such  beams would allow dedicated studies of 
 single particles effects at lower energy for
\begin{itemize}
     \item{new tracking detectors such as}
\begin{itemize}
     \item{micro-pattern gas detectors SiPM}
     \item{new (thin) pixel/strip sensor technologies}
     \item{new detectors for luminosity monitoring}
     \item{heavy fibers, new scintillating crystals;}
\end{itemize}
     \item{detailed effects of electromagnetic calorimeter measurements (very high resolution sampling at normal and low temperature);}
     \item{novel detector systems\,concepts,    etc.}
\end{itemize}
Detailed tests of detector samples and components for the upcoming High Luminosity LHC, nuclear physics experiments  or other colliders to follow could be performed at a PERLE testbeam. The beam energy would be low. A special application then may be to calibrate detectors
one would build for the physics with PERLE.

For a test-beam extension of the PERLE scope,
 the following aspects are important:
\begin{itemize}
     \item{ the extraction and shaping section has to be foreseen in the design ensuring the space and
	  elements necessary are available;}
     \item{a beam line enclosure with instrumentation;}
     \item{suitable shielding, transportation and escape routes have to be taken into account when
	  space requirements for the experimental setup are being discussed;}
     \item{Interlock system;}
     \item{Magnet control for momentum selection;}
     \item{Patch panels with pre-installed cables;}
     \item{Gas warning systems;}
     \item{Fast internet connection;}
     \item{light weight (state of the art) trigger setup; fast and precise.}
\end{itemize}
A strong community for an electron/photon user facility exists. A test beam use of PERLE would provide the host laboratory
with an extra attraction which  one may compare with 
 DESY's electron test beams.
%

An important consideration for building a facility such as PERLE
is the education and training of young scientists in the complexity of experimental 
 particle and nuclear physics. 
 For young physicists it is often difficult getting involved in all phases of HEP experiment, its  development and running, especially when engaged in the 
 large LHC experiments.
The preparatory phases for detectors are getting longer and usually only a few aspects can be studied by one person in detail.
The data taking periods of current experiments are longer and generations of students never get to work on  the/a real detector.

Test beam studies allow education in many respects as in 
the experimental preparation, trigger setup and evaluation, data acquisition, data taking (shifts, on-call), or software on track reconstruction or alignment.
A test beam configuration at PERLE appears attractive to consider indeed.

\clearpage 

\chapter{Design and Parameters}
\label{chap:DesignandParameters}
The PERLE facility aims at a maximum of \SI{1}{GeV} energy recovery demonstration of a recirculating SC linear accelerator. The test facility should serve as a test bed to gain quantitative and qualitative understanding of the electron beam recovery process. The accelerator development purposes of this test facility, as introduced above, are first, confirming the feasibility of the LHeC ERL design by demonstrating stable intense electron beams with the intended parameters (current, bunch spacing, bunch length); secondly, testing novel accelerator components such as a (polarized) DC electron gun, SC RF cavities, cryomodule design and feedback diagnostics; finally, experimental studies of the lattice dependence of stability criteria. The realisation of this facility will allow addressing several physics challenges such as maintaining high beam brightness through preservation of the six dimensional emittance, managing the phase space during acceleration and energy recovery, stable acceleration and deceleration of high current beams in CW mode operation. The facility design must also allow addressing other performance aspects such as longitudinal phase space manipulations, effects of coherent synchrotron radiation (CSR) and longitudinal space charge, halo and beam loss and microbunching instability. These issues could have sizeable impacts on machine performance in the region of the design parameter space. Thus a design emerges of a system that, in principle, needs to be flexible in supporting multiple operating points and indeed, provides a reasonable validation of the LHeC accelerator baseline.

\begin{figure}
\begin{center}
\includegraphics[width=0.9\textwidth]{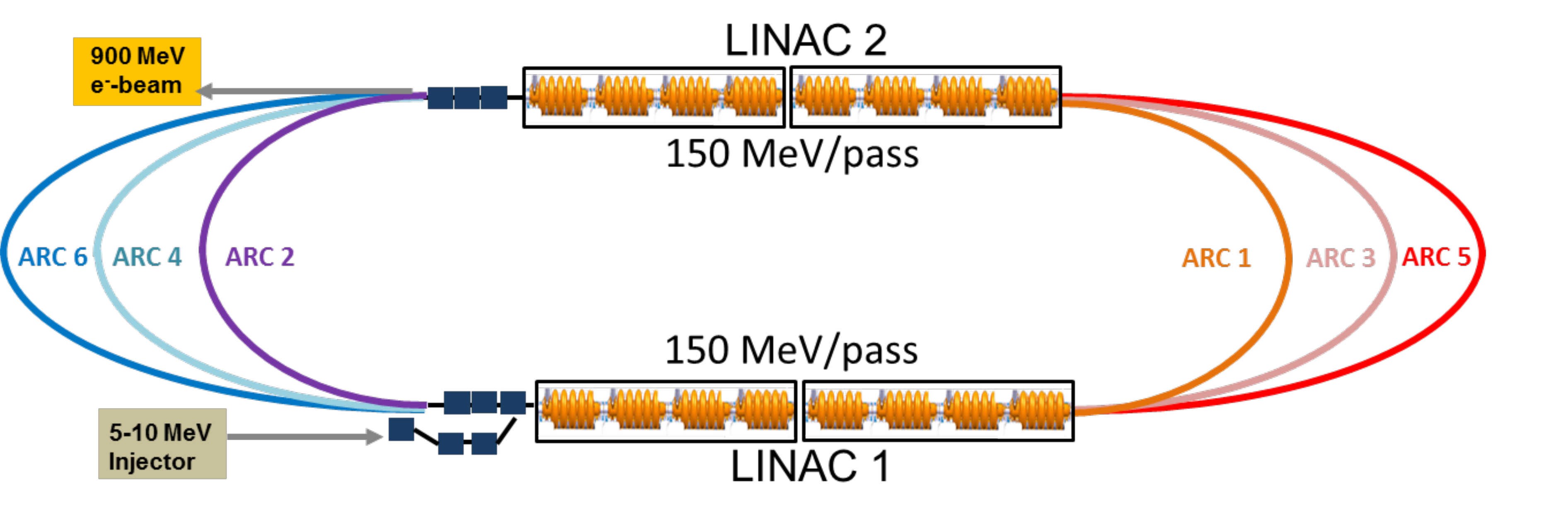}
\caption{PERLE configuration of two parallel linacs comprising two 4-cavity cryomodules each to achieve \SI{150}{MeV} acceleration per linac and \SI{300}{MeV} per pass. There are up to three passes.  There will be a pre-acceleration unit following the source to enter the ERL with relativistic electrons (\SI{>5}{MeV}).}\label{FIG_DESIGN_SKETCH}
\end{center}
\end{figure}

PERLE may be constructed in stages from initially \SI{150}{MeV} to nearly \SI{900}{MeV} in \SI{3}{steps}. The final baseline design of the ERL configuration (Fig.~\ref{FIG_DESIGN_SKETCH}) would consist of the following elements:
\begin{enumerate}
\item a \SI{5}{MeV} to \SI{10}{MeV} energy injector;
\item two \SI{150}{MeV} linacs each consisting of eight 5-cell SC structures;
\item optics transport lines including spreader regions at the exit of each linac to separate and direct the beams via vertical bending, and recombiner sections to merge the beams and to match them for acceleration through the next linac;
\item beam dump at $5-10$\,MeV.
\end{enumerate}

Each beam recirculates up to three times through both linacs to boost the energy to \SI{900}{MeV}. To enable operation in the energy recovery mode, after acceleration the beam is phase shifted by \SI{180}{^\circ} and then sent back through the recirculating linac at a decelerating RF phase. During deceleration the energy stored in the beam is reconverted to RF energy and the final beam, at its original energy, is directed to a beam dump. The set of main parameters incorporated into the ERL prototype injector is shown in Table~\ref{TABLE_DESIGN_PARAMS}.

\begin{table}
\begin{center}
\begin{tabular}{lc}
\hline
TARGET PARAMETER & VALUE\\
\hline
Injection energy [MeV] & 5-10\\
Maximum energy [GeV] &  1 \\
Normalised emittance $\gamma\varepsilon_{x,y}$ [mm mrad] & 6 \\
Average beam current [mA] &  15 \\
Bunch spacing [ns] & 25 \\ 
Bunch length (rms) [mm] &  3 \\
RF frequency [MHz] & 801.58 \\
Duty factor & CW \\
\hline
\end{tabular}
\caption{Basic Parameters of PERLE}\label{TABLE_DESIGN_PARAMS}
\end{center}
\end{table}

\section{System Architecture}
\begin{figure}
\begin{center}
\includegraphics[width=0.9\textwidth]{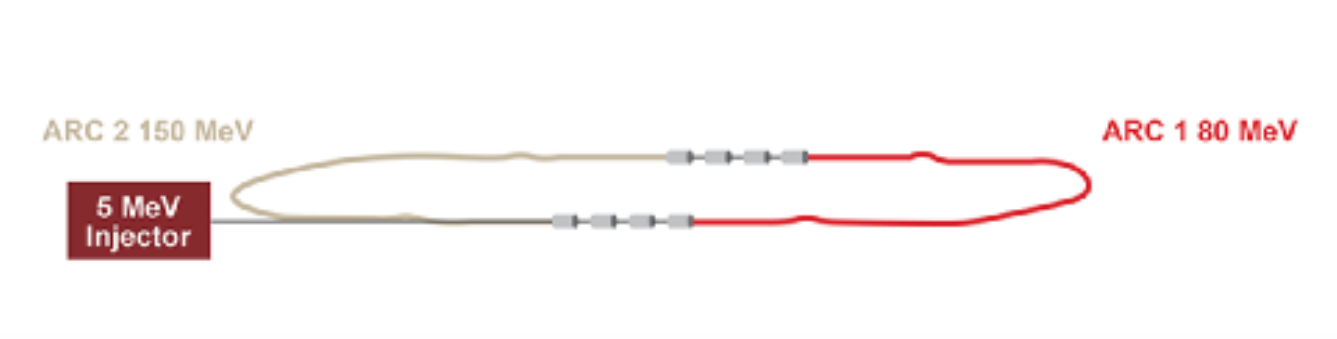}
\caption{The facility is designed in a modular way. This picture shows a Step 1 layout of two parallel cryomodules to achieve $\sim \SI{75}{MeV}$ acceleration per linac and a final beam energy of $155$\,MeV (or  half of it with just one initial cryomodule).}\label{FIG_DESIGN_STAGE1}
\end{center}
\end{figure}

\begin{figure}
\begin{center}
\includegraphics[width=0.9\textwidth]{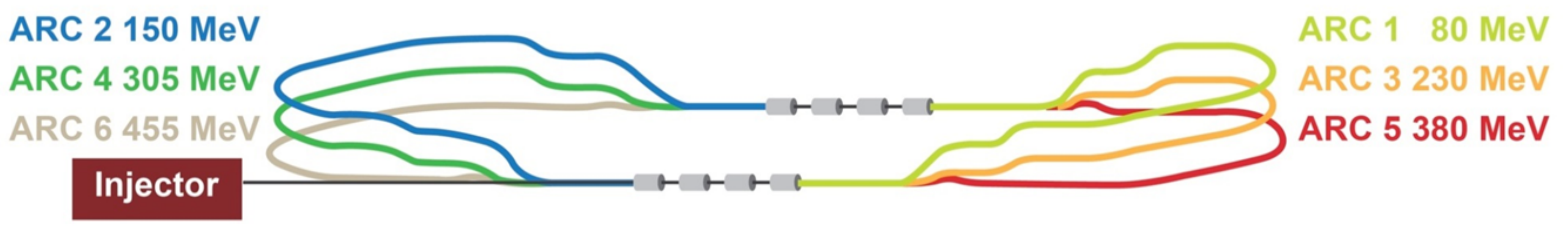}
\caption{A second phase with recirculation could feature three-pass operation to reach \SI{455}{MeV}.}\label{FIG_DESIGN_STAGE2}
\end{center}
\end{figure}

PERLE may be constructed in stages. A first phase would only use two 4-cavity cryomodules, minimally one. With
a single pass it could reach \SI{150}{MeV} and be used for injector studies and SC RF tests (Fig.~\ref{FIG_DESIGN_STAGE1}). A subsequent upgrade could be the installation of two additional arcs on each side to raise the beam energy up to \SI{450}{MeV} (Fig.~\ref{FIG_DESIGN_STAGE2}). This configuration accommodates for available space for implementation of feed-back, phase-space manipulations, and beam diagnostic instrumentation, giving the possibility of a full validation testing with energy recovery.
In phase 3, as shown above (Fig.~\ref{FIG_DESIGN_SKETCH}), four additional cavities in each linac will be added to permit energy recovery recirculation tests at full energy. The facility, in this final configuration, could represent, in principle, a smaller clone of the final LHeC project and could serve as a model for  a pre-accelerator/injector to the final \SI{60}{GeV} machine, see \ref{injectorLHeC}.

\section{Transport Optics}
Appropriate recirculation optics are of fundamental concern in a multi-pass machine to preserve beam quality. The design comprises three different regions, the linac optics, the recirculation optics and the merger optics. 

\begin{figure}
\begin{center}
\includegraphics[width=0.49\textwidth]{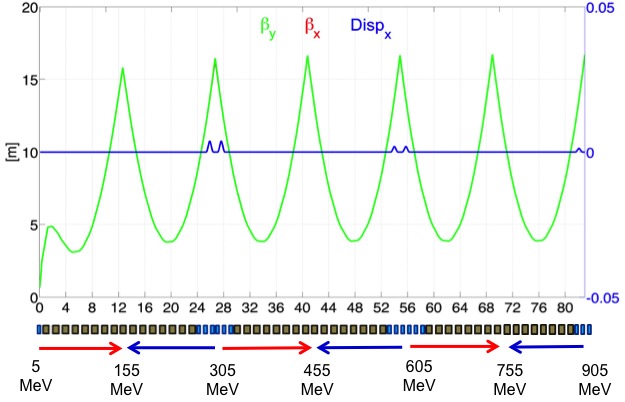}
\includegraphics[width=0.49\textwidth]{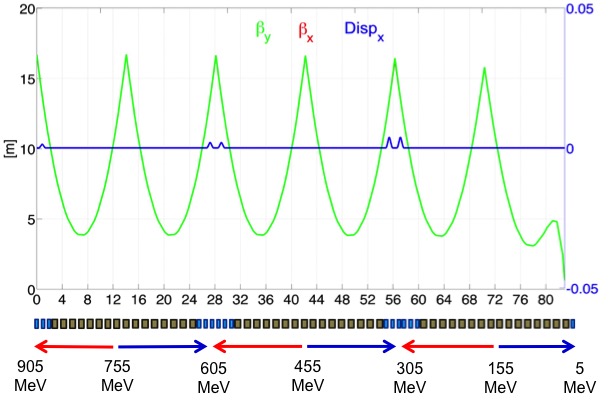}
\caption{ERL multi-pass linac optics. The requirement of energy recovery puts a constraint on the exit/entrance Twiss functions for the two linacs. Green and blue curves show, respectively, the evolution of the beta functions amplitude and the horizontal dispersion for Linac 1 (left) and Linac 2 (right). Red and blue arrows indicate the passages of acceleration and deceleration.}\label{FIG_DESIGN_LINACS}
\end{center}
\end{figure}

A concise representation of multi-pass ERL linac optics for all six passes, with constraints imposed on Twiss functions by sharing the same return arcs by the accelerating and decelerating passes, is presented in Fig.~\ref{FIG_DESIGN_LINACS}.

\begin{figure}
\begin{center}
\includegraphics[width=0.8\textwidth]{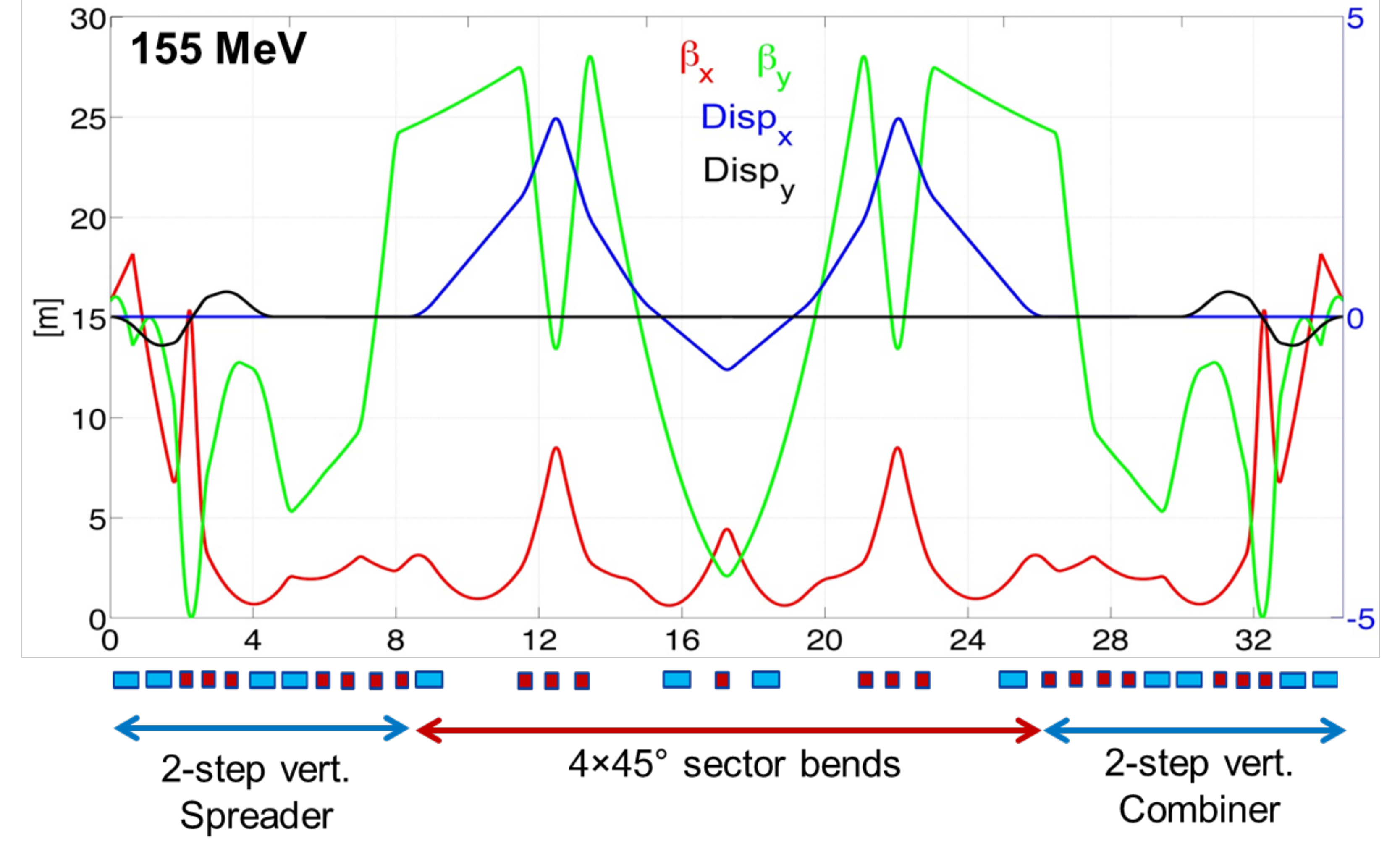}
\caption{Optics based on an FMC cell of the lowest energy return arc. Horizontal (red curve) and vertical (green curve) beta-functions amplitude are illustrated. Blue and black curves show, respectively, the evolution of the horizontal and vertical dispersion.}\label{FIG_DESIGN_ARC1}
\end{center}
\end{figure}

Due to the demand of providing a reasonable validation of the LHeC concept,
the system is oriented towards employing a Flexible Momentum Compaction (FMC) cell based lattice. Specifications require isochronicity, path length controllability, large energy acceptance, small higher-order aberrations and tunability. An example layout which fulfils these conditions is shown in Fig.~\ref{FIG_DESIGN_ARC1}, describing the lowest energy arc optics as an example. It includes a two-step achromat spreader and a mirror symmetric combiner to direct the beam into the arc. The vertical dispersion introduced by the first step bend is suppressed by the quadrupoles located appropriately between the two stages. The switchyards separate all 3 arcs into a 90 cm high vertical stack, the highest energy arc is not elevated and remains at the linac-level. A horizontal dogleg, used for path length adjustment and made of 3\---\SI{13}{cm} long dipoles, is placed downstream of each spreader providing a tunability of \SI{\pm 1}{cm} (\SI{10}{^\circ} of RF). 

\begin{figure}
\begin{center}
\includegraphics[width=0.4\textwidth]{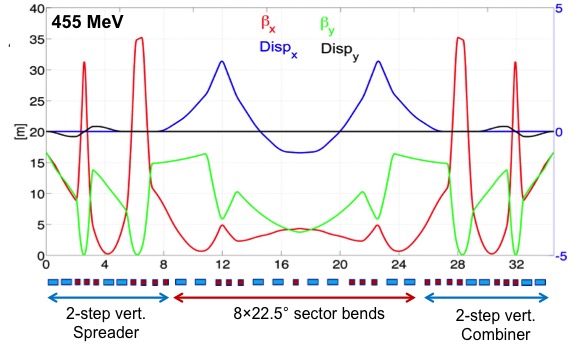}
\includegraphics[width=0.41\textwidth]{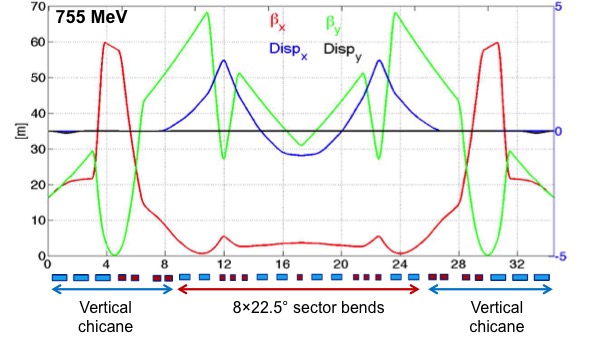}
\caption{Optics layout of the arcs at \SI{455}{MeV} and \SI{755}{MeV}. The arc at \SI{755}{MeV} is not elevated and remains at the linac-level, the spreader/combiner consists of a vertical chicane with \SI{60}{cm} long dipoles. Horizontal (red curve) and vertical (green curve) beta-functions amplitude are illustrated. Blue and black curves show, respectively, the evolution of the horizontal and vertical dispersion.}\label{FIG_DESIGN_ARC3_5}
\end{center}
\end{figure}

The recirculating arc at \SI{155}{MeV} is composed of 4\---\SI{70}{cm} long dipoles to bend the beam by \SI{180}{^\circ} and of a series of quadrupoles (two triplets and one singlet). A complete first-order layout for switchyards, arcs and linac-to-arc matching sections has been accomplished for all the arcs on both sides. Arc 3 and Arc 5 are presented in Fig.~\ref{FIG_DESIGN_ARC3_5}.

\begin{figure}
\begin{center}
\includegraphics[width=0.7\textwidth]{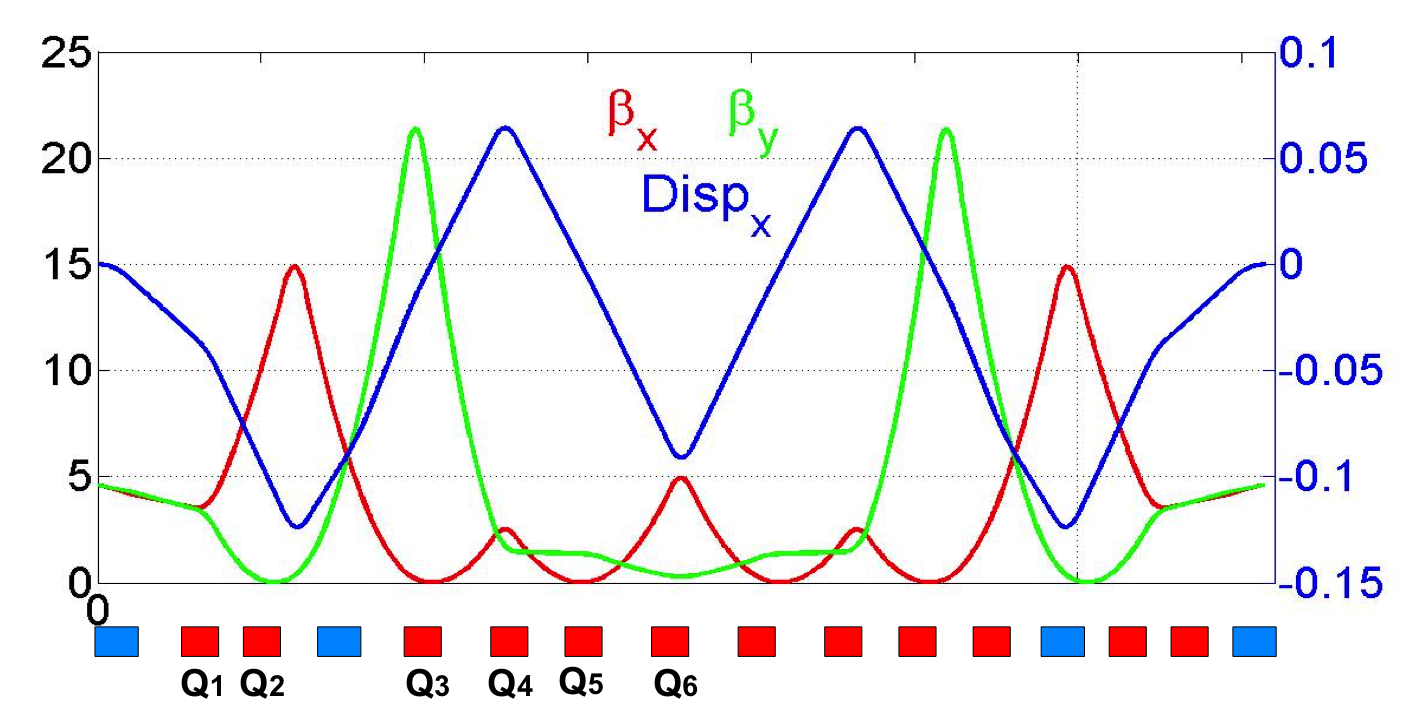}
\caption{Injection chicane optics at 5 MeV.}\label{fig20.png}
\end{center}
\end{figure}

Injection into the racetrack at \SI{5}{MeV} is accomplished through a rectangular chicane, which bypasses the arcs. The injection chicane is configured with four identical rectangular bends and 11 quadrupoles distributed in a mirror symmetric fashion, leaving six independent quadrupole gradients to control: betas and alphas at the beginning of the linac (4 parameters), momentum compaction (1 parameter) and the horizontal dispersion (1 parameter). The resulting chicane layout and optics are illustrated in Fig.~\ref{fig20.png}. The chicane optics features a horizontal achromat, by design, with tunable momentum compaction to facilitate bunch-length control and finally with Twiss functions matched to the specific values required by the linac (Fig.~\ref{fig20.png}). 

\section{Layout and Magnet Inventory}
The path of each pass is chosen to be precisely an integer number of RF wavelengths, except for the highest energy pass whose length is shifted by half an RF wavelength to recover the energy through deceleration. The total beam path for a full 3 pass accelerating cycle is around \SI{300}{m}. This leads to an approximate footprint of \SI{43x16}{m} of the ERL itself. Accurate values are presented in Table ~\ref{arcslength}.

\begin{table}
\begin{center}
\begin{tabular}{lc}
\hline
Segment & Length [m]\\
\hline
ARC 1 & 35.98\\
ARC 2  & 35.74\\
ARC 3  & 35.61 \\
ARC 4  & 35.74 \\ 
ARC 5 & 35.98 \\
ARC 6 & 34.43 \\
PASS 1  &  99.86\\ 
PASS 2 & 99.48 \\
PASS 3 & 98.55 \\
Total & 297.9\\
\hline
\end{tabular}
\caption{Beam path for a full 3 pass accelerating ERL.}\label{arcslength}
\end{center}
\end{table}

Diverse plausible optics layouts have been studied. A possible option would consist of arcs with identical configurations in order to have compact magnets stacked on top of each other.

A preliminary inventory of the magnets of the LHeC Test Facility lists:
\begin{itemize}
\item 40 bending magnets (vertical field);
\item 36 bending magnets (horizontal field) in the spreaders / combiners;
\item 114 quadrupole magnets;
\item 6 magnets in the injection / extraction parts;
\item a few magnets for path length adjustment.
\end{itemize}

\section{Bunch Recombination Pattern}
\label{bunchpat}
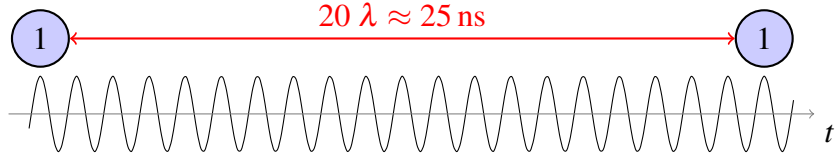
\begin{figure}
\begin{center}
\begin{tikzpicture}
  \draw plot[line width=2,domain=-0.03:10.03,smooth,samples=210] (\x,{0.5*sin(21*36*\x)});
  \draw[gray,->] (-0.3,0) -- (10+0.3,0) node[black, anchor=north west] {$t$};
  \node[draw, thick, fill=blue!20, circle] (n1) at (10/21/4,1) {1};
  \node[draw, thick, fill=blue!20, circle] (n12) at (10/21/4 + 20*10/21,1) {1} edge [<->,red,thick] node[above]{20 $\lambda \approx$ \SI{25}{ns}} (n1);
\end{tikzpicture}
\caption{Basic RF structure, without recirculation. Bunches are injected every \SI{25}{ns}.
The waves indicate the RF electromagnetic oscillations.} \label{FIG_REC_PATTERN_1}
\end{center}
\end{figure}

%
The bunch spacing at the injector, dump and delivered is \SI{25}{ns}, as shown in Fig.~\ref{FIG_REC_PATTERN_1}.
However, due to continuous injection and the recirculation, more bunches at different energies are interleaved in the linacs, appearing in periodic sequences.
The spreader and combiner design, employing fixed-field dipoles, do not pose timing constraints. For this reason the recombination pattern can be adjusted by simply tuning the length of the return arcs to the required integer number of $\lambda$.

In order to minimise collective effects, the arc lengths have been tuned avoiding to combine different bunches in the same bucket, 
like it would happen if the full turn length was an integer number of $20\lambda$.
On the contrary, the lattice is adjusted to achieve a nearly constant bunch spacing.

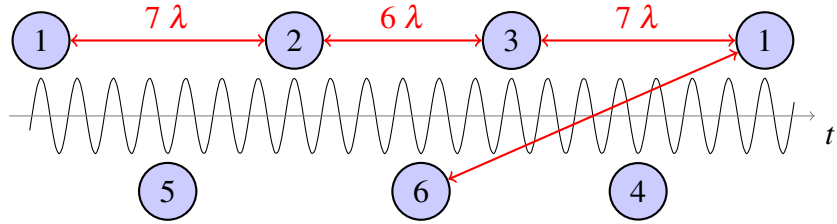
\begin{figure}
\begin{center}
\begin{tikzpicture}
  \draw plot[line width=2,domain=-0.03:10.03,smooth,samples=210] (\x,{0.5*sin(21*36*\x)});
  \draw[gray,->] (-0.3,0) -- (10+0.3,0) node[black, anchor=north west] {$t$};
  \node[draw, thick, fill=blue!20, circle] (n1)  at (10/21/4,1) {1};
  \node[draw, thick, fill=blue!20, circle] (n2)  at (10/21/4 + 7*10/21,1) {2} edge [<->,red,thick] node[above]{7 $\lambda$} (n1);
  \node[draw, thick, fill=blue!20, circle] (n3)  at (10/21/4 + 13*10/21,1) {3} edge [<->,red,thick] node[above]{6 $\lambda$} (n2);
  \node[draw, thick, fill=blue!20, circle] (n12) at (10/21/4 + 20*10/21,1) {1} edge [<->,red,thick] node[above]{7 $\lambda$} (n3);

  \node[draw, thick, fill=blue!20, circle] at (3/4*10/21 + 16*10/21,-1) {4};
  \node[draw, thick, fill=blue!20, circle] at (3/4*10/21 + 3*10/21,-1) {5};
  \node[draw, thick, fill=blue!20, circle] (n6) at (3/4*10/21 + 10*10/21,-1) {6};
  \path[<->,red,thick] (n12) edge (n6);
\end{tikzpicture}
\caption{When the recirculation is in place, the linacs are populated with bunches at different turns (the turn number is indicated). 
The recombination pattern shown maximises separation inside the RF structure between the low energy bunches (at the first and sixth turn).}\label{FIG_REC_PATTERN_2}
\end{center}
\end{figure}

Special care has been taken to select a pattern that maximises the distance between the lowest energy bunches inside the RF structure:
the ones at the first and the last turn, as shown in Fig.~\ref{FIG_REC_PATTERN_2} and summarised in Table~\ref{TAB_REC_PATTERN_1}. 
This comes from the fact that, with a nearly constant $\beta$ function, the kicks from HOMs are more disruptive at lower rigidities, thus, if two low energy bunches follow each other, the BBU threshold current can be reduced.

\begin{table}
\begin{center}
\begin{tabular}{cc}
\hline
Turn number & Total pathlength \\
\hline
1 & $n\times 20 \lambda + 7 \lambda$\\
2 & $n\times 20 \lambda + 6 \lambda$\\
3 & $n\times 20 \lambda + 3.5 \lambda$\\
\hline
\end{tabular}
\caption{Summary of the total path lengths of each turn of the ERLF design.}\label{TAB_REC_PATTERN_1}
\end{center}
\end{table}

\section{End-to-end Beam Dynamics Simulations}
Tracking simulations have been performed initially with the tracking code elegant \cite{Borland2000elegant}, to investigate single-bunch effects as the coherent synchrotron radiation (CSR) and the impact of multipolar field components, and later with PLACET2 \cite{Pellegrini2015PLACET2}, to verify the recombination pattern and asses the BBU threshold current.

\subsection{Single-bunch end-to-end}
\begin{figure}
\begin{center}
\includegraphics[width=0.8\textwidth]{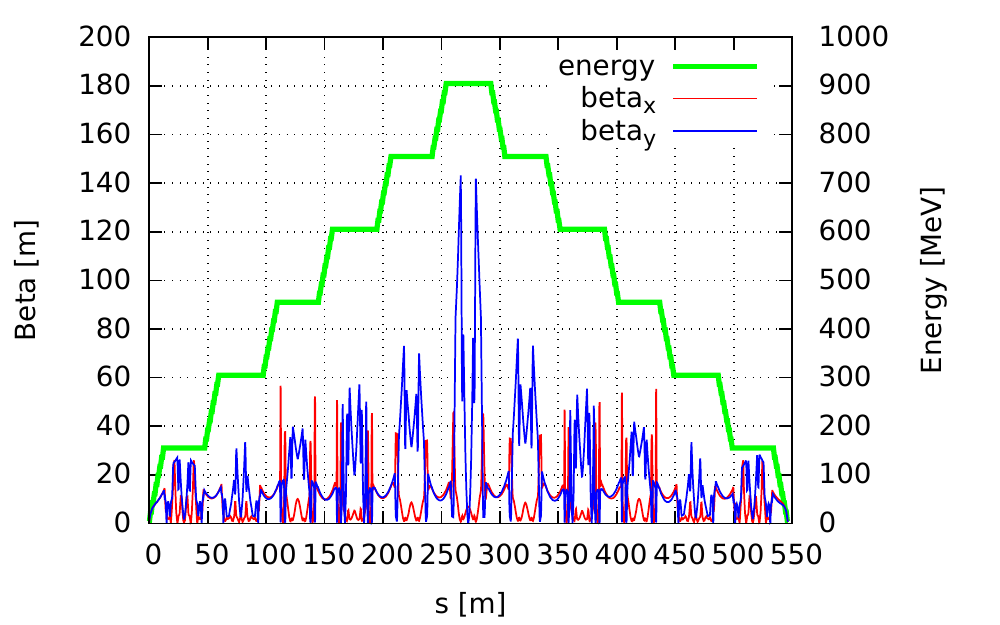}
\caption{Energy and twiss parameter tracked with PLACET2}\label{FIG_PLACET2_TWISS}
\end{center}
\end{figure}
PLACET2 is a tool that allows to describe the whole machine without unrolling the lattice and computes the element phases according to the beam time of flight.
The $\beta$ functions and the energy profile shown in Fig.~\ref{FIG_PLACET2_TWISS} are obtained following a test bunch into the lattice from the injector to the dump.
The energy profile shows that the lengths of the arcs are properly tuned to obtain the maximum acceleration and deceleration.
The regularity and the symmetry of the $\beta$ functions, validate the matching of all the arcs in the presence of strong RF-focussing from the linacs.

\begin{figure}
\begin{center}
\includegraphics[width=0.49\textwidth]{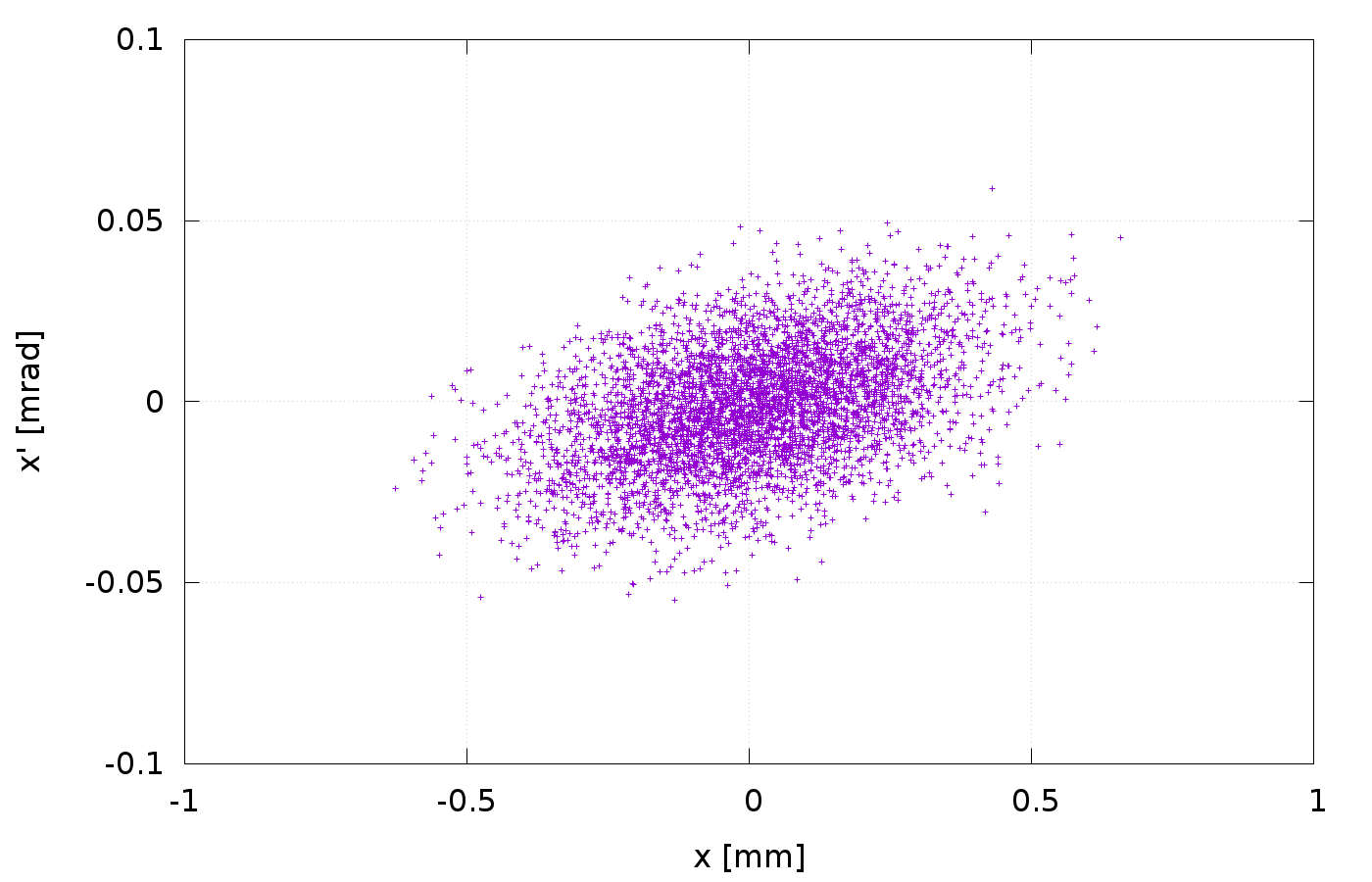}
\includegraphics[width=0.49\textwidth]{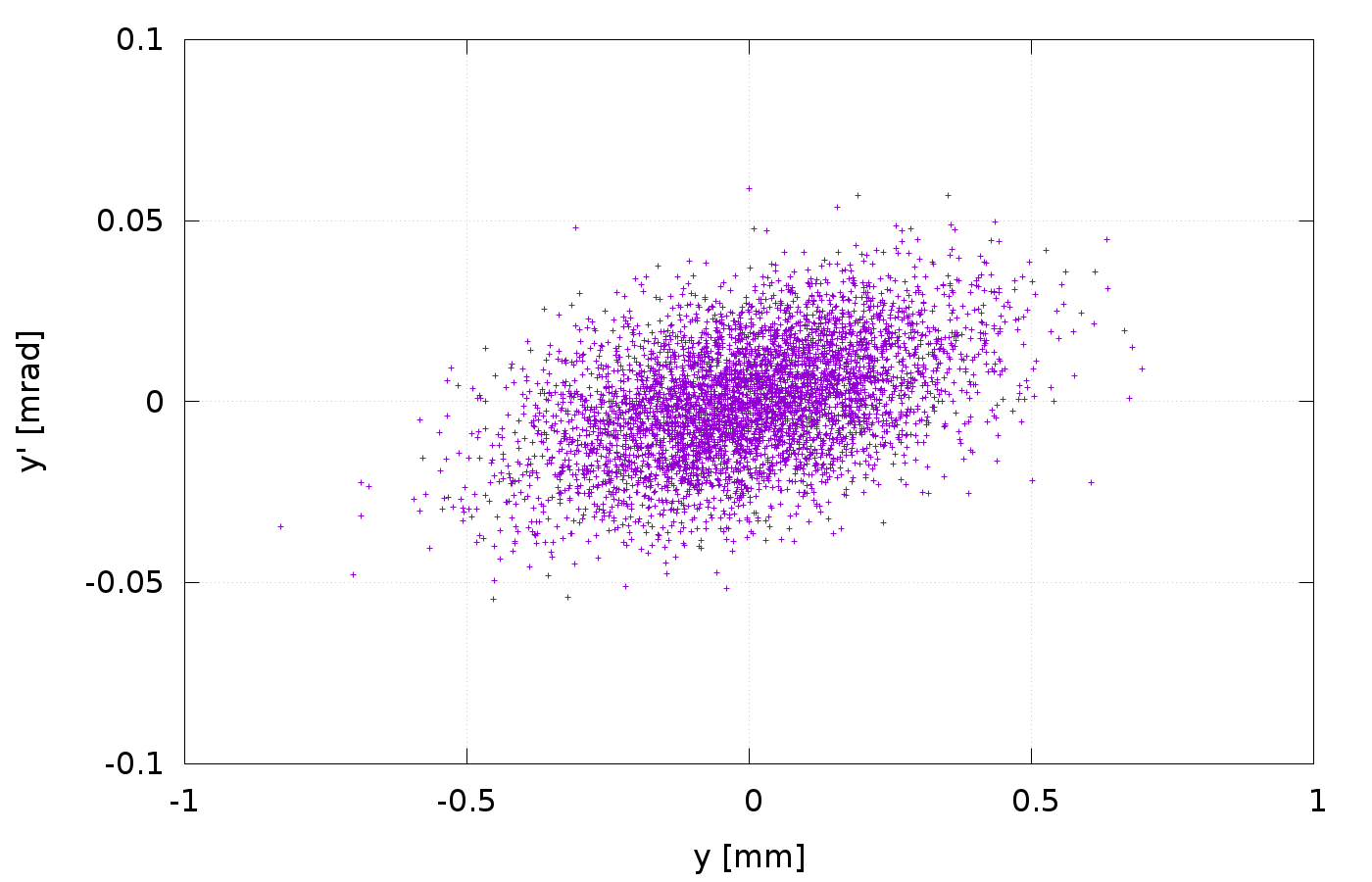}
\caption{Horizontal and vertical phase space at \SI{900}{MeV}.}\label{FIG_DESIGN_TPS}
\end{center}
\end{figure}

Figure~\ref{FIG_DESIGN_TPS} shows the transverse phase space at \SI{900}{MeV}: the plots show the emittance preservation, and in particular the absence of non linearities. Collective effects such as coherent synchrotron radiation and short-range wake fields are not included, however analytical computations predict a small impact.

\begin{figure}
\begin{center}
\includegraphics[width=0.49\textwidth]{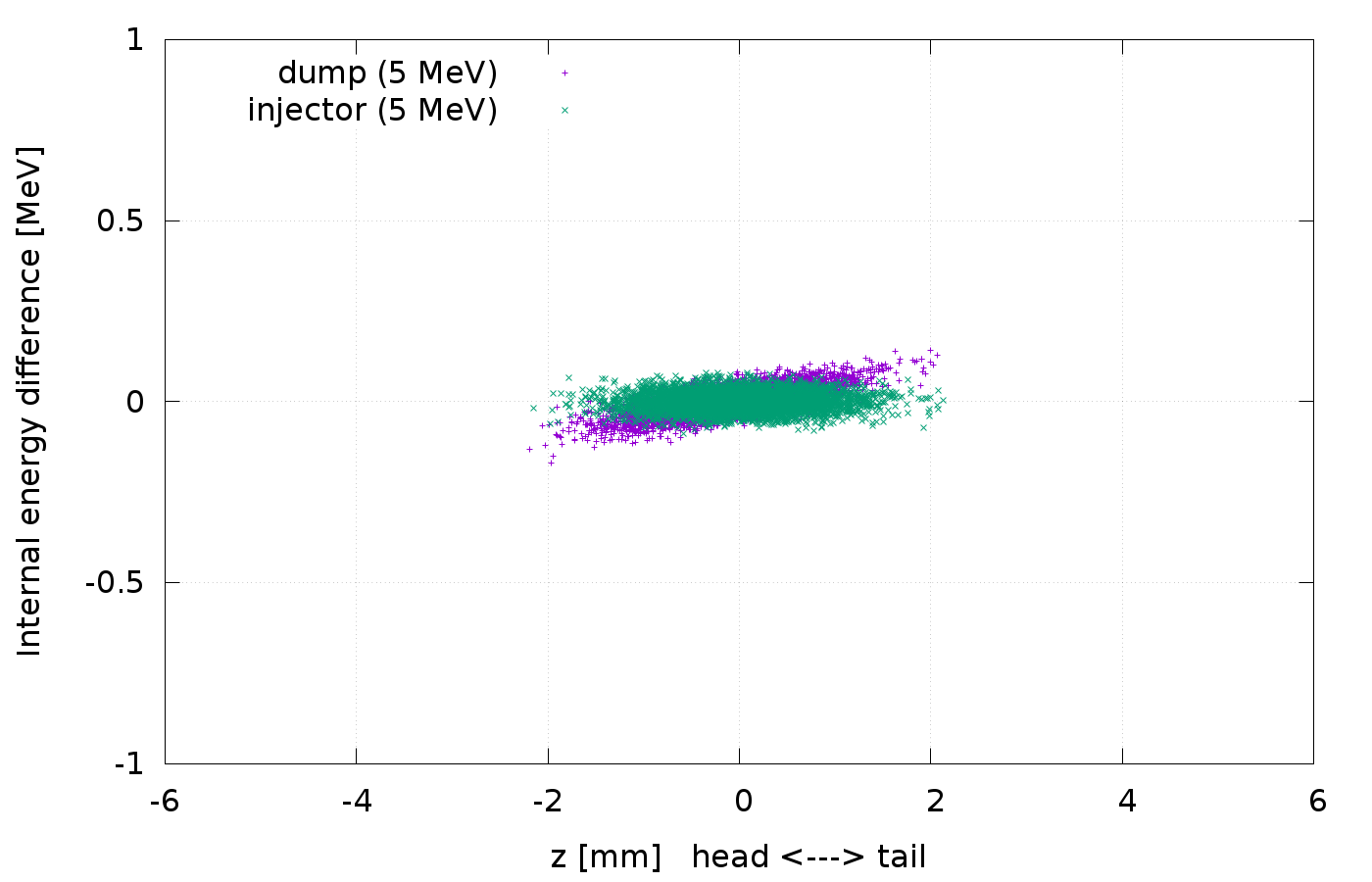}
\includegraphics[width=0.49\textwidth]{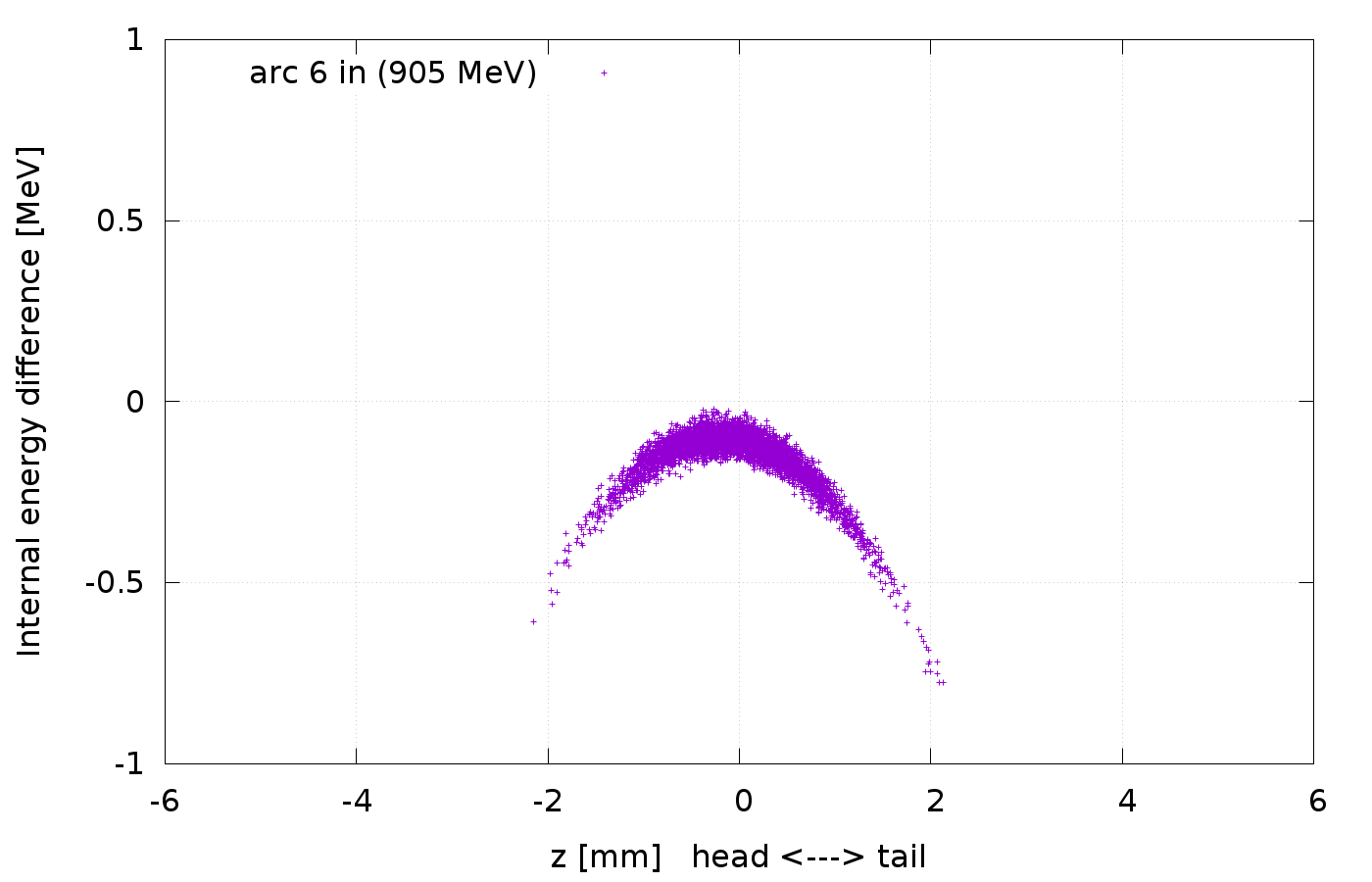}
\caption{Longitudinal phase space at injector/dump (left) and at \SI{900}{MeV} (right).}\label{FIG_DESIGN_LPS}
\end{center}
\end{figure}

Comparing the longitudinal phase space at injector and at dump (see Fig.~\ref{FIG_DESIGN_LPS}) one can note that the bunch length is well preserved, proving the isochronicity of the whole lattice. A small energy chirp is present at the dump, which shall be removed with a fine tuning of the arc lengths. Figure~\ref{FIG_DESIGN_LPS} (right) shows the longitudinal phase space at \SI{900}{MeV}. While the curvature induced by the RF can be seen, the total energy spread remains extremely contained (below $0.01~\%$).

\subsection{Multi-bunch tracking and BBU}
\begin{figure}
\begin{center}
\includegraphics[width=0.49\textwidth]{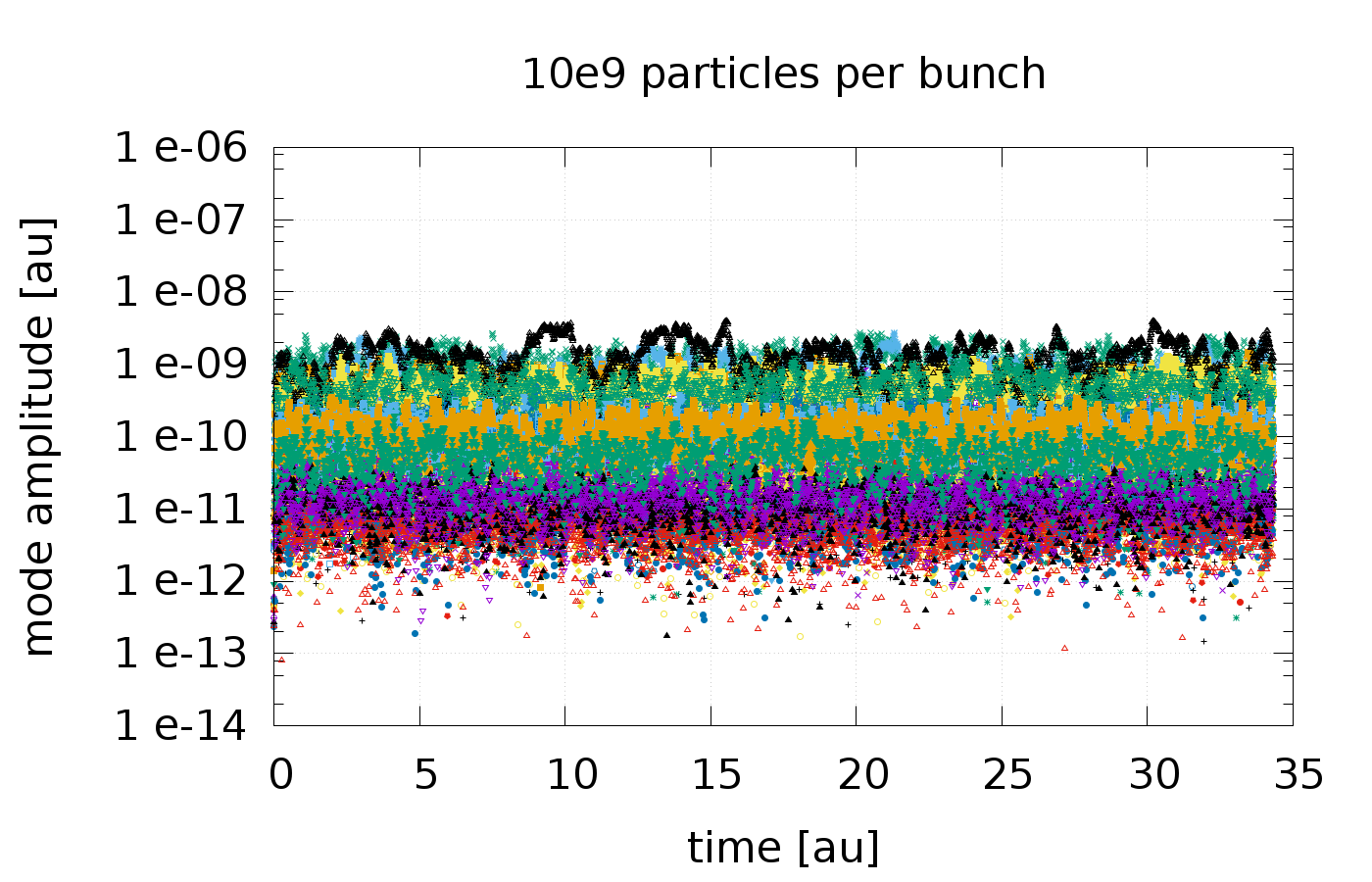}
\includegraphics[width=0.49\textwidth]{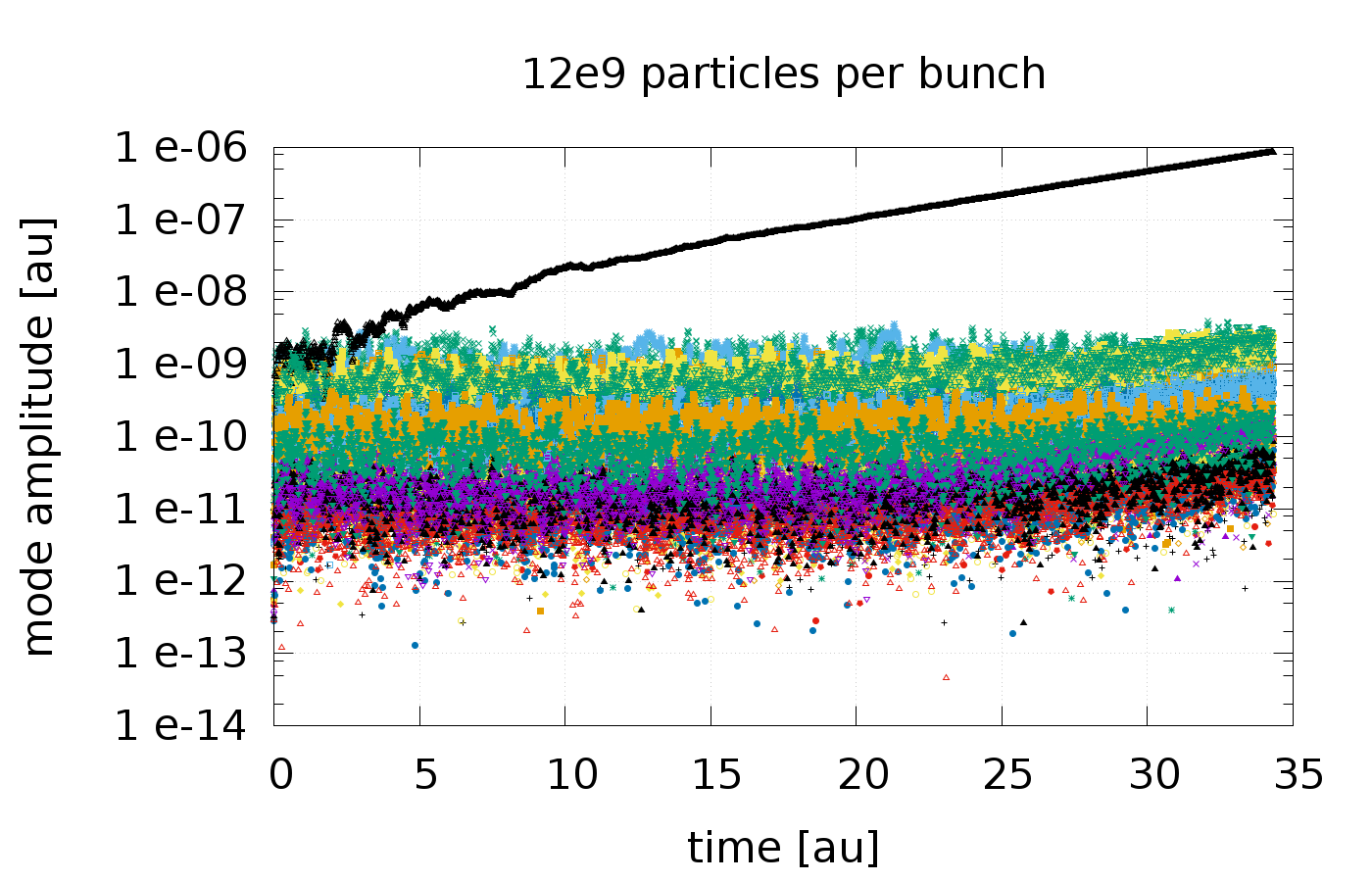}
\caption{Evolution of the amplitudes of the dipole modes for two different charges per bunch.}\label{FIG_PLACET2_BBU}
\end{center}
\end{figure}
PLACET2 is capable of tracking many bunches simultaneously in the lattice preserving their time sequence everywhere in the machine.
This allowed to verify the bunch recombination pattern and assess multi-bunch effects in a realistic operational scenario.

Estimations of the BBU threshold current have been performed using the major 26 dipole modes of the SPL cavity design, scaled to \SI{802}{MHz}.
A 6D distribution of 100 macro-particles per bunch has been used and tens of thousand of bunches have been tracked, simulating the continuous operation.
The statistical fluctuations of the positions of the bunch centroids are enough to excite the HOMs without the need of further perturbations.

A Gaussian spread has been introduced in the frequencies of the cavity HOMs assuming a detuning factor of \SI{1e-4}{}. It has been verified that for the final design stage, including a total of \SI{16}{cavities}, different detuning seeds lead to similar results.  

The plots in Fig.~\ref{FIG_PLACET2_BBU} show the amplitudes of the HOMs in one of the cavities as many bunches pass by.
One can see that when the bunch charge is increased from \SI{1.6}{nC} to \SI{1.9}{nC} a mode starts to build up in the vertical plane leading to an instability. Note that this bunch charge is more than \SI{5}{times} the one foreseen for operation.

\clearpage 

\chapter{Components}

\section{Source and Injector}
\label{sec:sourceandinjector}
The injector of PERLE needs to deliver beams with an average current of O(10)\,mA (with the possibility of future upgrades to deliver polarised electrons) and an energy of about 
$5$\,MeV. Bunches with a charge of $320$\,pC or higher follow with a repetition rate of $40.1$\,MHz ($20^{th}$ subharmonic of the ERL RF frequency $801.6$\,MHz).  The parameters of the required beam are summarised in Table ~\ref{TABLE_DESIGN_PARAMS}.
%

In principle, there are several possibilities to meet these specifications. As the requirement to normalised emittance is rather modest, it can be delivered with a grid modulated thermionic gun followed by a multi stage bunching-accelerating structure, similar to the one realised at   ELBE \cite{elbe}.   This choice, however, will rule out any future upgrade to deliver polarised electrons. Photocathode guns, where electrons are emitted from the photocathode illuminated with laser light, are more flexible in terms of the beam charge and temporal structure and allow operation with both polarised and unpolarised photocathodes. Photocathode guns utilise different accelerating technologies ranging from DC to superconducting RF, but presently only DC technology may be considered as mature and applicable to PERLE.  DC guns successfully operate at different ERL facilities \cite{HernandezGarcia:2005xs,Jones:2011zza,Obina:2016ywe}. The injector experiment at Cornell University demonstrated an average current of $52$\,mA with a GaAs photocathode and of $65$\,mA with a Cs2KSb photocathode \cite{Bazarov:2012zzb}. 

\begin{figure}
\begin{center}
\includegraphics[width=0.98\textwidth]{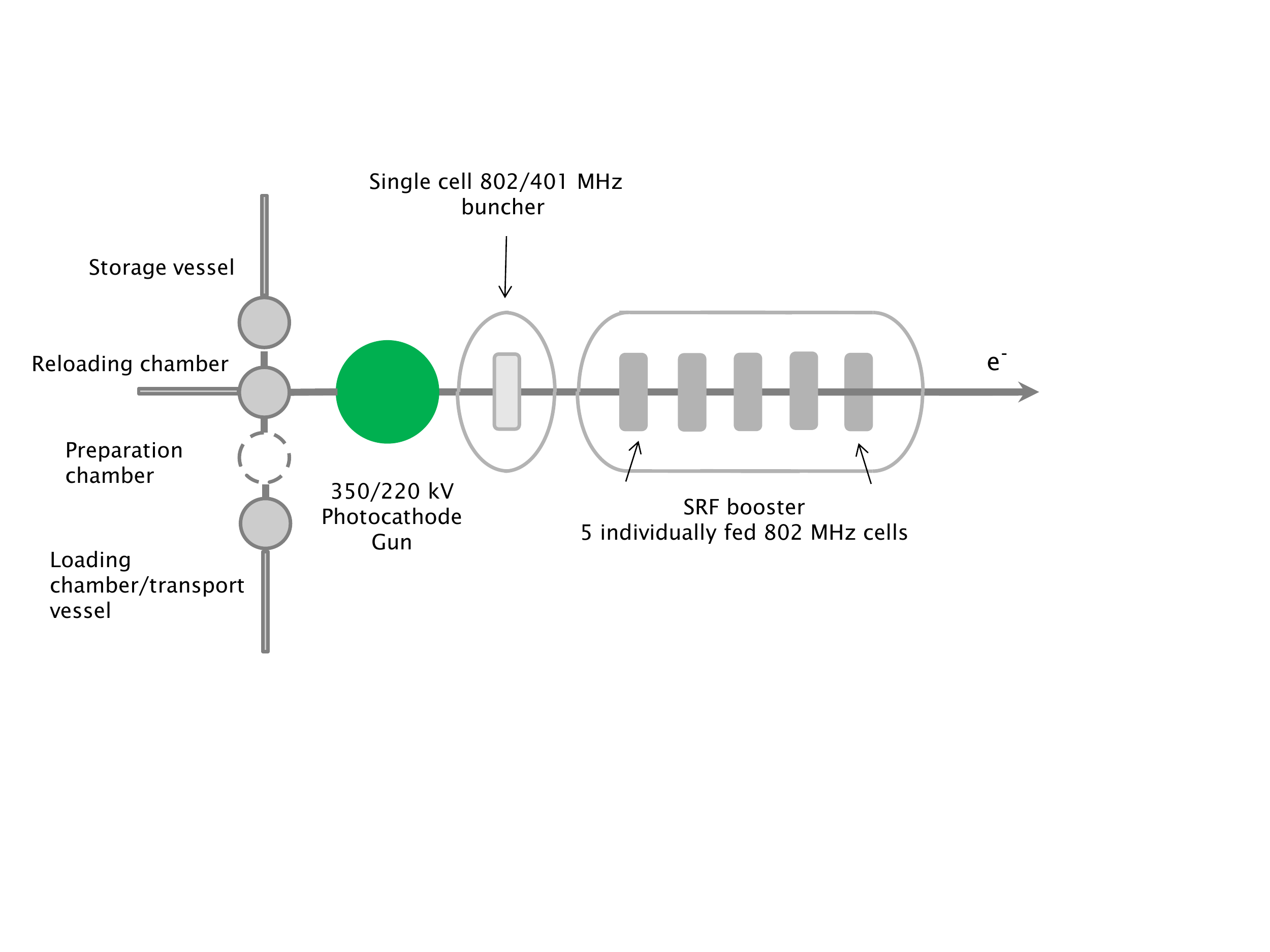}
\vspace{-2.5cm}
\caption{General layout of the photoinjector for PERLE (see text).}
\label{fig28.pdf}
\end{center}
\end{figure}

DC photocathode guns are widely used for production of polarised electrons because of their possibility to reach extra high vacuum conditions with a pressure of less than $10^{-11}$\,mbar. That is required for providing long lifetime of polarised photocathodes with typical oxygen dark lifetime $2\cdot10^{-8}$ mbar$\cdot$s. This vacuum is also sufficient for operation with antimonite based photocathodes with dark lifetime of $10^{-5}$ mbar$\cdot$s which are considered as a source of unpolarised electrons.  In addition, modern GaAs based photocathodes have reasonable quantum efficiency of $\sim1\%$ and are able to produce electron beams with polarisation of higher than 85$\%$ \cite{Adderley:2007zz,Aulenbacher:2011zz}. For PERLE a photoinjector schematic is considered as shown in Fig. ~\ref{fig28.pdf}. It comprises a DC photocathode gun surrounded by a well-developed photocathode delivery/production infrastructure, a single cell buncher cavity which compresses the beam at the exit of the gun, and a booster which accelerates the beam to $\sim$5 MeV.

\subsection{Photocathode - sources of electrons}
Physical parameters of the beam, delivered by the photocathode, are essentially defined by the gun. It also dictates the parameters of the drive laser. Photocathodes are typically characterised by their quantum efficiencies (Q.E.), defined as the ratio of extracted electrons to incident photons, its dependence on the energy of incident photons and characteristics of photocathode material. The last parameter defines the laser wavelength which should be used to extract the beam. Difference between energy of incident photons and work function defines initial energy of emitted electrons. In combination with the angular distribution of emitted electrons it determines the initial beam emittance~\cite{Flottmann1997}. 
 
Originally, in DC guns for ERL application, GaAS photocathodes were used illuminated with laser light with a wavelength of $532$\,nm~\cite{Siggins:2000ya,Gerth:2004mv}. These photocathodes are usually activated to the surface state close to Negative Electron Affinity (NEA) in the gun with Caesium dispensers. This procedure was difficult to properly control and thus does not allow reaching high quantum efficiency, typically few percent. Another problem of GaAs photocathodes is the requirement to ensure high vacuum conditions in the gun and poor lifetime due to back ion bombardment which does not allow reaching high average current for reasonable long time. More recent designs at Cornell University, Daresbury Laboratory~\cite{Militsyn:2010zza} and JAEA-KEK~\cite{Nishimori:2011zz} proposed activation of the photocathodes in a dedicated preparation facility directly connected to the gun and to replace photocathode in the gun as operating photocathode degrades. GaAs photocathodes prepared separately following this approach reached maximum Q.E. of 20$\%$ at operational wavelength, but did not solve the problem of lifetime. 

More robust photocathodes based on Sb are less sensitive to vacuum conditions and to back ion bombardment. Pioneering experiments at Boeing~\cite{Dowell:1993pc}, and the University of Twente \cite{twente1997}, at Brookhaven Laboratory~\cite{Burrill:2005hn}, TJNAF \cite{Mammei:2013cxa}, and Cornell University~\cite{Bazarov:2012zzb} demonstrated the possibility to obtain a reasonable Q.E. for Sb-based photocathodes at a level of 5-10$\%$ and, most importantly, their ability to deliver a high current for a substantial period of time. 

For delivery of polarised electrons, GaAs based photocathodes still remain the only choice. So far, maximum demonstrated current of polarised electrons is at the level of $5$\,mA  \cite{BastaniNejad:2011yp} while the possibility to reach level of $20$\,mA needs to be investigated.  Main parameters of photocathode families principally applicable for PERLE are shown in Table ~\ref{Photocathodematerials}. It can be seen that if the requirements to the laser for unpolarised beam are modest, the production of polarised electrons demands a yet high laser power.  However, this higher laser power leads to thermal desorption resulting in a deterioration of vacuum and reduction of the photocathode lifetime. Cooling down of the photocathode during operation should be taken into account at the gun design.

\begin{table}
\begin{center}
\begin{tabularx}{\textwidth}{lllllll}
\hline
Material&Typical &Work &Observed &Laser &Observed &Obs. \\*[-1ex] 
&oper.$\lambda$ & function&Q.E.&power &max &lifetime \\*[-1ex]
&&&&for 20 mA&current\\
\hline
Sb-based &532 nm&1.5-1.9 eV&4-5\%&4.7 W &65 mA &Days\\*[-1ex] 
unpolarised&&&&at Q.E.=1\%&[Cornell]\\*[-1ex] 
&&&&&& rep.\\
\hline

GaAs-based &	780 nm&	1.2 eV  &	0.1-1.0\%	&31.8 W  &	5-6 mA&	Hours \\*[-1ex] 
polarised&&at NEA state&&at Q.E.=0.1\%&[JLAB]& \\
\hline
\end{tabularx}
\caption{Characteristics of photocathode materials available for PERLE}\label{Photocathodematerials}
\end{center}
\end{table}

\subsection{Photocathode gun}
The main decisive parameter of a DC photocathode gun is its operational voltage.  It defines the energy of electrons at the exit of the gun and the 'rigidity of the beam'. This operational voltage also dictates the electric field on the photocathode which defines maximum emission density and, as a result, the beam emittance which may be estimated as
\begin{equation}
\epsilon_n=\sqrt{\frac{qkT}{2\pi\epsilon_0E_cmc^2}}.
\end{equation} 
Here, $q$  denotes the bunch charge,
$T$ the photocathode effective temperature, with $kT$  referred to as 
the photocathode's intrinsic energy,
$E_c$ is the electric field on the photocathode surface at bunch launching
and $m$ the mass of the electron.
 The traditional approach to design guns for ERLs for driving FELs demands that the gun operation voltage should be as high as possible to a reach minimal beam emittance. Maximum operational voltage of $500$\,kV with a field of $10$\,MV/m has been demonstrated at the gun developed at JAE for the cERL project at KEK~\cite{Nishimori:2014dja}. However, a very high cathode field leads to the risk of field emission, especially from photocathode materials with low work function like GaAs activated to negative electron affinity (NEA) state.  As to polarisation, it is worth noting that the field emitted electron 'dissolve' photo-emitted electrons and effectively  decrease the polarisation of the beam. A lower voltage is also more convenient for spin manipulation. The optimal values of gun voltage and cathode field should therefore be properly selected at the design stage. A dual operation mode of the gun, at high voltage for unpolarised photocathodes and at low voltage for polarised photocathodes, may not be excluded. Considering these aspects as well as a demonstrated stable operation at other 
 facilities~\cite{Burrill:2005hn,Militsyn:2010zza,Obina:2016ywe}, a choice of the maximum operation voltage of $350$\,kV seems reasonable. 

\begin{figure}
\begin{center}
\includegraphics[width=.98\textwidth]{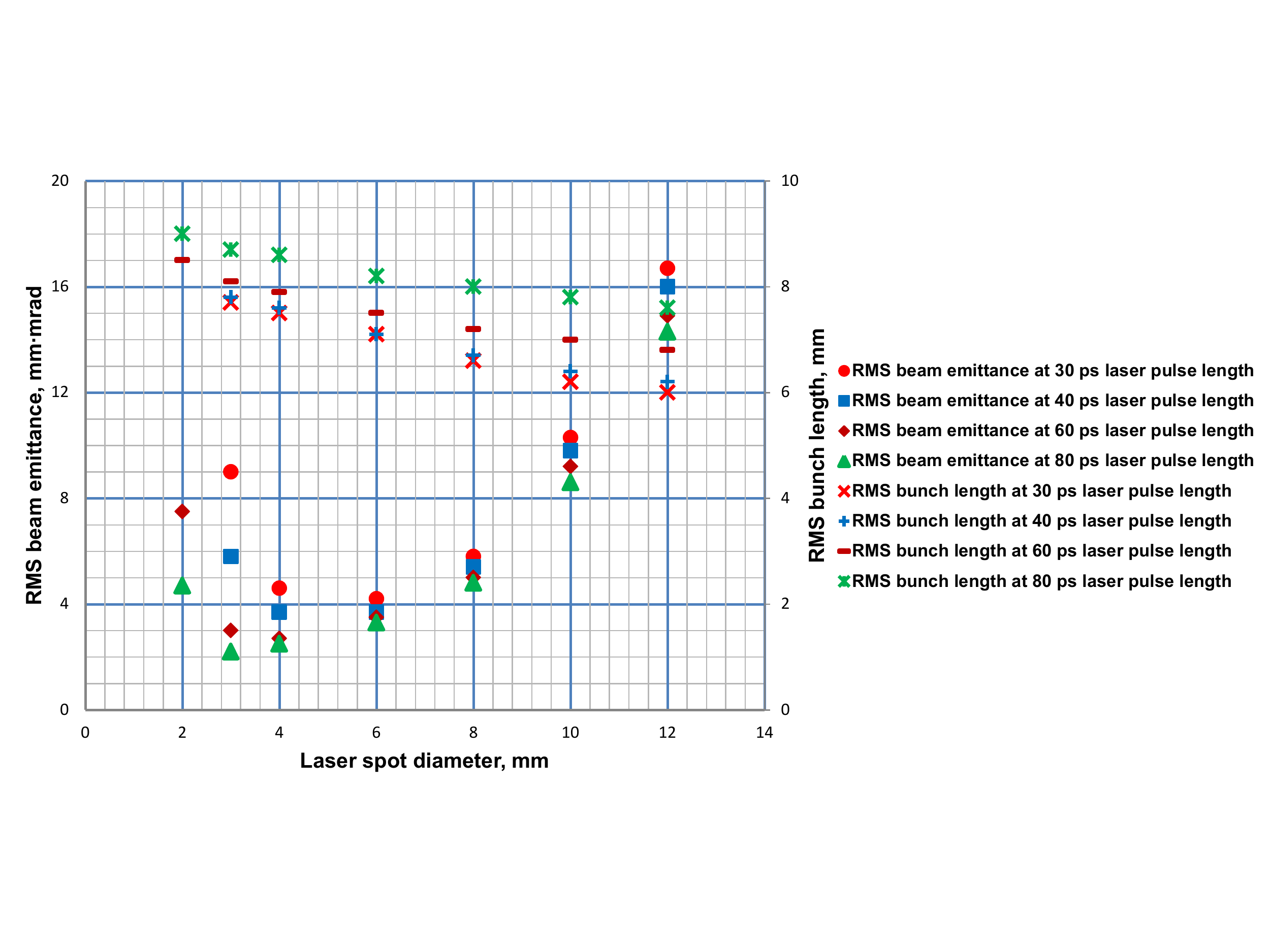}
\vspace{-1.cm}
\caption{Dependence of the calculated normalised RMS emittance and RMS length of 
$300$\,pC bunches at the exit of a modified  $350$\,kV JLAB-DL type gun with Cs$_3$Sb photocathode on laser spot diameter at different laser pulse length.}
\label{fig29.pdf}
\end{center}
\end{figure}

In order to get preliminary estimates required on the drive laser system to deliver beam with parameters required for PERLE, the performance is calculated of a $350$\,kV gun 
with a JLAB-DL electrode system operating with Cs3Sb photocathode  
(Fig.\,\ref{fig29.pdf}). Simulations have shown that an optimal beam emittance of 2 $\pi$mm-mrad  can be obtained with illumination of the photocathode with a laser pulse with hat top spatial distribution with a diameter of $3$\,mm and a flat top laser pulse with a length of $80$\,ps. The RMS bunch length at $1$\,m from the photocathode is $8.5$\,mm ($36$\,ps) which only slightly depends on the laser pulse length. 
\subsection{Buncher and booster}
Once emerged from the gun, the electron beam begins to elongate due to the space charge repulsion. To longitudinally compress the bunch to the required $3$\,mm a compensation energy chirp should be introduced which is typically done with an RF buncher. In order to provide linear energy modulation the frequency of the buncher should be selected to have bunch flight time at the buncher shorter than 10$^{\circ}$ of its RF phase. For the bunch charge of $320$\,pC which has an RMS buncher flight time of $36$\,ps the required frequency should be less than 775 MHz. Further increase of the bunch charge leads to an increase of the bunch flight time and may require even lower buncher frequency. Practically attractive is $400.8$\,MHz - the first sub-harmonic of the PERLE default frequency.  Further gradual beam compression and acceleration can be provided with a booster consisting of a series of single cell $801.6$\,MHz cavities with individual coupling and control of amplitude and RF phase. As the energy transferred to the beam in the injector booster to reach $5$\,MeV is $60$\,kW and is not recovered, the precise number of cavities is defined by the maximum power which may be loaded into a single cavity with the coupler. Assuming that maximum coupler power is $20$\,kW the booster should consist of at least four cavities. Taking into account that the first two cavities are operated essentially off-crest and at low field as well as a required contingency in case of increasing injector energy, the number of the cavities should be increased to five. \\  
\vspace{-1cm}
\subsection{Summary on source and injector}
An analysis of the current scientific and technological level of the high average current electron sources for ERLs allows us to conclude that an unpolarised electron source with beam parameters required for PERLE may be built in a relatively short time. This would best be based on a $350$\,kV DC photocathode gun operated with Sb-based photocathodes followed by a buncher and superconducting booster consisting of five independently fed and controlled RF cavities. A design of a high current polarised electron source requires more investigation but is  considered to be a second step for PERLE. A baseline scheme, delivering an average current  $2 - 4$\,times less than in the unpolarised regime may be realised on the basis of an unpolarised source operating with a family of GaAs photocathodes and reduced DC gun at an operational voltage to 200 kV.


\section{Cavity Design}
PERLE will be a low to medium energy facility 
in several stages from 150-450-900 MeV for both technology validation
and a versatile test bench for high average current applications. 
This section will outline some key aspects of the linac cavity design 
and its optimization. Table~\ref{FIG:CAV} lists the cavity configurations
for the three phases of the ERL facility.
\subsection{Choice of operating frequency}
The choice of frequency and gradient is important for any project and depends on a range of factors. It is definitely not a one-size-fits all situation. For large projects, the total cost is dominated by a few competing items such as RF power, cryogenics, structure costs (e.g. modules) and conventional facilities (tunnel, surface buildings, penetrations, etc.). Each of these has a frequency and gradient dependence and depends on the choice of underlying technology assumed. In general the overall cost optimum is a balance between linear costs (such as structure and tunnel) which increase as the gradient is lowered and the machine gets longer, and quadratic terms such as RF power and cryogenic capacity, which increase as the gradient is increased but result in a shorter machine. The result is a rather broad cost minimum allowing some flexibility in the choice of frequency and gradient to accommodate other factors. There are various cost models in use or under development but in general the optimum frequency for this type of machine is somewhere between a few hundred MHz and one GHz. Below this range the structures become very expensive and above this range RF power costs increase. As has been extensively studied in the conceptual design of the LHeC the
frequency needs to be significantly below a GHz also for avoiding adverse effects 
due to beam breakup instability~\cite{Abelleira2012}.
For compatibility with the LHC, a harmonic of 
$200$\,MHz is highly desirable. A frequency of $801.58$\,MHz
 is a convenient harmonic\footnote{Note that $801.58$\,MHz is the $20^{th}$ harmonic of the bunch repetition frequency, and, since $20$ is not an integer multiple of $3$, the bunches of the three re-circulations cannot be equally spaced; this is discussed in more detail in Section\,\ref{bunchpat} above.}
  that is close to the estimated cost optimum and also compatible with other systems currently in use or under development at CERN~\cite{ERK,CALAGA1,CALAGA2}. 
The optimum gradient range is also quite wide, ranging from around 
$10$ to $20$\,MV/m depending on assumptions about the temperature and $Q_0$ that can be reliably expected. In general for a large machine the lowest reasonable gradient should be adopted to maximise reliability and minimise the chances of field emission. However, for a small machine like PERLE, at least in the first phase, the cost optimum may favour a higher gradient. 
\begin{table}[h]
\begin{center}
\begin{tabular}{|p{4cm}|ccccc|}\hline
\textbf{Parameter} & \textbf{LHeC}& \textbf{PERLE $\Phi_1$} &
& \textbf{PERLE $\Phi_2$}  & \textbf{PERLE $\Phi_3$} \\ \hline
Energy  [GeV]             &   60   & 0.15   & & 0.45      & 0.90    \\
Cells/Cavity               &            &  & 5              &&       \\
Gradient [MV/m]               &            &  & 18              &&       \\
Cav/Cryomodule        & 4-8       & 4      & & 4        & 4      \\
\# of Cryomodules/linac     &  44-22          & 1    &   & 1        & 2  \\ 
\# of Turns                &  3         &  1    & &   3    &   3 \\
RF Power/cavity [kW]        &   & & 5-50 & & \\ 
\hline
\end{tabular}
\caption{Design choices for the cavities and cryomodules for the LHeC and
different stages of PERLE. The default frequency is chosen to be $801.58$\,MHz,
see text. All stages of PERLE here considered, as well as the LHeC, are configured with two linacs.}
\label{FIG:CAV}
\end{center}
\end{table}

\subsection{Design considerations}  
The maximum accelerating gradient is primarily
limited by the CW power dissipated on the cavity walls. 
Due to the quadratic dependence, a medium  accelerating gradient
with the lowest surface resistance (high $Q_0$) at moderate to high 
gradients is required. 
The number of cells per cavity is a compromise between a reasonable ``real estate gradient" while reducing the probability of trapped modes. 

The salient feature of an energy recovery linac, at least in CW operation, is the continuous transfer of stored energy from the cavity to the accelerated beam and simultaneously the transfer of (almost equal) energy from the decelerated beam back into the cavity. To first order, the power fed into the cavity through the fundamental power coupler (FPC) from the power source is equal to the power losses in the cavity walls, which can be extremely small. Another formulation of this feature is that the net beam loading at the fundamental frequency is zero in spite of a large beam current. As a consequence, the excitation of HOMs, notably at frequencies where accelerated and decelerated beam currents are not in anti-phase, will be dominating the design.

\subsubsection{Initial design choices}
The choice of five cells per cavity  is  retained from  technical 
arguments derived in Ref.~\cite{CALAGA2}. 
The standard parameterisation for elliptical cavities is 
used~\cite{PAOLO}. Fig.~\ref{FIG:CAV} shows the envelope of the 
scaled five-cell cavity
with a large iris aperture diameter of $150$\,mm, 
scaled from an existing $704$\,MHz design.
\begin{figure}
\centering
\includegraphics*[width=0.8\textwidth]{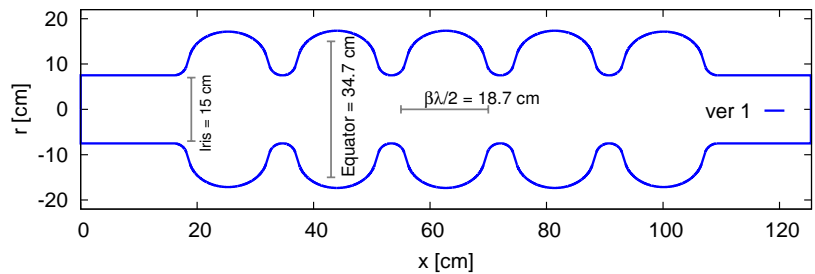}
\vspace{-0.5cm}
\caption{Envelope of the first proposal~\cite{CALAGA2} for a five-cell ERL cavity 
at $802$\,MHz.}
\label{FIG:CAV}
\end{figure}
%

%
Detailed parametric scans were carried out to further optimise the
aperture choice from the scaled version~\cite{CALAGA3}. Some key RF
parameters such as the ratio of B$_p$/E$_p$, R/Q, cell-to-cell coupling
for the fundamental and higher order modes, frequency dependence of 
the fundamental mode and HOMs were studied.
A first optimisation aimed at minimising the integrated longitudinal loss factor, which is a measure for the power lost into well-damped HOMs for very short bunches; for a beam current of $40$\,mA, the $150$\,mm diameter aperture (version 1) would result in a total HOM power of the order of $35$\,W.

The geometrical scans performed are
used as guidance considering both fundamental mode and HOMs.
An increase in aperture to $160$\,mm from  version 1
and adapting the other geometrical parameters leading to an 
optimum B$_p$/E$_p$ ratio, is a reasonable choice. 
This design will be referred to as version 2. An alternative ``low-loss''
like design was also considered; it is described below in Sect.\,\ref{seccavop}.

Relevant RF parameters for the mid-cell and five-cell geometries
are listed in Table~\ref{TAB:RFPARAM} and compared to the 
initial scaled version. 
\begin{table}[h]
\begin{center}
\begin{tabular}{|p{5.2cm}|cc|} \hline
\textbf{Parameter}  & \textbf{Ver 1 (Scaled)} & \textbf{Ver 2} \\ \hline
Frequency [MHz]       & 801.58 & 801.58  \\ 
Number of cells       & 5  & 5     \\ 
Active cavity length [mm] & 935  & 935 \\ 
Voltage [MV]           & 18.7  & 18.7  \\  
E$_p$ [MV/m]           & 45.1  & 48.0  \\ 
B$_p$ [mT]            & 95.4  & 98.3  \\ 
R/Q [$\Omega$]        & 430 & 393 \\ 
Cell-cell coupling (mid-cell)    & 4.47\%  & 5.75\%\\  
Stored Energy  [J]    & 154   & 141 \\ 
Geometry Factor  [$\Omega$]    & 276   & 283 \\ 
Field Flatness        & 97\% & 96\%\\  \hline
\end{tabular}
\caption{RF parameters of five-cell geometry for version 2
compared to that of the scaled initial version. }
\label{TAB:RFPARAM}
\end{center}
\end{table}

\subsubsection{Impedance spectra}
\begin{figure}
\centering
\includegraphics*[width=0.7\textwidth]{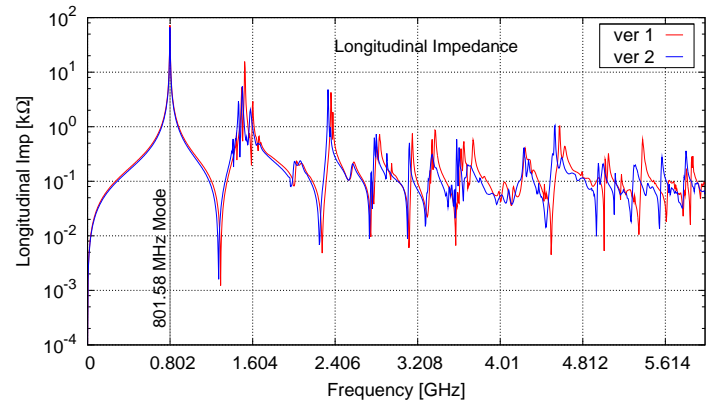}
\caption{The impedance spectra for the longitudinal modes as a 
function of frequency compared between the initial two versions 1 and 2.
The vertical grid shows harmonics of the fundamental mode.}
\label{FIG:IMP}
\end{figure}

\begin{figure}
\centering
\includegraphics*[width=0.7\textwidth]{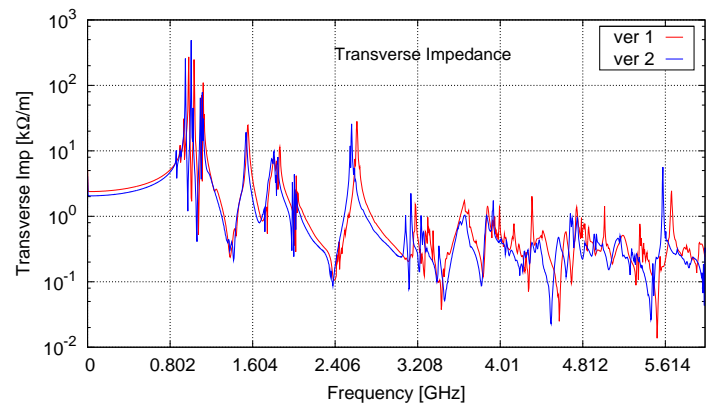}
\caption{Impedance spectra for the transverse modes as a function 
of frequency compared between the two versions. The vertical grid shows harmonics
of the fundamental mode.}
\label{FIG:IMP2}
\end{figure}

The longitudinal impedance spectrum calculated in time domain for both versions
 are shown 
in Fig.~\ref{FIG:IMP}. This first two to three 
monopole pass-bands pose the highest impedance and do not easily 
propagate into the beam pipes requiring targeted HOM couplers to damp 
them to sufficiently low values.
In the transverse plane, see Fig.~\ref{FIG:IMP2}, a few passbands of interest with primarily 
the two first bands (TE$_{11}$ and TM$_{11}$) being at least an order 
of magnitude higher than the rest. Similar to the longitudinal plane,
transverse impedances at 
frequencies above $2.8$\,GHz are significantly smaller in impedance 
and above the cutoff of the beam tube. 

Detailed simulations with loop-like coaxial HOM 
couplers are underway to determine the level of damping achieved for the
lowest order HOMs which pose the highest risk.

\subsubsection{Loss factors and  HOM power}
\begin{figure}
\centering
\includegraphics*[width=0.7\textwidth]{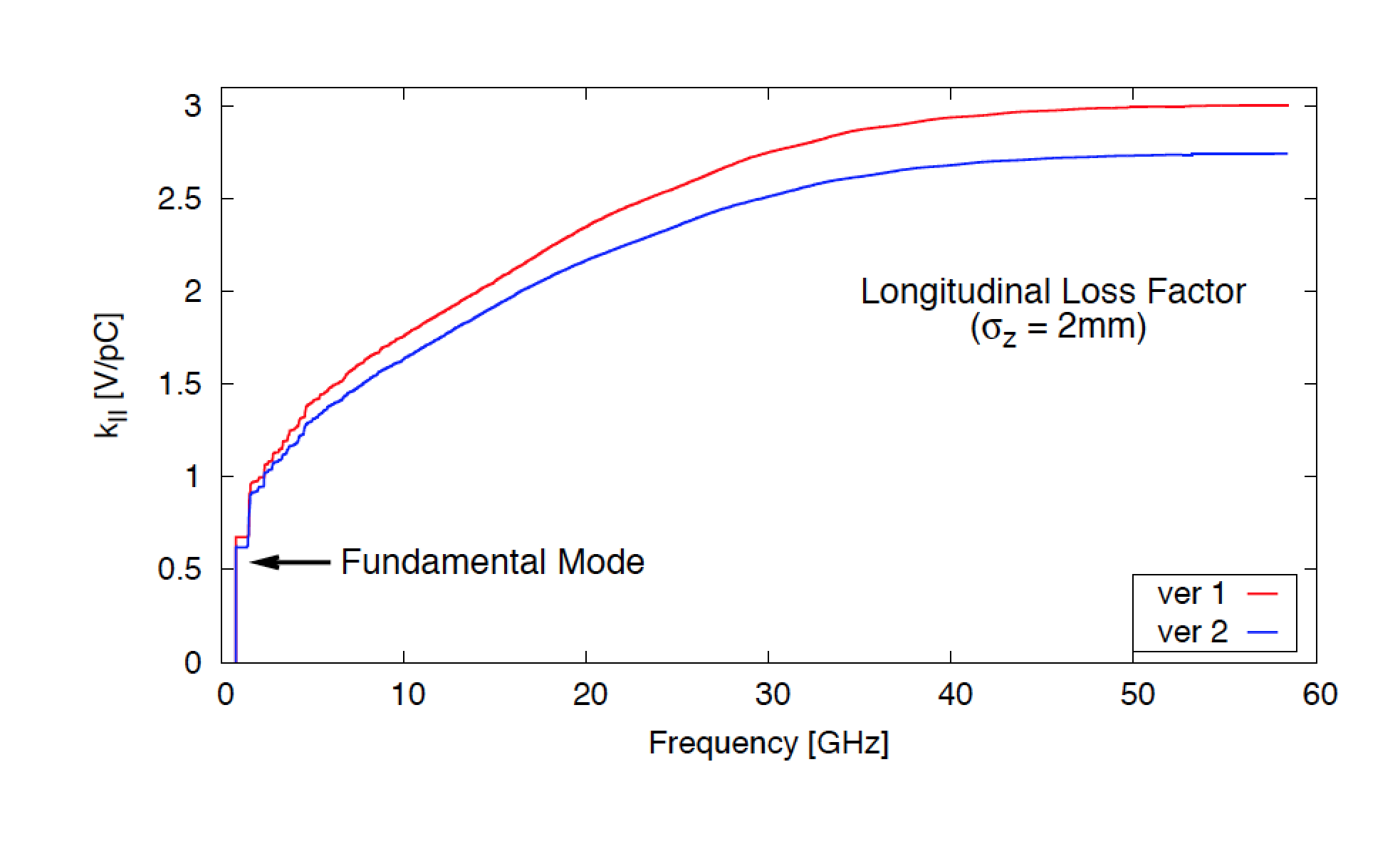}
\caption{Integrated longitudinal loss factor for the two initial versions as a function 
of frequency, for an assumed bunch length of $2$\,mm.}
\label{FIG:LF}
\end{figure}

The very small bunch length can excite frequencies well up to 
$50$\,GHz or above. This is characterised by the 
longitudinal loss factor $k_{||}$. Fig.~\ref{FIG:LF} shows the
frequency dependence of the
integrated loss factors for the initial two versions of the cavity. 


In addition to HOM damping, the induced HOM power from the short 
bunches is of the order of $35$\,W for the nominal bunch charge of 
$0.32$\,nC and average beam current of $40$\,mA, for three passes. This level
of power can easily be handled by loop-coupled couplers. However, resonant
excitation of a HOM can easily lead to powers in the $1 -2$\,kW range (assuming
$R/Q = 50$ $\Omega$ and Q$_{ext} = 10^4$). Therefore, the couplers will 
have to be designed to handle this power and impose the condition of 
HOM impedance to not exceed $500$\,k$\Omega$ for the longitudinal modes.
For transverse modes, single and multi-bunch simulations have to be
carried out to determine the acceptable damping levels. 
The effect of the transition sections using tapers and bellows
is already discussed in Ref.~\cite{CALAGA2}.

\subsubsection{External $Q$ and power requirements}
\begin{figure}
\centering
\includegraphics*[width=0.7\textwidth]{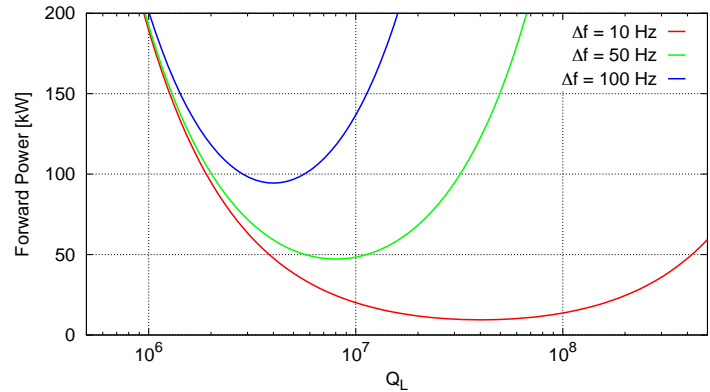}
\caption{Forward power as a function of the loaded $Q$, for $Q_L \simeq Q_{ext}$, 
of the cavity for different detunings and zero beam loading.}
\label{FIG:POWER}
\end{figure}

Considering the steady state condition of recirculating 
beams and energy recovery only, the beam loading can be assumed
to be small. Then the input RF power required to maintain 
the cavity voltage is directly proportional to the peak detuning,
see Fig.\,\ref{FIG:POWER}.

A realistic $Q_{ext} \sim 10^7$ with a corresponding power of $50$\,kW 
will allow for sufficient margin during transients. At these power levels 
and frequency range, standard UHF television IOTs become an attractive 
and robust option. 

\subsection{Cavity optimisation}
\label{seccavop}
The cavity cell shape should be carefully optimised to balance accelerating mode efficiency with HOM damping needs (loaded Q$\text{'}$s) and HOM power extraction (HOM frequencies relative to the high current lines in the beam spectrum), as well as mechanical and cleaning considerations. Shapes such as the JLab ERL high-current profile \cite{Rimmer12th} and BNL3 ERL \cite{Wencan2011} cavity are good examples.
Starting from these so-called ``Low-Loss" shapes, which feature cavity shapes with a steep wall angle down to $0^\circ$, led to the cavity optimisation described here. The low-loss type profile (vertical wall) and contoured irises produce moderate surface magnetic and electric field enhancements normalised to the accelerating gradient; the vertical walls also are the main difference compared to the initial designs with larger inner diameter describe above. This is a one - die design, meaning all the cell cups are produced from the same profile with the end cells simply being trimmed shorter to tune for field flatness.

 Extracting HOM power from the cavities to room temperature absorbers must be considered in the cryomodule design (see below). Very effective HOM damping can be achieved by absorbers on the beamline either side of the cavity, providing the beam pipe is sufficiently enlarged to allow the dangerous HOMs to propagate. These, however, consume valuable space and the absorbers must be thermally isolated from the cold beamline components. The JLab waveguide damping scheme \cite{Rimmer12th} avoids this by taking the HOM power out sideways to warm loads but is probably overkill for the LHeC requirements.
As already indicated above, loop-coupled HOM dampers, possibly similar to the LHC type mounted on the ends of the cavity close to the end cell, will be sufficient.
An example of the implementation of these couplers is described in detail in 
Sect.\,\ref{seccryomod} below. Many other configurations are of course possible. For this type of coupler, the HOM power is removed via a cable to a warm termination. This also allows easy monitoring of the HOM signals for diagnostic purposes.

\begin{figure}
\hspace{16mm}
\includegraphics[width=0.73\textwidth]{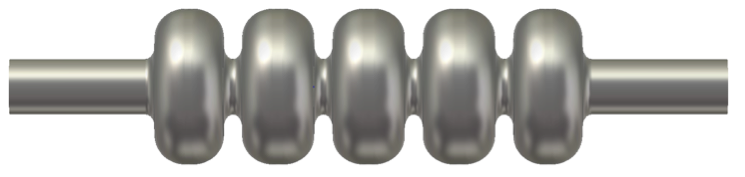}
\begin{center}
\includegraphics[width=0.8\textwidth]{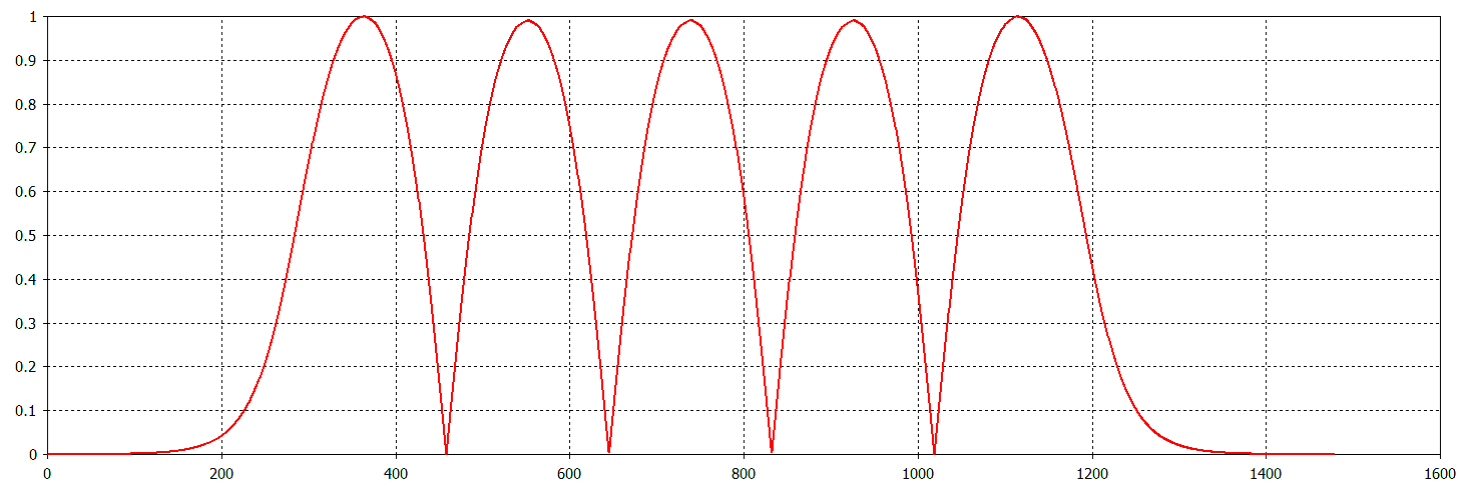}
\caption{Cavity design (single-die, iris ID=tube ID) $801.58$\,MHz (top); 
Axial field on axis (bottom).}\label{fig35.png}
\end{center}
\end{figure}

Fig.~\ref{fig35.png} shows a potential candidate cavity shape optimised for the PERLE and LHeC applications, it uses a median iris diameter (= tube) of $130$\,mm. The main parameters of the selected shape are listed in Table~\ref{tabcaverk}, comparing it to a subset of the shapes investigated in this study with iris diameters varying from $115$ to $160$\,mm and limited to solutions with equal iris and tube diameters.
\begin{table}
\begin{center}
\begin{tabular}{|p{3.5cm}|c|c|c|c|c|} \hline
Parameter&Unit& Jlab$_1$ & Jlab$_2$ & CERN$_1$ & CERN$_2$ \\
\hline 
Iris & mm & 115 & 130 & 150 & 160 \\
Frequency & 	MHz &	802	& 802 &	801.58 & 	801.58  \\
L$_{active}$ &	mm	& 922.14	& 917.911	& 935	& 935 \\
$R/Q = V_{eff}^2/(\omega W)$ &	$\Omega$	& 583.435 &	523.956 &430 &	393 \\

Integrated $k_{loss}$ &  V/pC	& 3.198	& 2.742	& 2.894	& 2.626 \\
(R/Q)/cell & $\Omega$ &	116.687	& 104.7912	& 86 &	78.6 \\
G & $\Omega$ & 273.2 & 	274.717 &	276 &	283 \\
(R/Q) $\cdot$ G /cell & $\Omega^2$ &		31877	& 28788 &	23736	& 22244 \\
Equator diameter &	mm	& 323.1 &	328.0 &	350.2 &	350.2 \\
Wall angle  & degree & 0 & 0& 14 & 12.5 \\
$E_{pk}/E_{acc}$ && 		2.07	&2.26 &	2.26 &	2.40 \\
$B_{pk}/E_{acc}$ &	$10^{-9} s/m$ & 	4.00	& 4.20 & 	4.77 &	4.92 \\
$k_{cc}$ & \% &		2.14	&3.21	& 4.47	& 5.75 \\
$N^2 / k_{cc}$ & & 		1168 &	778 &	559 &	435 \\
cutoff $TE_{11}$ &	GHz &	1.53 &	1.35 &	1.17 &	1.10 \\
cutoff $TM_{01}$ & 	GHz &	2.00 &	1.77 &	1.53 &	1.43 \\
$E_{acc}$ & 	MV/m &	20.3 &	20.4 &	20.0 &	20.0 \\
$E_{pk}$ &	MV/m	& 42.0 &	46.1 &	45.1 &	48.0 \\
$B_{pk}$ &	mT	& 81.1 & 	85.5 &	95.4 &	98.3 \\
\hline
\end{tabular}
\caption{Parameters of a subset of cavity shapes studied during the cavity optimisation. Each cavity has 5 cells and a nominal effective voltage of
$18.7$\,MV.}\label{tabcaverk}
\end{center}
\end{table}

\begin{figure}
\begin{center}
\includegraphics[width=0.8\textwidth]{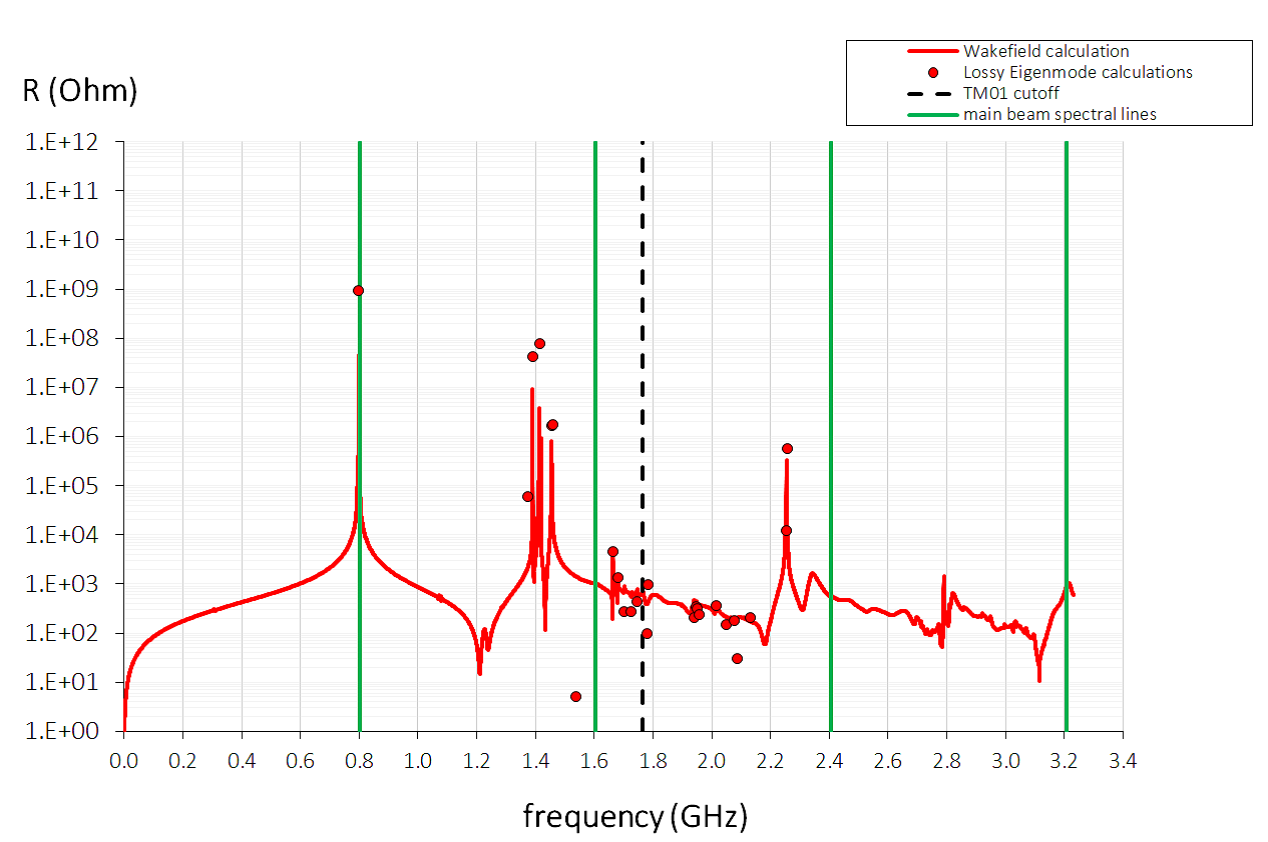}
\caption{
Impedance spectrum for the longitudinal modes as a function of frequency of the low-loss cavity design with iris diameter of $130$\,mm 
(compare Fig.\,\ref{FIG:IMP}).}
\label{fig3nine}
\end{center}
\end{figure}

Normalised to $\lambda$, the beam tube and iris diameter of the selected solution are slightly larger than the TESLA or CEBAF upgrade (LL) shapes, but smaller than the original CEBAF (OC) or JLab high-current (HC). This allows good cell-to-cell coupling for HOM damping and reduced sensitivity to fabrication errors, while preserving high shunt impedance for the operating mode for good efficiency. The outer part of the cell profile is tuned to keep harmful HOMs far away from beam harmonics.
 Figure\,\ref{fig3nine}
shows the monopole spectrum of the cavity calculated from a long-range wakefield simulation with matched terminations on the beam pipes but no other HOM absorbers (similar to Fig.\,\ref{FIG:IMP} above). Note that modes below the beam tube cutoff are unresolved and their final amplitudes and their $Q$'s will depend on the HOM damping configuration, but all modes are well separated from the RF harmonics. 

\begin{figure}
\begin{center}
\includegraphics[width=0.8\textwidth]{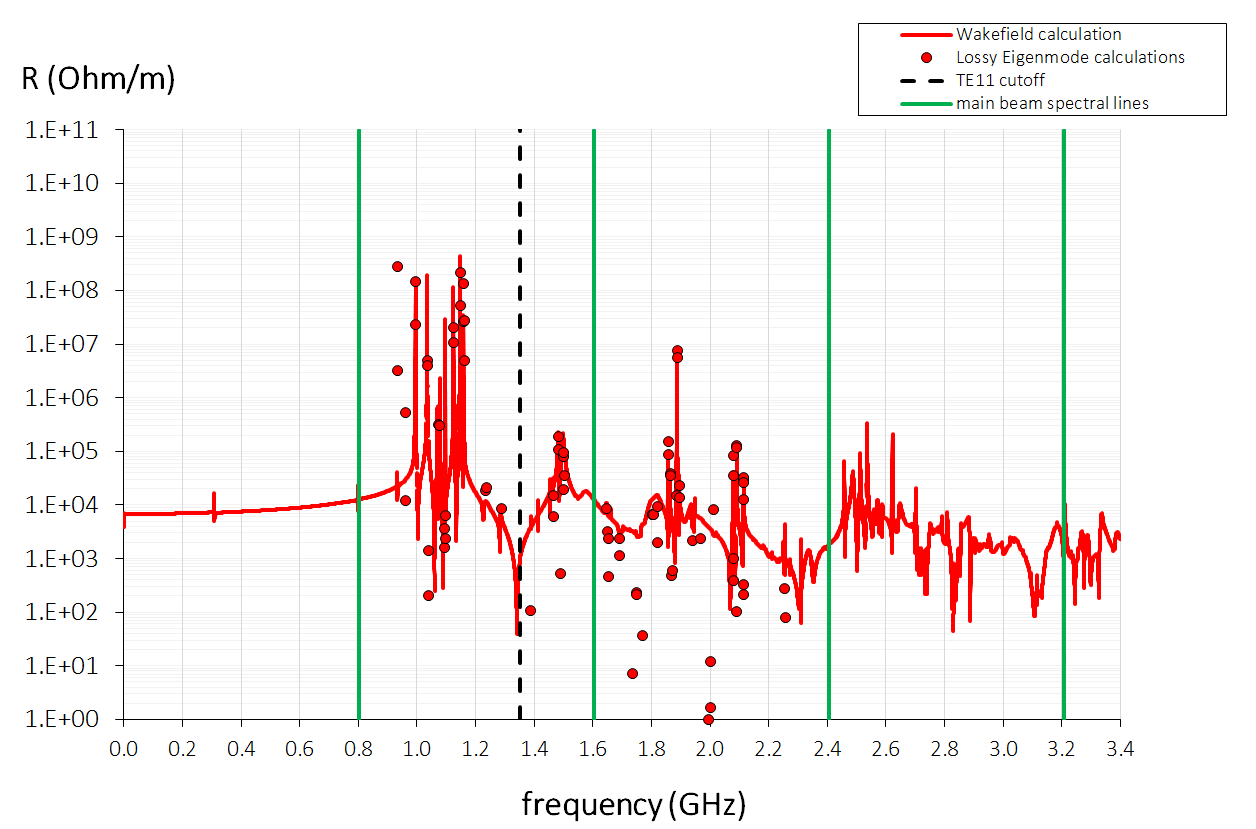}
\caption{
The impedance spectrum for the transverse modes as a function of frequency of the low-loss cavity design with iris diameter of 130 mm (compare Fig.\,\ref{FIG:IMP2}).}
\label{fig3ten}
\end{center}
\end{figure}

 Figure\,\ref{fig3ten}
shows the dipole spectrum, which is similarly well separated from harmful frequencies. The low-loss type profile (vertical wall) and contoured irises produce moderate surface magnetic and electric field enhancements, normalised to the accelerating gradient. This is a one-die design, meaning all the cell cups are produced from the same profile with the end cells simply being trimmed shorter to tune for field flatness.

\subsection{Summary on the cavity design}
The first scaled version of the $802$\,MHz ERL cavity was further 
optimised. Moderate improvement of the HOM performance was obtained with a 
small increase in aperture with the consequence of about 10\% decrease  in the fundamental mode R/Q. Given the short bunches and moderately high currents, version 2 (Jlab$_2$ in Tab.\,\ref{tabcaverk}) is considered as a baseline towards realising
a first prototype. Detailed studies including the fundamental power 
coupler and HOM couplers are ongoing to finalise the cavity geometry and  the optimum placement of the couplers. 


\section{Cryo Module}
\label{seccryomod}
\begin{figure}
\begin{center}
\includegraphics[width=0.8\textwidth]{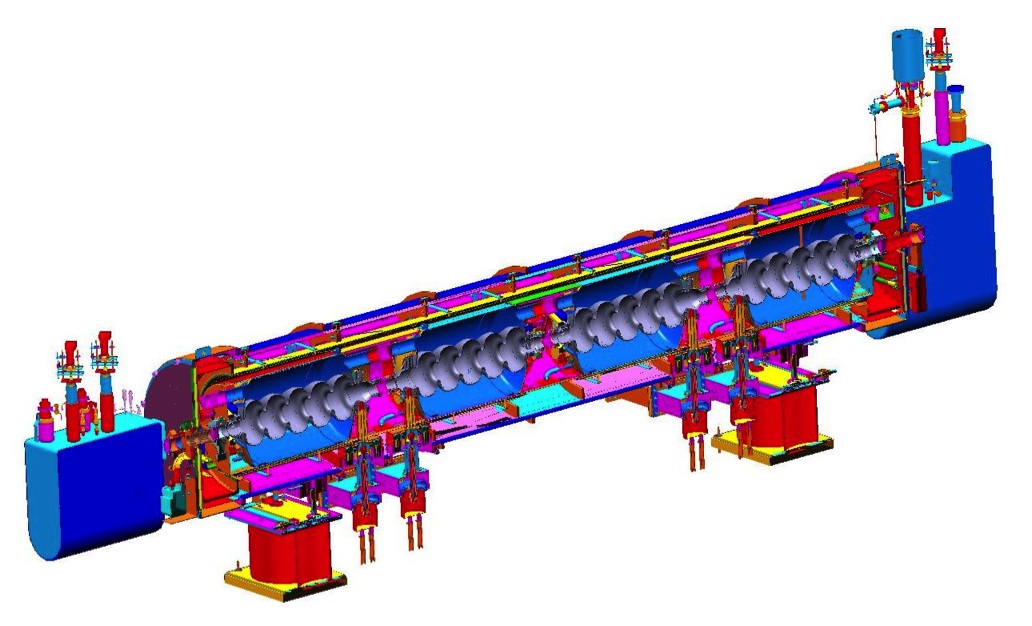}
\caption{SNS high $\beta$ module adapted to house $\beta$ =1 5-cell cavities for LHeC.}\label{fig39.png}
\end{center}
\end{figure}

\begin{figure}
\begin{center}
\includegraphics[width=0.8\textwidth]{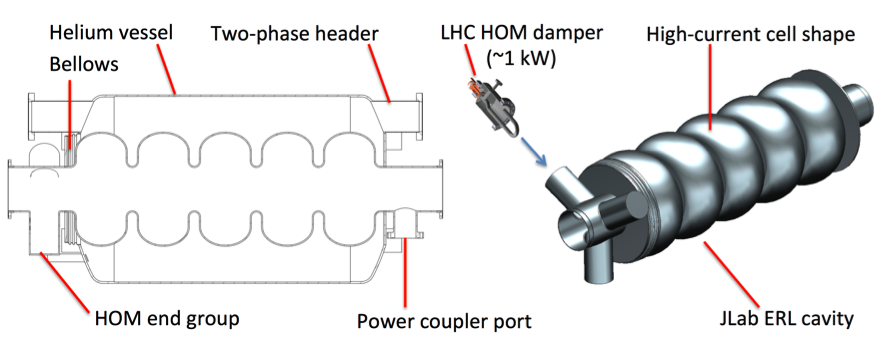}
\caption{Concept for cavity and helium vessel arrangement.}\label{fig40.png}
\end{center}
\end{figure}

\begin{figure}
\begin{center}
\includegraphics[width=0.8\textwidth]{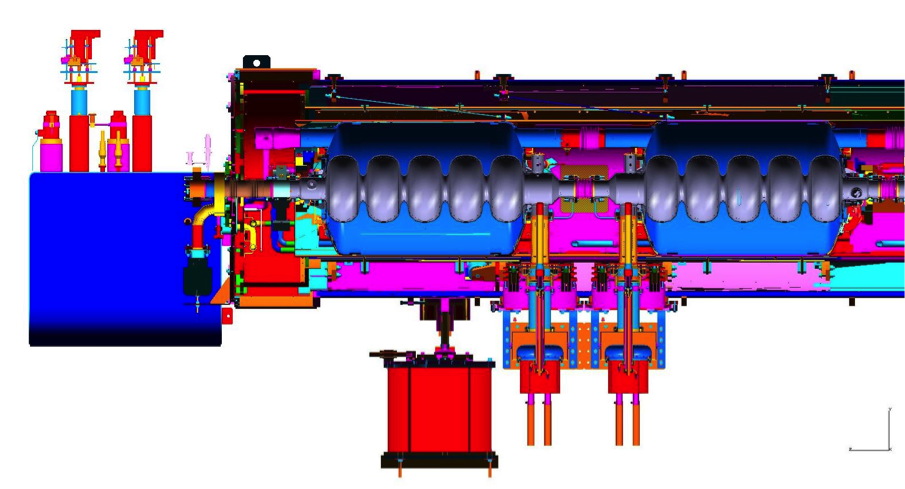}
\caption{Cavity, coupler and end can detail view.}\label{fig41.png}
\end{center}
\end{figure}

PERLE comprises up to four cryo modules each containing four $802$\,MHz five-cell cavities. A convenient concept for these may be developed by adapting the four-cavity SNS high beta cryo module designed by JLab~\cite{Schneider2001}, to accommodate 5-cell $\beta$=1 cavities, as is shown in Fig.~\ref{fig39.png}. Since the cavities are almost the same length as the original $805$\,MHz 
 $\beta= 0.81$ 6-cells, no major changes to the module would be required. This design uses a single, large volume helium vessel for each cavity, 
 Fig.~\ref{fig40.png}, with the vessels connected by a two-phase pipe to allow gas and liquid to pass freely along the module. No separate gas return or two-phase pipes are needed.  At the ends of the module this header is connected to supply and return end cans that contain the bayonet connections, valves, reliefs, etc., Fig.~\ref{fig41.png}. The valve boxes are offset from the centerline of the module to accommodate short warm interconnecting sections between the modules for magnets, vacuum pumps, correctors, BPM$\text{'}$s etc. Each helium vessel has an end-mounted, Saclay-type tuner~\cite{Hogan2001} and there are bellows between the cavities that minimise mechanical cross talk during tuner operation. On the other end of each cavity, there is a coaxial fundamental power coupler~\cite{Wilson2001} developed from the Tristan design at KEK. The cavities are suspended from a warm space-frame by low conductivity rods. The couplers are at longitudinal fixed points in the support scheme so only have to accommodate radial motion during cool down. This is achieved with an external warm bellow in the top hat connection. There are no cold bellows or indeed any bellows in the RF section of the coupler. For SNS, the cold part of the outer conductor is trace cooled with counter-flowing helium gas to minimise the heat load to $2$\,K. This gas flow is controlled by a separate dedicated valve. This active cooling may not be required for LHeC. The module could also be adapted to use an LHC type or other proven coupler.
 
The helium vessel may be titanium like the SNS modules or stainless steel like the CEBAF 12 GeV upgrade modules. For Titanium, a NbTi transition piece is used adjacent to the end irises to connect the helium vessel to the cavity and titanium bellows are used. For stainless steel, a Nb to stainless brazed joint can be used and the vessel bellows and piping can all be stainless steel. Care must be taken to avoid introducing permeable or magnetic material close to the cavity.
Fig.~\ref{fig40.png} shows a concept with provision for three such couplers mounted symmetrically on the end group to share the damping duties without introducing any dipole perturbation to the cavity mode or any asymmetry between damping of different dipole mode orientations. Many other configurations are of course possible. For this type of coupler the HOM power is removed via a cable to a warm termination, or taken outside the module where it can be monitored for diagnostic purposes.

%
The measured static loads at $2$\,K of the SNS type cryo-module were typically less than the $28$\,W budget, and shield static load was less than the $200$\,W budget at $\sim 50$\,K (inlet $40$\,K, outlet up to $80$\,K). For LHeC the dynamic loads of the CW cavities will be much higher than the pulsed SNS cavities.  For standard Nb material at $2$\,K dynamic heat loads of $30 - 40$\,W per cavity at $18.7$\,MV/m with $Q_0 \sim 2 \cdot 10^{10}$ may be expected. Thus the maximum dynamic load per module may approach $160$\,W,  with total $2$\,K load less than $190$\,W. This is well within the capacity of the helium circuit and end cans. Advances in surface treatment such as nitrogen or titanium doping, use of ingot niobium, Nb$_3$Sn or other improvements may significantly lower this number.

\begin{figure}
\begin{center}
\includegraphics[width=0.8\textwidth]{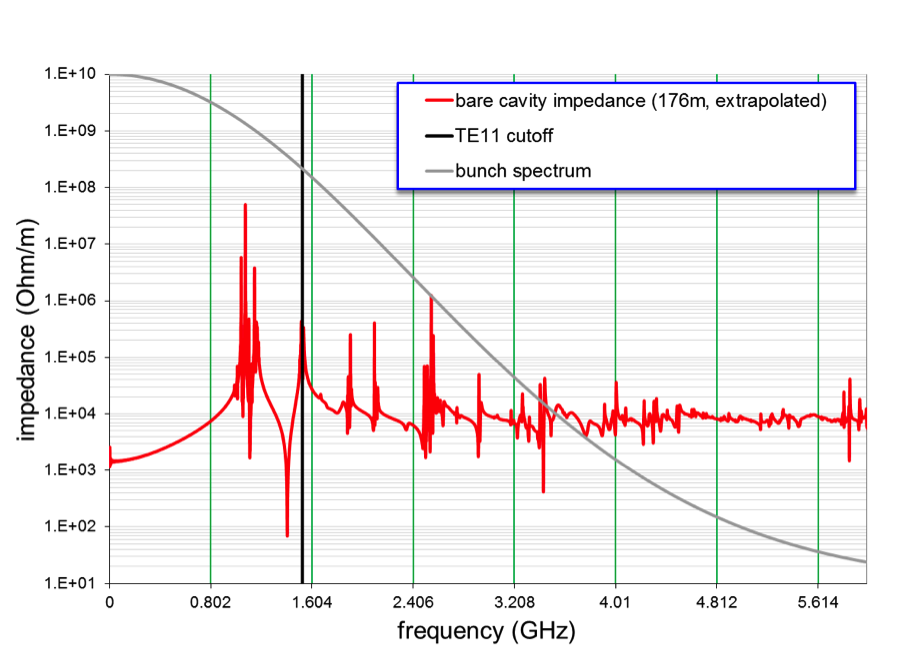}
\caption{Dipole HOM spectrum of the bare cavity. All harmful modes are well separated from RF harmonics. Impedances of modes below cut-off are unresolved and will be determined by the HOM damping configuration.}\label{fig42.png}
\end{center}
\end{figure}

The SNS cryo-module is therefore a convenient model for PERLE and could be adapted with minimal changes to host the new $802$\,MHz 5-cell $\beta = 1$ cavities. A new concept \cite{Rimmer2015} using many of the design features of this module, as well as attractive features of other JLab designs, is being developed for the JLab Electron Ion Collider~\cite{Lin2015}.
Features of that module might also be considered for an eventual LHeC production cryo-module. A simple cavity design has been developed that is a favourable balance between good HOM properties and good operating efficiency. Further refinement and optimisation of these concepts is expected in the near future.

\begin{figure}
\begin{center}
\includegraphics[width=0.8\textwidth]{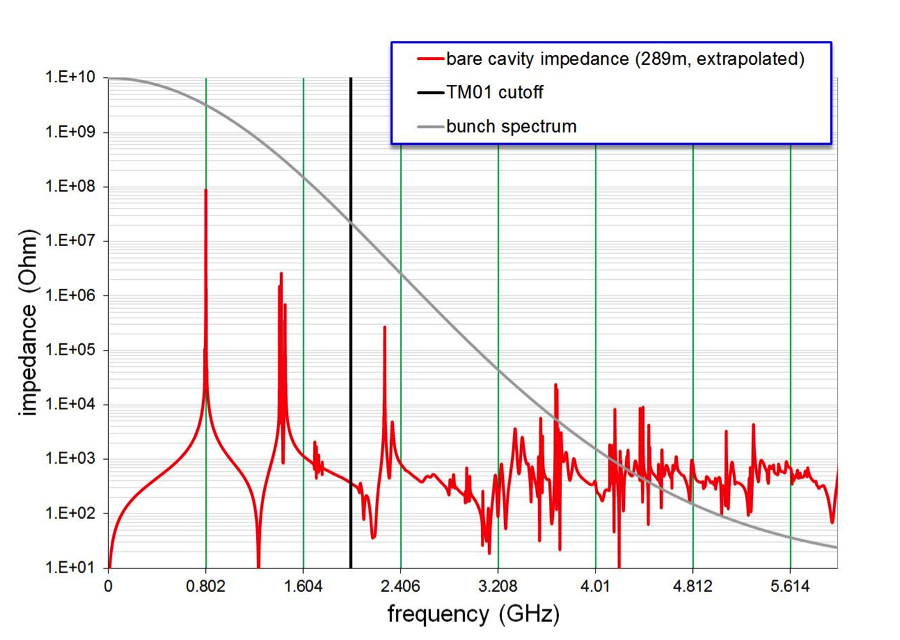}
\caption{Monopole HOM spectrum of the bare cavity. All harmful modes are well separated from RF harmonics. Impedances of modes below cut-off are unresolved and will be determined by the HOM damping configuration.}\label{fig43.png}
\end{center}
\end{figure}


\clearpage

\section{Arc Magnets}
The inventory of the main magnets for PERLE lists:
\begin{itemize}
	\item{40 bending magnets (vertical field)}
	\item{114 quadrupole magnets}
	\item{Bending magnets (horizontal field) in the spreaders and combiners}
	\item{Quadrupoles in the spreaders / combiners and in the injection / extraction parts}
\end{itemize}

A sketch of the arcs is given in Figs.~\ref{fig:arc_1_magnets} to \ref{fig:arc_6_magnets}, together with the main characteristics of the bending magnets and quadrupoles. The regions of the spreaders and combiners are not considered here, as these will need a dedicated analysis in view of the limited space available. In all cases, the vertical full gap of the dipoles is taken as 40 mm, and a similar dimension is taken for the horizontal extent of their good field region. Also the quadrupoles feature the same aperture throughout the arcs, which is fixed at 40 mm diameter.

\begin{figure}[H]
  \vspace{20mm}
	\centerline{\includegraphics[trim = 20mm 130mm 25mm 110mm, clip]{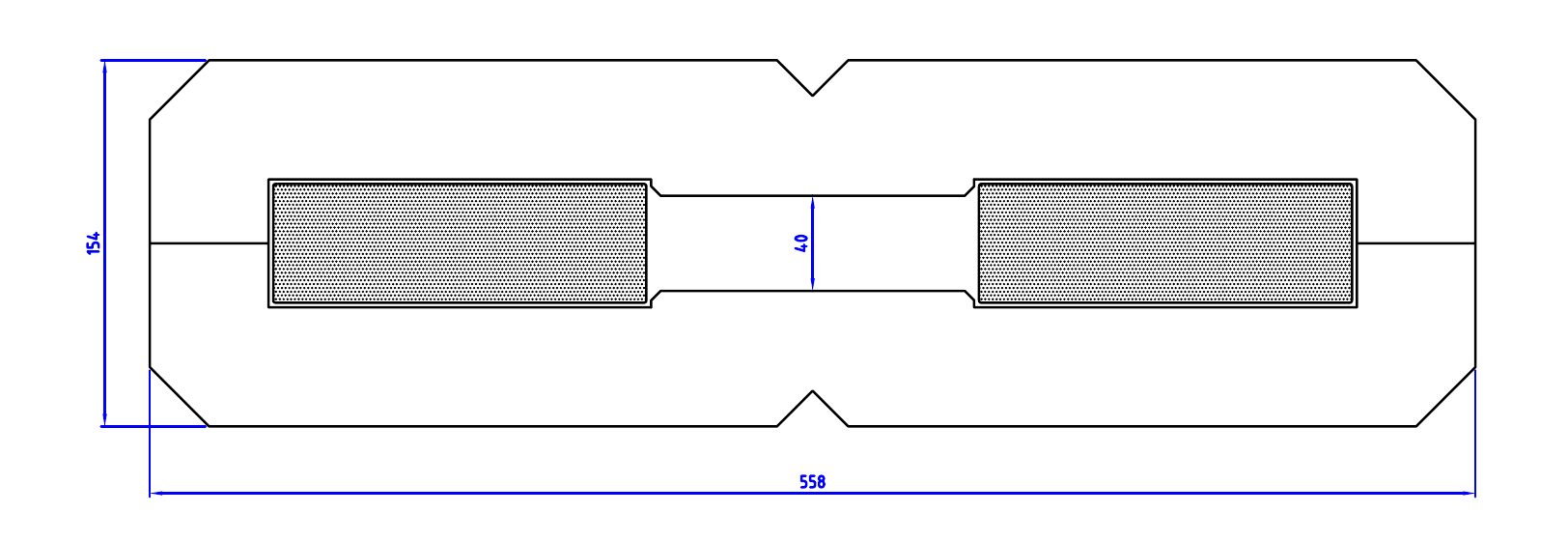}}
	\caption{Preliminary cross-section of bending magnets.}
  \label{fig:bending_magnet}
\end{figure}

In the lowest energy arcs, i.e. arc 1 and 2, there are four dipoles, with a $45^\circ$ bending angle. The higher energy arcs have on the other hand eight dipoles of $22.5^\circ$ each. Two families of bending magnets are then proposed: one to cover arcs 1 and 2, and another for arcs 3 to 6. The same cross-section could be used for both, though they would differ in terms of length and curvature radius. In both cases a curved construction is assumed, with possibly machined yokes. A tentative cross-section is shown in Fig.~\ref{fig:bending_magnet}. An H type yoke is proposed, rather narrow in the vertical direction, to minimize the vertical distance between the arcs. The dimensions could be further reduced -- in particular horizontally -- after an iteration on the required field quality. The coils will need to be designed as part of an overall optimization, including the power converters. The shaded area in Fig.~\ref{fig:bending_magnet} refers to 6-7 A/mm$^2$ of current density at the maximum field of 1.31~T of arc~6. 

\begin{figure}[H]
  \vspace{20mm}
  \centerline{\includegraphics[trim = 20mm 125mm 25mm 110mm, clip]{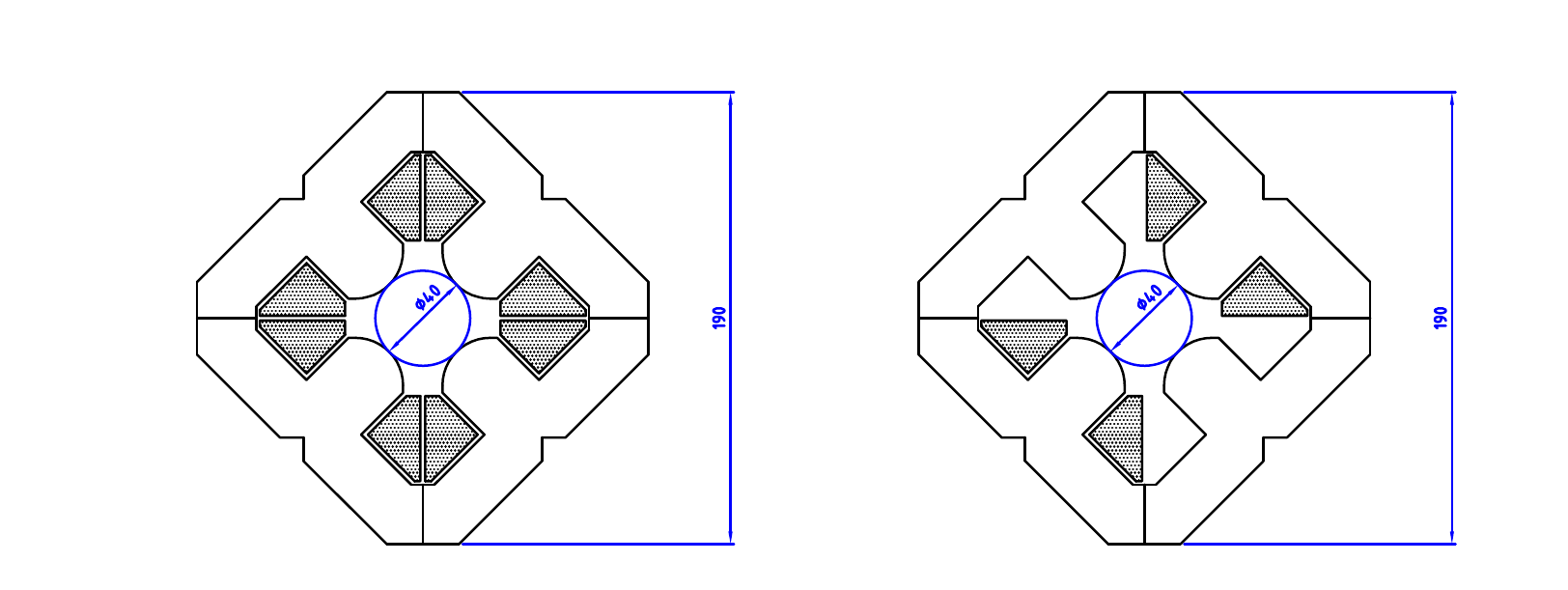}}
	\caption{Preliminary cross-section of quadrupole magnets.}
	\label{fig:quad_magnet}
\end{figure}

\begin{figure}[H]
  \vspace{30mm}
	\centerline{\includegraphics[trim = 30mm 72mm 25mm 70mm, clip]{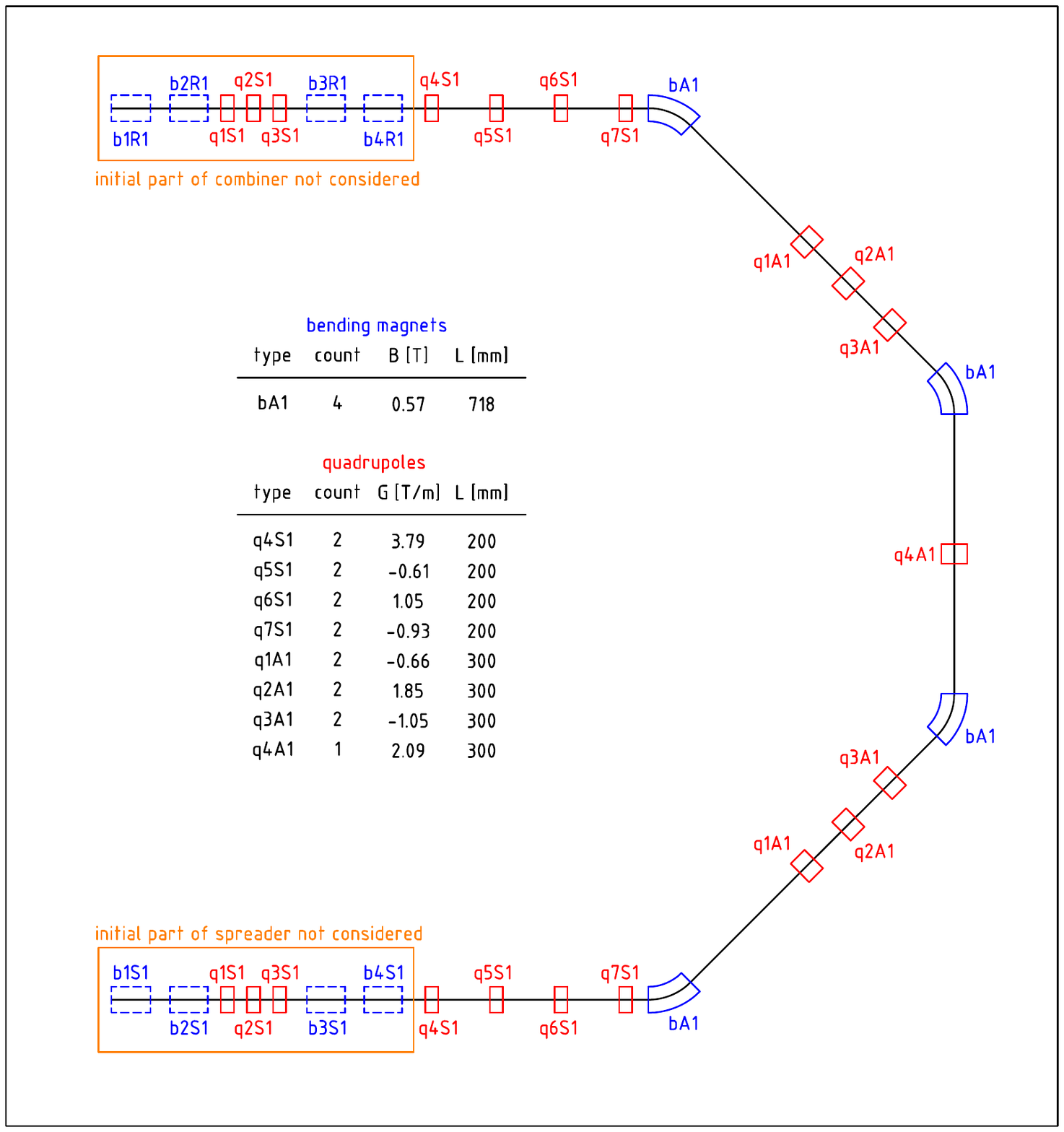}}
	\vspace{15mm}
	\caption{Arc 1 and main magnets, where $b$ denotes bending 
	and $q$ quadrupole magnets.}
	\label{fig:arc_1_magnets}
\end{figure}

\begin{figure}[H]
  \vspace{30mm}
	\centerline{\includegraphics[trim = 30mm 72mm 25mm 70mm, clip]{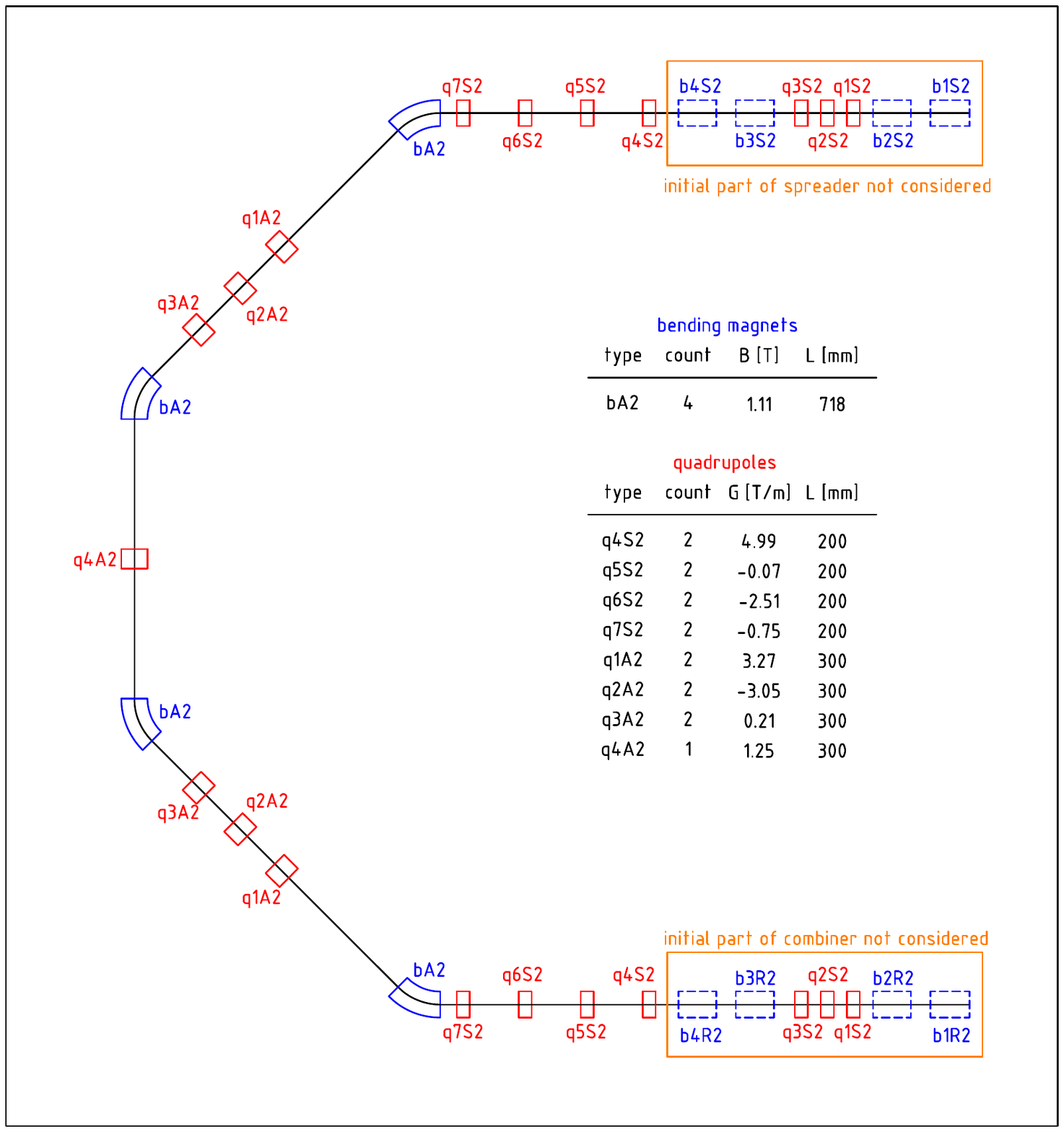}}
	\vspace{15mm}
	\caption{Arc 2 and main magnets.}
	\label{fig:arc_2_magnets}
\end{figure}

\begin{figure}[H]
  \vspace{30mm}
	\centerline{\includegraphics[trim = 30mm 72mm 25mm 70mm, clip]{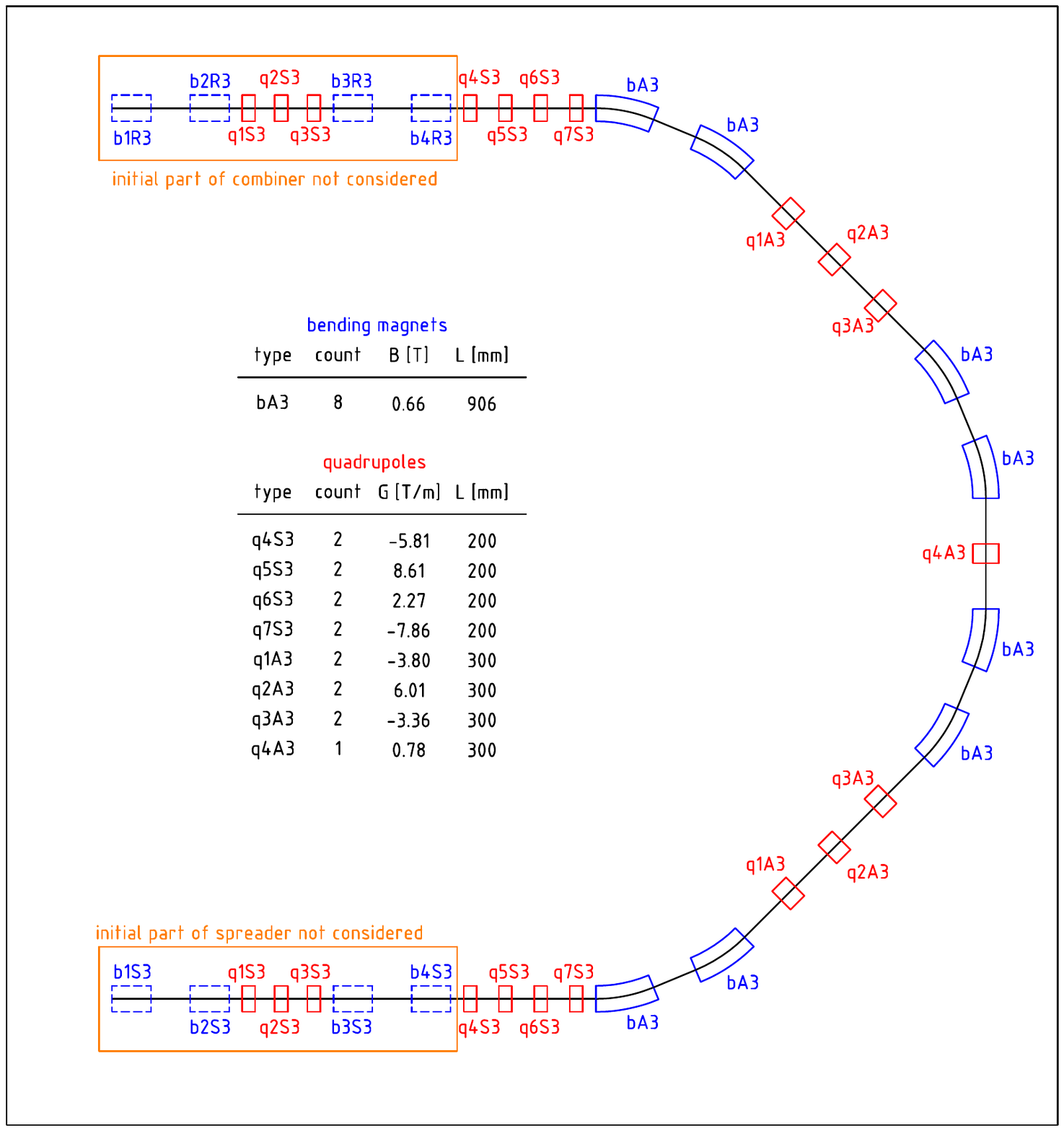}}
	\vspace{15mm}
	\caption{Arc 3 and main magnets.}
	\label{fig:arc_3_magnets}
\end{figure}

\begin{figure}[H]
  \vspace{30mm}
	\centerline{\includegraphics[trim = 30mm 72mm 25mm 70mm, clip]{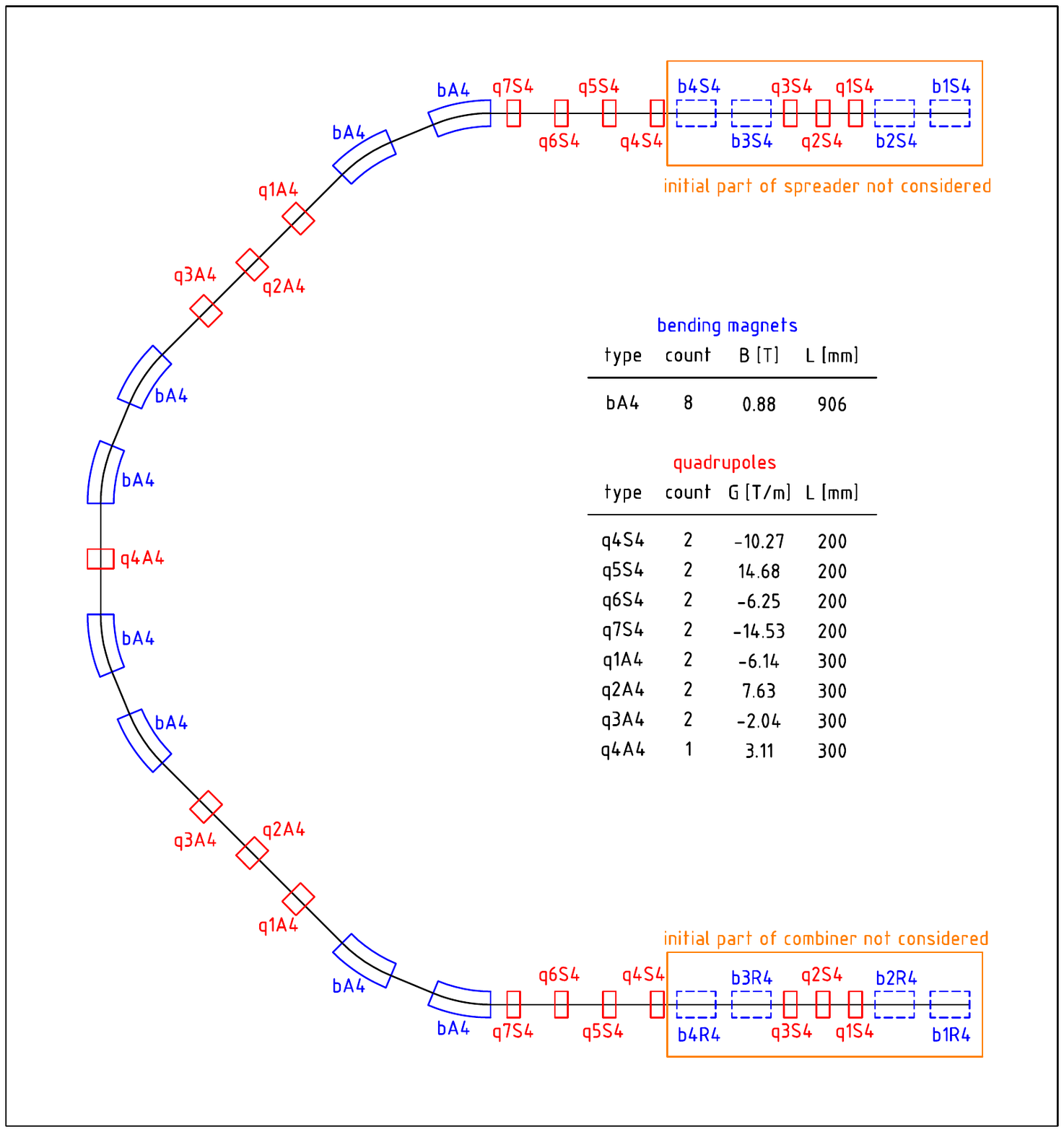}}
	\vspace{15mm}
	\caption{Arc 4 and main magnets.}
	\label{fig:arc_4_magnets}
\end{figure}

\begin{figure}[H]
  \vspace{30mm}
	\centerline{\includegraphics[trim = 30mm 72mm 25mm 70mm, clip]{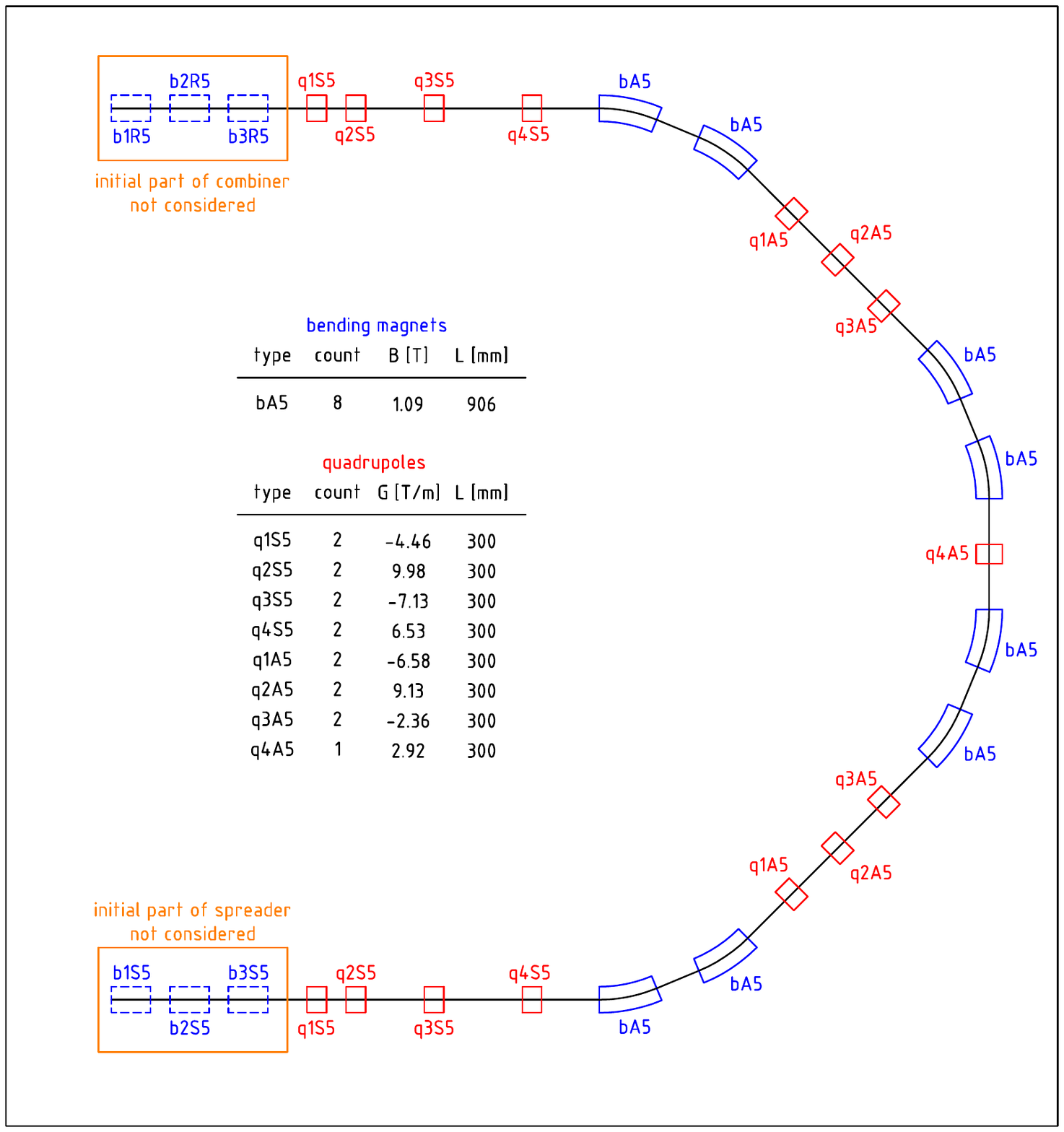}}
	\vspace{15mm}
	\caption{Arc 5 and main magnets.}
	\label{fig:arc_5_magnets}
\end{figure}

\begin{figure}[H]
  \vspace{30mm}
	\centerline{\includegraphics[trim = 30mm 72mm 25mm 70mm, clip]{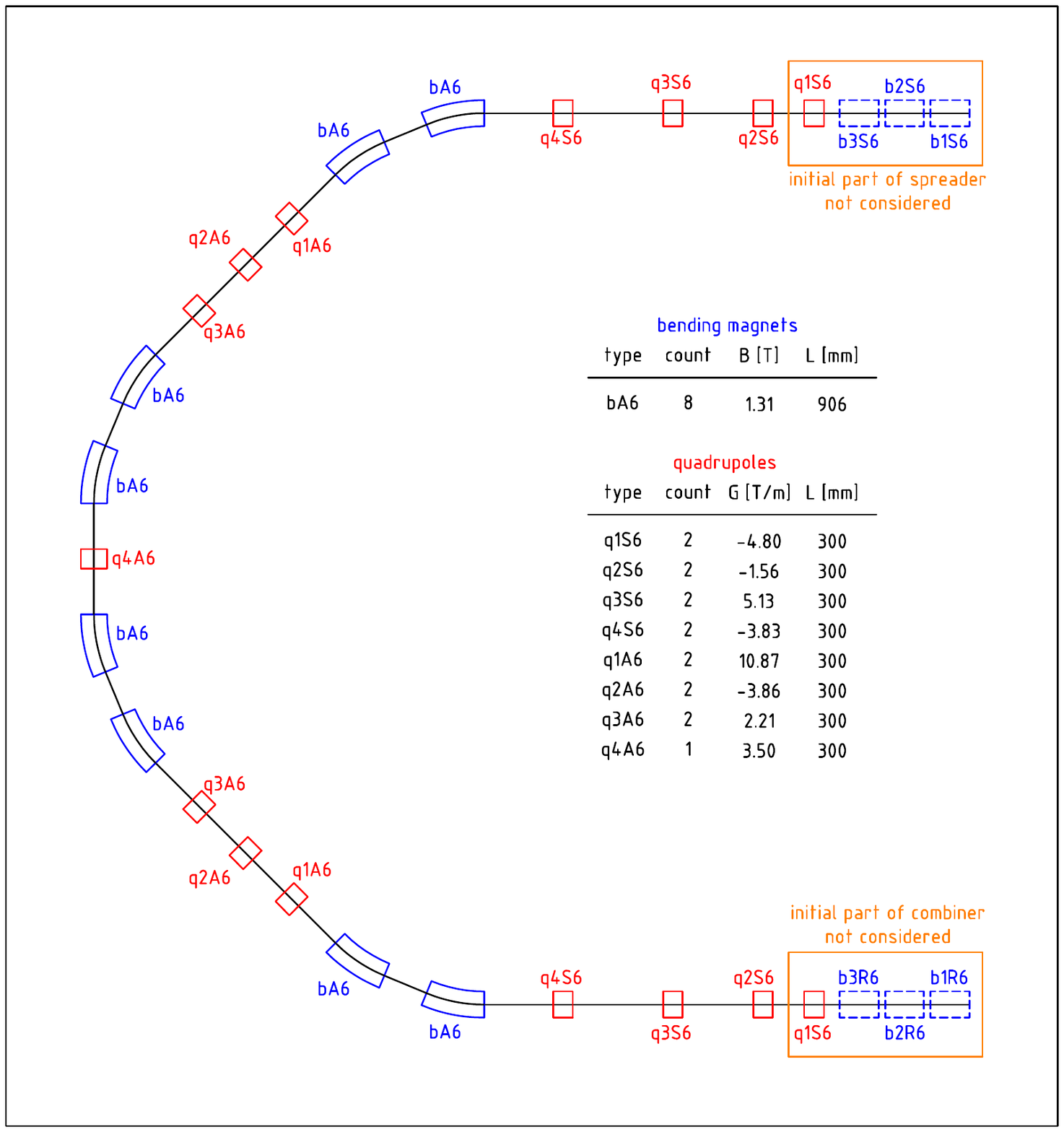}}
	\caption{Arc 6 and main magnets.}
	\label{fig:arc_6_magnets}
\end{figure}

While the dipole strenghts simply scale across the arcs, this is not the case for the quadrupoles, as each arc has a different optics. Table~\ref{tab:quads} summarizes the maximum and minimum integrated gradients as well as pole tip fields for the quadrupoles. This is based on the two lengths -- 200 and 300~mm -- currently specified in the lattice, as in Figs.~\ref{fig:arc_1_magnets} to \ref{fig:arc_6_magnets}. This results in a quite wide range of integrated gradients and pole tip fields. Moreover, some quadrupoles are rather weak. This prompts an iteration with the optics, which needed to be refined after a full design of the bending magnets including the edge effects. The possibility of making families, grouping by gradient or length or both, would need to be considered. Two preliminary cross-sections are shown in Fig.~\ref{fig:quad_magnet}. Since the aperture is the same throughout the arcs, an option could be to keep the same iron design, though to have only 2 instead of 4 coils for the weaker quadrupoles. The impact of this asymmetry on the field uniformity is rather minor, about $2\cdot10^{-4}$ at 2/3 radius on the skew octupole in 2D. As for the main bending units, the coils could be water cooled (for compactness) and they will need to be designed as part of the overall optimization, including the power converters, the magnet manufacturing cost and the operational scenarios, considering for example different baseline optics. The shaded area in Fig.~\ref{fig:quad_magnet} corresponds to 7-8 A/mm$^2$ of current density at maximum gradient. More exotic designs -- for example a flat quadrupole with an open magnetic circuit -- could, if needed, provide a more compact design in the vertical direction, though the stray field would need to be properly addressed.

\begin{table}[!h]
  \centering
  \begin{tabular}{|c|cccc|}
  \hline
		 & $|GL|_{max}$ & $|GL|_{min}$ & $|B_{pole}|_{max}$ & $|B_{pole}|_{min}$ \\ 
		 \hline
		arc 1 & 0.76 & 0.12 & 0.076 & 0.012 \\ 
		arc 2 & 1.00 & 0.01 & 0.100 & 0.001 \\ 
		arc 3 & 1.80 & 0.23 & 0.172 & 0.016 \\ 
		arc 4 & 2.94 & 0.61 & 0.294 & 0.041 \\ 
		arc 5 & 2.99 & 0.71 & 0.200 & 0.047 \\ 
		arc 6 & 3.26 & 0.47 & 0.217 & 0.031 \\ 
		\hline
  \end{tabular}
	\caption{Summary of integrated gradients and pole tip fields of quadrupoles, in T.}
	\label{tab:quads}
\end{table}	

A further analysis will address in detail the magnets in the spreaders and combiners regions. Furthermore, a set of vertical / horizontal dipole correctors will most likely need to be added. According to their strength and field uniformity tolerances, these correctors could be combined with some of the quadrupoles in a hybrid design. 
Path length adjustments, mainly from seasonal contraction and expansion effects, amounting to an expected O(1)\,cm correction, may be addressed via dog legs in the arcs. 
Finally, multiple aperture magnets could be analyzed as part of an overall cost optimization, though much could depend on the staged construction of the facility.

\clearpage

\clearpage

\clearpage

\section{Dumps and Transfers}

\begin{figure*}[ht]
   \centering
      \includegraphics*[width=120mm]{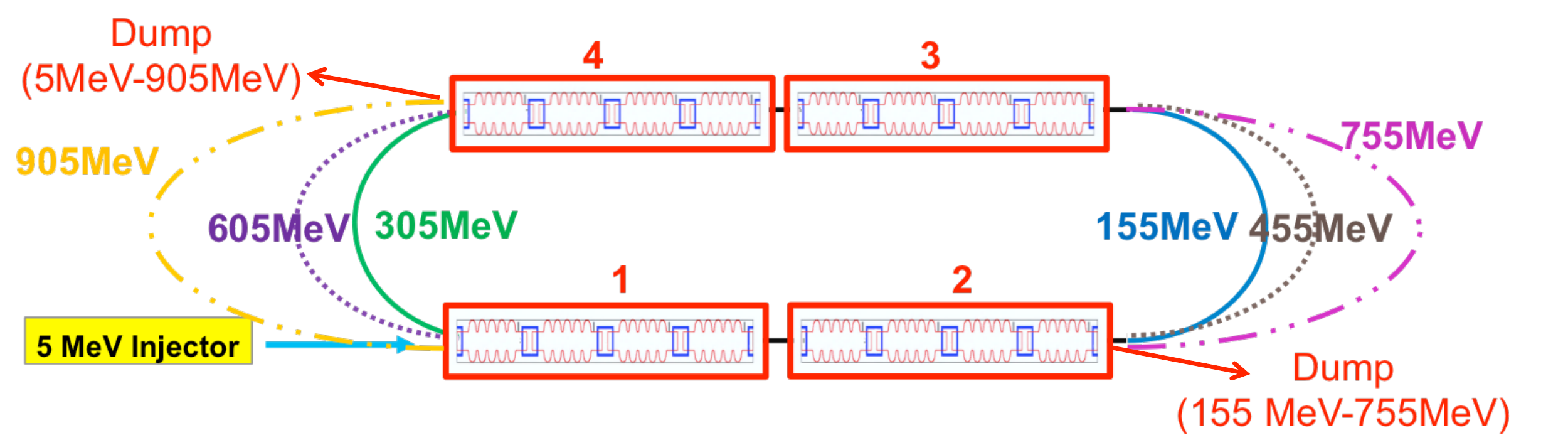}
   \caption{Top view of PERLE and the transfer lines-to dump systems for nominal operation and beam setup at the different energies.}
   \label{Figure1}
 \end{figure*}

The nominal operation of PERLE foresees to continuously dump the decelerated 5 MeV electron beam; this corresponds, for a  current of 12.8 mA, to a constant power deposition of 64~kW on the beam dump. The possibility of dumping the beams at all the different energies during the setup period is considered. In this case a system of Transfer Lines (TL) and a beam dump has to installed at the end of each Linac as shown in Figure\,\ref{Figure1}. 

\subsection{Operational dump}

\begin{figure*}[ht]
   \centering
      \includegraphics*[width=80mm]{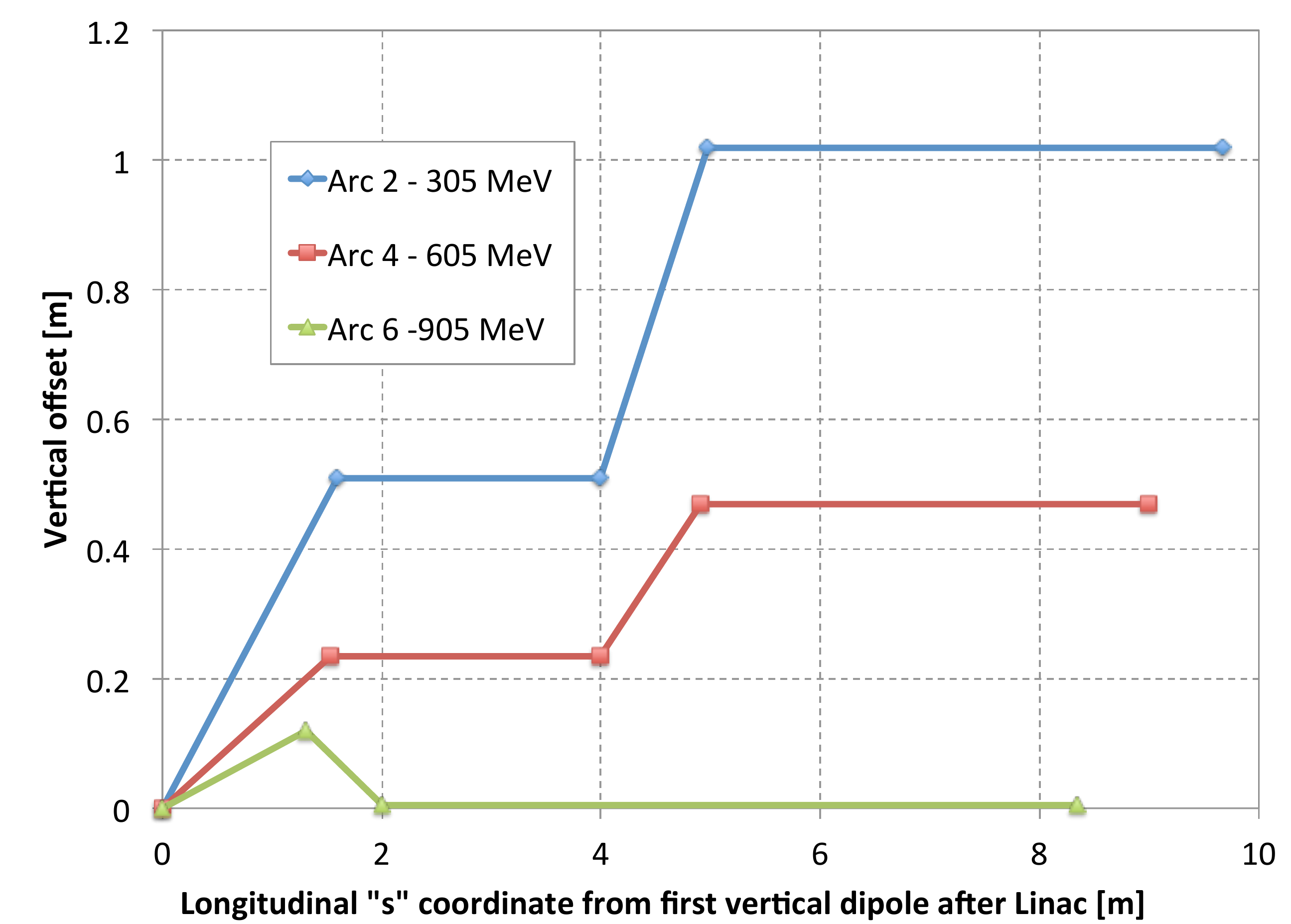}
   \caption{Schematic view of the vertical spreader which directs the 305 MeV, 605 MeV and 905 MeV beams towards the respective superimposed arc.}
   \label{Figure2}
 \end{figure*}

Two options are investigated for the operational beam dump. In the first case no additional magnet has to be installed in the main lattice. A 0.66~m long dipole (SBEND) with a 0.906~T magnetic field acts as a spectrometer and separates vertically the different energy beams to direct them towards the respective superimposed arc (Fig.\,\ref{Figure2}).     

\begin{figure*}[ht]
   \centering
      \includegraphics*[width=120mm]{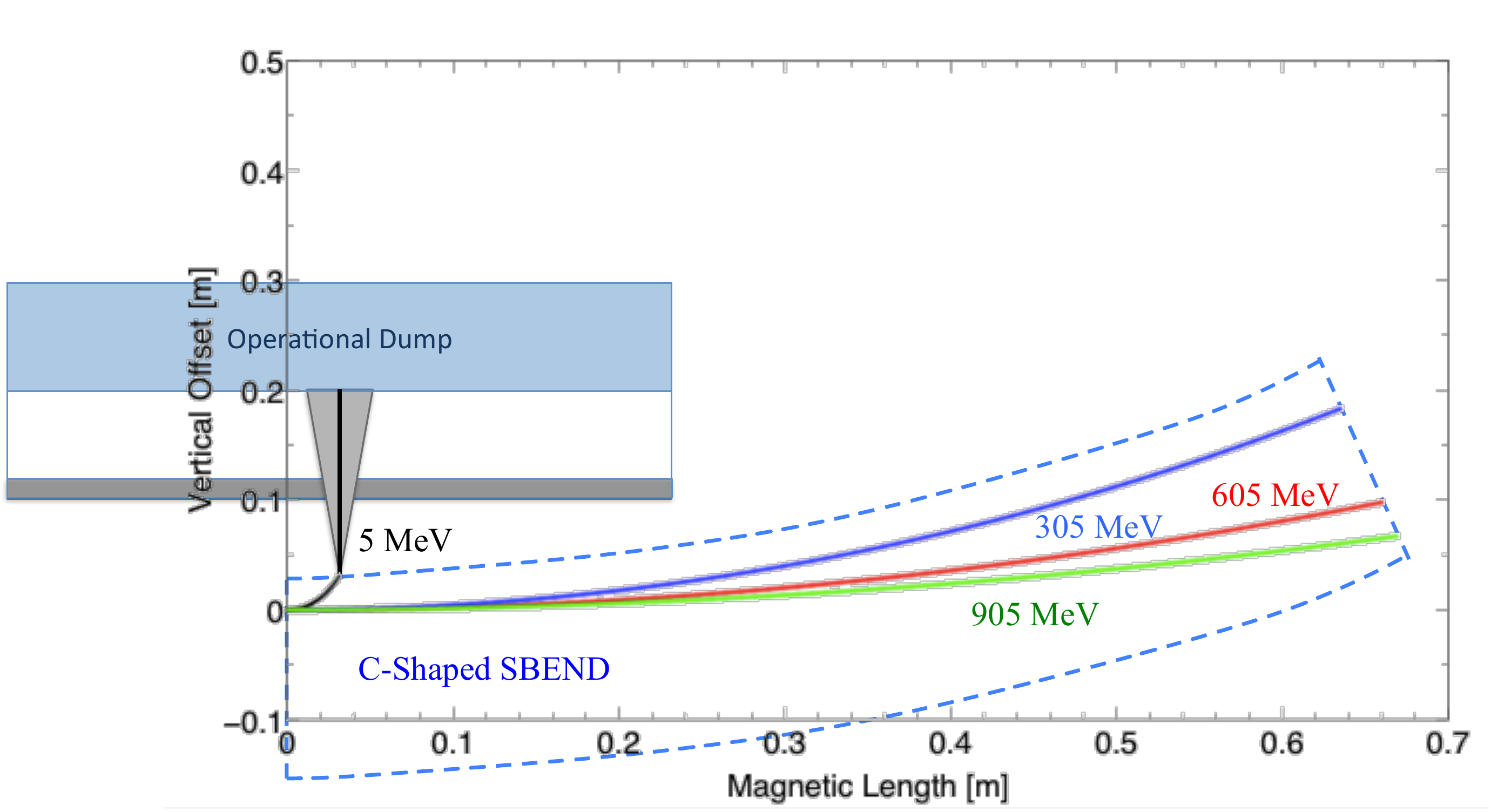}
   \caption{The first dipole of the vertical spreader is a C-Shaped SBEND which allows to extract the 5 MeV beam from the magnetic field region (between the dashed blue lines) towards the vertical dump.}
   \label{Figure3}
 \end{figure*}

This magnet can be used to deflect the 5~MeV beam towards a vertical beam dump as shown in Fig.\,\ref{Figure3}. A C-shaped dipole has to be used to host a T-shaped vacuum chamber. The 5~MeV beam gets a deflection of about 90$^{\circ}$ in 3~cm and is extracted from the magnetic field region. Due to the strong edge effects and the low energy, the beam size increases rapidly and the 3~$\sigma$ envelope has a radius of 65~mm (for a normalised emittance of 10 mm mrad) at a height of 10~cm from the Linac axis; here the vertical dump has to be installed (Fig.\,\ref{Figure3}). Due to the low energy no window can be installed at the entrance of the dump system.
The beam continues diverging in vacuum before hitting the dump material. A low Z material, like Carbon, can be used to limit the backscattering and the weight of the dump block which has to have a size of indicatively 0.4~m~$\times$~0.1~m$\times$~0.1 m (length, width and thickness respectively). 
For an incident energy of 5~MeV, about 1-1.5\% of the electrons are scattered back from Carbon. The corresponding fraction of energy (or power) which is backscattered is a bit less as the electrons deposit part of their energy before being scattered back. For a 64~kW electron beam one can estimate roughly 0.6~kW backscattered from the Carbon dump.
To further reduce the backscattering towards the recirculating beam, a thin layer of a heavier material should be installed at the entrance of the dump, provided that a free hole is left for the passage of the beam. 
Detailed studies are needed to assess the feasibility of the proposed design (including a cooling system and additional shielding), evaluate potential integration conflicts (especially for the replacement of the underneath dipole) and the real impact of the backscattering on the recirculaitng beam quality. Moreover detailed tracking studies in a real 3D field  have to be performed to check the effect of the strong fringe fields on the electron beam.

\begin{figure*}[ht]
   \centering
      \includegraphics*[width=100mm]{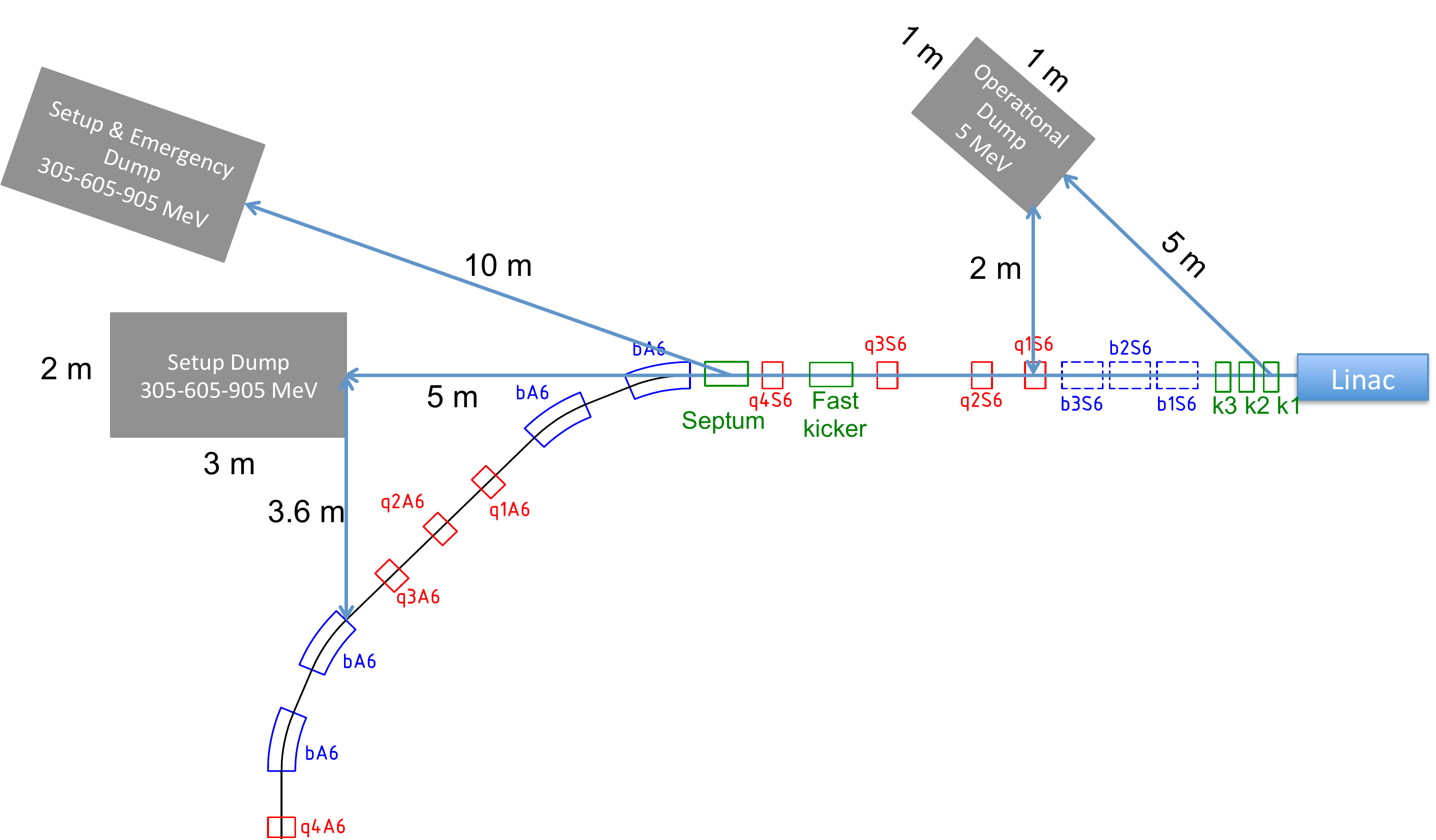}
   \caption{The transfer lines to the operational, setup and setup\&emergency beam dumps are shown with respect to the 905 MeV beam arc.}
   \label{Figure4}
 \end{figure*}

\begin{figure*}[ht]
   \centering
      \includegraphics*[width=100mm]{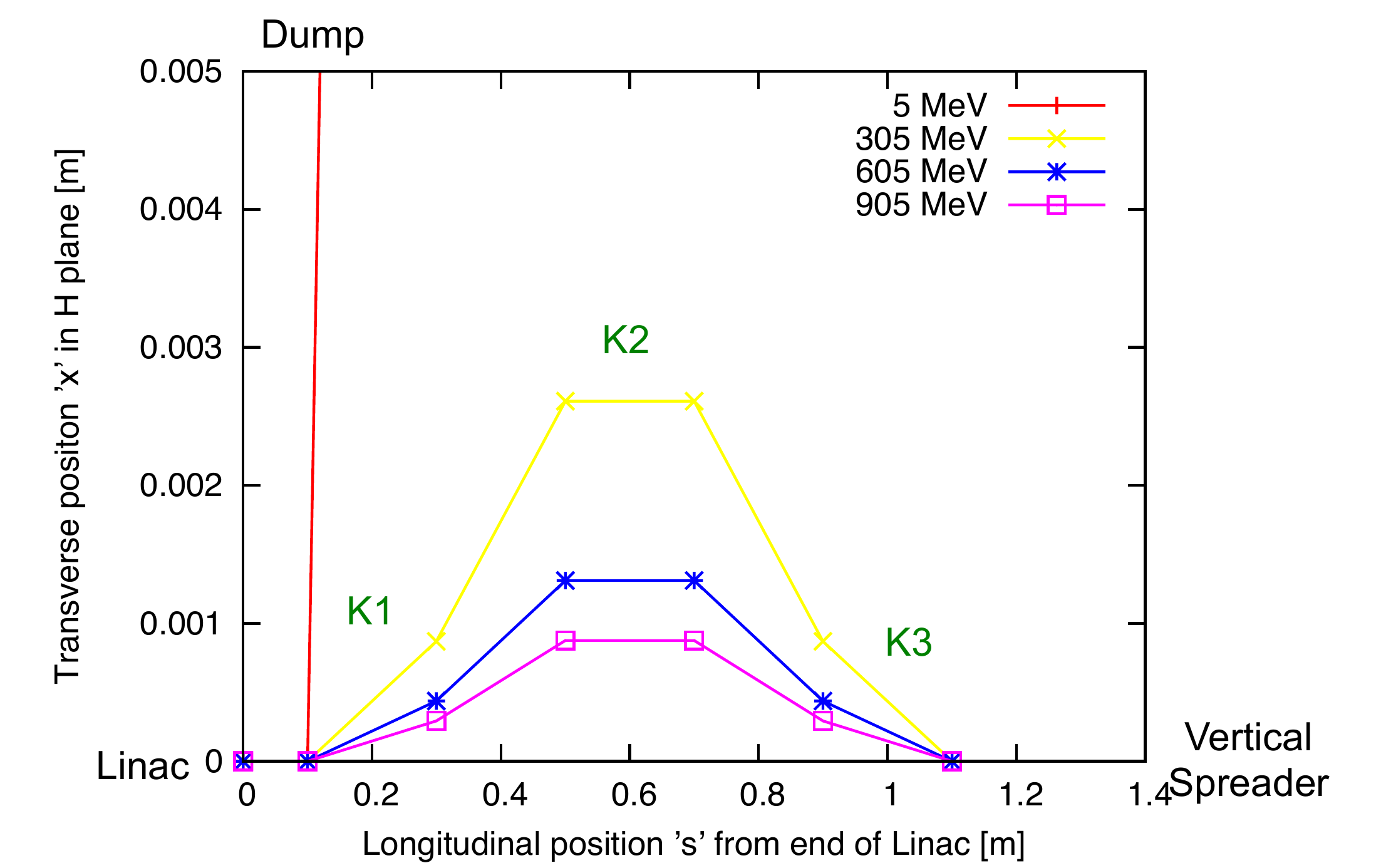}
   \caption{Horizontal trajectory of the different energy beams before the vertical spreader. The 5~MeV beam is extracted to the dump while the other beams are brought back to the reference trajectory. The bump and the dump are directed towards the outside of the ERL facility.}
   \label{Figure5}
 \end{figure*}

The second option foresees the installation of three additional small dipoles in the 1.42~m drift between the end of the Linac and the start of the vertical spreader (k1, k2 and k3 in Fig.\,\ref{Figure4}). The first dipole has a magnetic length of 0.2~m, a magnetic field of 0.044 T and kicks the 5 MeV beam by 30$^\circ$ to extract it horizontally towards the beam dump. After a 5~m drift line the beam is dumped against a cylinder of graphite (20~cm radius and 10~cm long). Also in this case a cooling system and a surrounding shielding have to be foreseen. A clearance of ~2 m is obtained between the main lattice and the shielding assuming a shielding transverse size of 1~m.
Since k1 is operated in DC mode, all the beams are slightly affected by its magnetic field. The two remaining magnets are thus used to bring the other energy beams back on to the reference trajectory before the vertical spreader (Fig.\,\ref{Figure5}). 
All three magnets have the same magnetic length, and the magnetic field is 0.088~T (with opposite polarity) and 0.044~T for k2 and k3 respectively. Preliminary studies were performed to check the impact of the proposed bump on the optics. The horizontal dispersion can be closed to $1.6 \cdot 10^{-7}$\,m while the $\beta$ functions at the entrance of the first dipole of the vertical spreader differ by $15$\,\% with respect the nominal optics; no further optimisation was attempted.

\subsection{Setup dumps}

During the commissioning period of  PERLE, and in general during the beam setup, it is important to be able to dump the beam at the different energies. The easiest solution is to keep switched off the first horizontal dipole of the arc corresponding to the energy of interest and let the beam go straight towards the dump (Fig.\,\ref{Figure4}). 
This dipole has to have a C-shape to allow the installation of a Y chamber for the recirculating and the extracted beam. The minimum bending angle of 22.5$^\circ$ guarantees enough clearance between the next dipole and the vacuum chamber of the extracted beam. 
If the dipoles of the arc are powered in series they can all be switched off during the setup period. Also in this case the line to the dump, one per each energy, corresponds to a 5~m drift. 
The $\beta$ function at the dump is about $50$\,m corresponding to a minimum beam size for the most energetic beam of 238~$\mu$m. 
In order to limit the energy deposition and the activation of the dump materials, the setup should be performed with a reduced intensity. In Table\,\ref{Table1}, the current corresponding to a power deposition of 64~kW at the different energies is shown.
\begin{table}[hbt]
   \centering
   \begin{tabular}{|c|c|c|}
       \hline
       Energy [MeV]  & Current [mA] & electrons per bunch/$10^7$ \\
       \hline
       5    & 12.8  & 200\\
       155 & 0.41 & 6.5\\
       305 & 0.21 & 3.3\\
       455 & 0.14 & 2.2\\
       605 & 0.11  & 1.7\\
       755 & 0.08 & 1.3\\
       905 & 0.07 & 1.1\\
       \hline
   \end{tabular}
    \caption{Current and number of electrons per bunch (25 ns bunch spacing) corresponding to a constant power deposition at the beam dump of 64 kW for the different energies of PERLE and assuming 
    an initial current of $12.8$\,mA. }
   \label{Table1}
\end{table}

The dump system will consist of three superimposed blocks of graphite with a radius of 20~cm and a maximum length of 1.2~m (for the 950~MeV beam) to absorb also the secondary showers. Additional shielding has to be envisaged and a total occupancy of 2~m$\times$3~m has to be considered around the dump blocks.

\subsection{Emergency dumps}

\begin{figure*}[ht]
   \centering
      \includegraphics*[width=120mm]{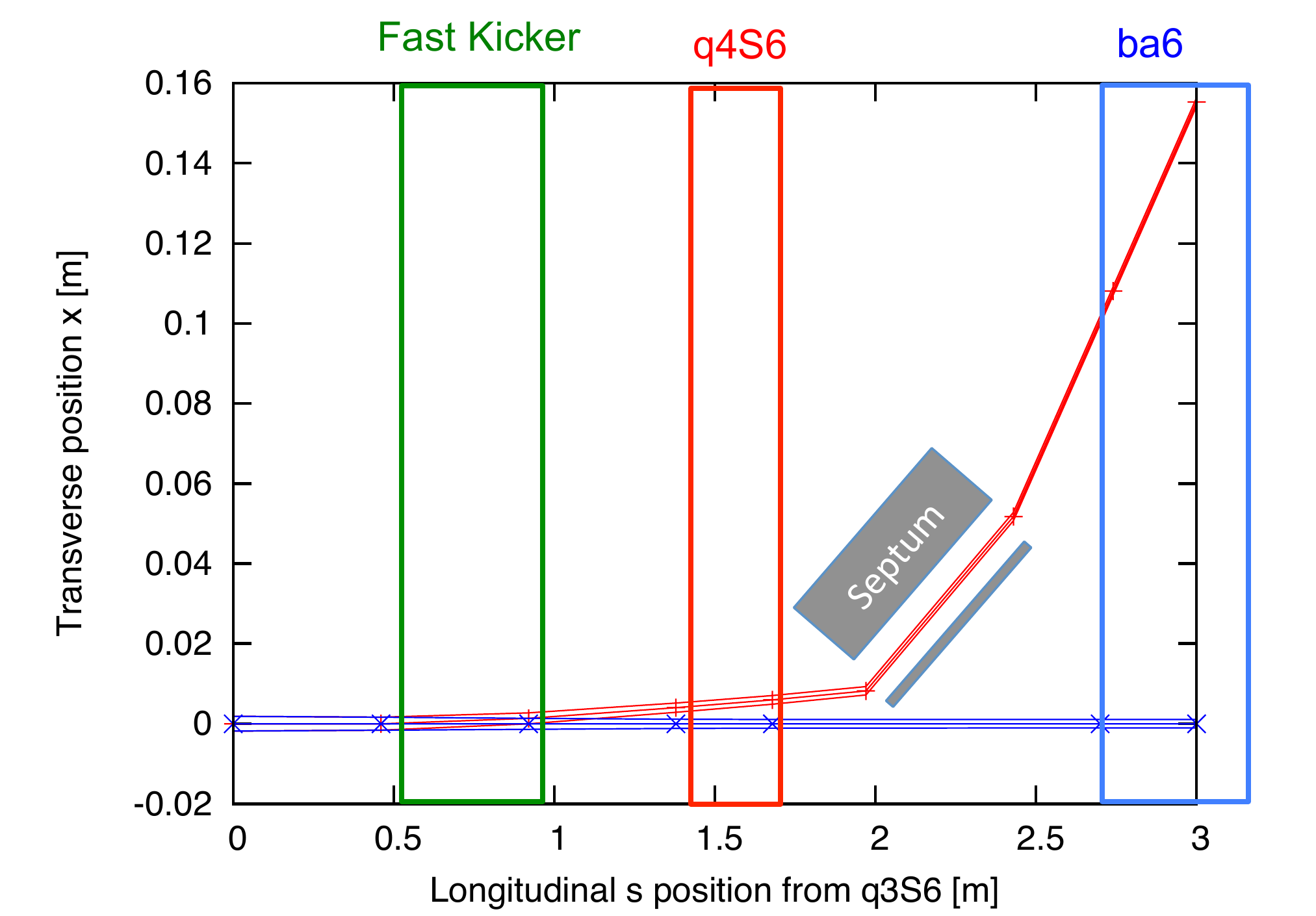}
   \caption{Fast extraction system for the emergency dump of the 905~MeV beam}
   \label{Figure6}
 \end{figure*}

Up to now only  DC magnets have been considered. In the eventuality that the setup dumps have to be also used as emergency dumps, fast kickers have to be included in the lattice.
 The CW operation mode and the 25~ns bunch spacing require a rise time $t_m$~=~23~ns to allow for some jitter. A system impedance Z of 25~$\Omega$ is assumed, and a rather conservative system voltage U of 60~kV.
Assuming a full horizontal and vertical opening of 40~mm, the magnetic length of the fast kickers has to be 0.46~m and the gap field 0.038~T. 
One extraction system per each each energy has to be installed after the vertical spreader when the beams are fully separated. Preliminary studies were carried out only for the 905~MeV beam but analogous considerations hold for the other energies. A fast horizontal kicker is installed between the last two quadrupoles before the arc (q3S6 and q4S6 in Fig.\,\ref{Figure4}). The beam is deflected outwards by the kicker and goes through the 40~mm diameter of the defocusing quadrupole (q4S6) getting an additional kick. A horizontal Lambertson septum, placed 0.5~m before the first arc dipole (ba6), extracts the beam towards the dump line (Fig.\,\ref{Figure6}). A clearance of 6~mm between the recirculating and the extracted beam envelope is obtained at the septum with the proposed configuration. 
The ba6 dipole has to be C-shaped (the present H-shaped design and the size of the magnet are not compatible with a fast extraction system due to the limited available space in the lattice) and the magnetic field free region is assumed to start at 70~mm from the main axis. Additional 30 mm are considered for the beam pipe of the extracted beam.
A 0.5~m long septum with a 1.1~T magnetic field provides a kick of 174~mrad and thus an offset of 108~mm at the ba6, in agreement with the specifications. 

In order to limit the energy deposition at the emergency dumps, the interlock system has to stop the injector and pulse the kickers of all the different arcs simultaneously. This limits the maximum number of dumped bunches to 7 (bunches contained in one arc and one Linac). A kicker flattop of 166~ns is needed to fit all the bunches in and a fall time of 23~ns is assumed.

The energy and power deposition at the dumps for different energies are summarised in Tab.\,\ref{Table2}.  
\begin{table}[hbt]
   \centering
   \begin{tabular}{|c|c|c|}
       \hline
	Energy [MeV]  & Energy deposition [J] & Power deposition [MW]  \\
       \hline
       155 & 0.35 & 2.09\\
       305 & 0.68 & 4.12\\
       455 & 1.02 & 6.14\\
       605 & 1.36  & 8.16\\
       755 & 1.69 & 10.2\\
       905 & 2.03 & 12.2\\
       \hline
   \end{tabular}
     \caption{Energy and power deposition when dumping seven bunches of $2 \cdot 10^9$ electrons on the emergency dumps.}
   \label{Table2}
\end{table}
The transfer lines to the dump have to be $\sim$10~m long and a defocusing quadrupole has to be installed at $\sim$4~m in order to increase the beam size at the dump and reduce the energy density (for the 905~MeV beam, a 0.6~mm~$\times$~0.4~mm beam size can be achieved using a quadrupole identical to q4S6). A block of superimposed kickers can be envisaged to align vertically the different energy beams at the dumps and reduce the transverse occupancy of the dump/shielding block.

\subsection{Test facility}

The possibility of using the PERLE transfer lines to perform quench and damage tests of superconducting magnets and cables is explored. 
The fast extraction system of the emergency dumps is used to extract only the number of needed bunches in the shadow of the nominal ERL operation. The length of the kicker waveform has to be extended up to 0.1~s (the risk of flashovers has to be carefully evaluated) to fulfil the test requirements. 

In this case the lines have to include a triplet to vary the focal point and the beam size at the focal point. The different energy lines are recombined and a system analogous to the one used at the entrance of the Linacs is used. Steering magnets and a matching insertion are included as well. In total the line can be up to 30~m long and additional 10-20~m have to be considered for the test samples and the downstream beam dump. The parameters in Tab.\,\ref{Table2} are used for the dump design. It is assumed that the beam setup is done with a reduced intensity, the full intensity beams will then be dumped on the samples. For further analysis,
more detailed optics studies have to be performed, the dynamic range of the magnets  and potential RP issues  to be evaluated.


\section{Photon Beam Production}

\subsection{Optical system}

Depending on the electron-beam time-structure, various optical systems capable 
to produce high gamma-ray fluxes are nowadays available.
On the one hand, for bunch trains of low repetition rate, non-linear 
\cite{shverdin2010high} or passive \cite{dupraz2014design} optical recirculators may be 
used (e.g. ELI-NP-GS \cite{ELI-NP-GBS:TDR}:  trains of 32 bunches separated by 16 ns 
at a repetition rate of 100 Hz). The related laser system has to provide the 
maximum pulse intensity allowed by the foreseen spectral density (e.g. 
ELI-NP-GS: 400 mJ at 100 Hz for green 515 nm light, 14 $\mu$m transverse spot 
size of the intensity profile and 3 ps longitudinal pulse width). 
 On the other hand, for CW electron bunches of repetition rate $\gtrsim 10$ MHz, 
 Fabry-Perot cavities~\cite{kogel} ({\it i.e.} optical resonators) 
 may be used \cite{ladon,higs,ruth,KEK4mirror}. This is the technical solution envisaged for the PERLE photon beam facility. 
 
\begin{figure}[htbp]
  			\centering
  			\includegraphics[width=12.5cm]{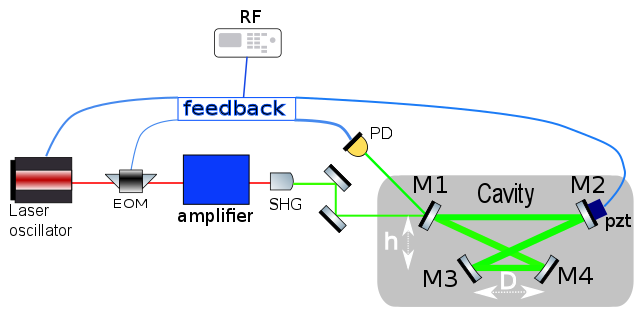}
  			\caption{Simplified scheme of a four mirror cavity locked to an 
amplified laser oscillator. Planar ($M_1$ and $M_2$) and concave ($M_3$ and 
$M_4$) mirrors are shown along with the electro-optic modulator (EOM) used to 
build the feedback error signal from the reflected signal (photodiode PD) and 
a piezo-electric transducer (PZT) fixed on $M_2$ to synchronize the cavity 
round trip frequency to the accelerator RF.}
  			\label{fig:cavity}
  		\end{figure}

  Fabry-Perot cavities consist of a sequence of high reflectivity mirrors (see 
  Fig.\,\ref{fig:cavity}). When the laser beam frequency satisfies resonance conditions (see 
  \cite{jones2001stabilization} for pulsed beams), the power is enhanced at most by a factor 
  $G=F/\pi$ inside the cavity (in practice laser/cavity spatio-temporal mode mismatches can reduce this factor by several dozens of percent). 
  The cavity finesse $F$ depends on mirror losses 
  and reflection coefficients. However, the higher the cavity enhancement factor the 
  narrower the optical resonance $\Delta\nu/\nu=\lambda/(LF)$, where 
  $\nu=c/\lambda$ is the laser frequency and $L$ the cavity optical round-trip 
  length. Dedicated laser cavity feedback is needed to preserve the resonance 
  conditions \cite{PDH,jones2001stabilization}. Experimentally, a cavity with $F\approx 
  28000$ ($G\approx 9000$) for picosecond pulses and with $L=4$~m was 
  demonstrated by some of us in \cite{LAL28000}.

    \begin{table}[h]
    \centering
    \caption{Expected laser beam and cavity parameters.}
    \label{laser_cavity}
    \begin{tabular}{l|l|l|}
    \cline{2-3}
      & $\lambda=1030$~nm &  $\lambda=515$~nm   \\
      \hline \multicolumn{1}{|l|}{ Laser beam average power  (W)} & 200 & 100 (200)  \\
      \hline \multicolumn{1}{|l|}{  Laser beam time FWHM (ps)} & 1-10  & 1-10  \\
      \hline \multicolumn{1}{|l|}{Cavity beam waist  ($\mu$m)} & 60 & 60   \\
       \hline \multicolumn{1}{|l|}{Cavity beam intensity spot size ($\mu$m)} & 30 & 30   \\
      \hline \multicolumn{1}{|l|}{ Cavity beam Rayleigh length (mm)}& 22.0 & 11.0   \\
      \hline \multicolumn{1}{|l|}{Cavity finesse } & 28000 & 28000   \\
     \hline \multicolumn{1}{|l|}{Cavity stacked average power (kW)}  & >600 & >300 (>600)   \\ \hline
    \end{tabular}
    \end{table}

  The power that can be stored inside the cavity is limited by thermal effects 
  and mirror coating damage threshold. An average power of 670 kW 
 (for 10 ps pulses and 250 MHz repetition rate) was obtained 
 \cite{pupeza_power} for intra-cavity high-harmonic attosecond pulse 
 experiments \cite{harmo}. Concerning Compton experiments, 50 kW 
 was recently demonstrated by some of us on the ATF electron ring of KEK 
 \cite{aurelien}. A 35.68~MHz cavity ($L\approx8.4$~m) designed for storing 10 
 ps pulses of average power above 600~kW is presently under development at LAL 
 by some of us for the Compton X-ray machine ThomX \cite{thomX}. This is a 
 similar optical cavity that is needed for the PERLE photon beam facility. 
 Besides, a CW laser beam of 700~kW will also be stored in the VIRGO 
 interferometer in a near future \cite{VIRGO}. There is thus a global effort to achieve 
 stable and routinely operating cavities in high average power regime. One 
 should also mention that developments on long $L\approx30$~m monolithic and 
 high finesse cavity are also on-going \cite{long-cavity}. 
   
  Mode properties (wave front profile, polarization) of optical cavities 
  solely depend on their geometries. Specific optical designs must then be 
  supplied to fulfill the requirements of Compton experiments 
  \cite{nous-polar,dupraz_abcd}. Following the arguments of  
  Ref.\,\cite{dupraz_abcd}, one must consider planar four-mirror cavities made of at 
  least two concave reflective surfaces for the ERL SCRF photon beam facility 
  (see Fig. \ref{fig:cavity}). The distance between the two planar mirrors ($M_1$ and 
  $M_2$) can be adjusted to lock the cavity round-trip frequency to the accelerator 
  radio-frequency while the distance between the two concave mirrors ($M_3$ 
  and $M_4$) can be varied to tune the laser beam spot size at the IP. This 
  geometry has been successfully tested at the ATF \cite{KEK4mirror}. Eventually, with a 
  careful design of the high reflectivity mirror coating, the mode 
  polarization of a planar four-mirror cavity can be freely tuned.   
  
  The laser source is of prior importance for high finesse cavities. One must 
  start from a low phase noise mode-locked oscillator and then amplify the 
  signal using the chirped pulse amplification technique \cite{CPA}. The laser 
  amplifier system is also of prior importance because it must not induce 
  additional phase noise (e.g. AM/PM coupling via non linear processes) while 
  providing stable and long term operations. Considering a repetition rate of 
  40~MHz and picosecond pulses, the most mature and powerful technology is 
  based on Ytterbium-doped diode-pumped fibres. Reasonably low noise laser 
  mode-locked oscillators are commercially available at this wavelength 
  (around 1030 nm) and amplifiers with up to 
  an average power of 830~W \cite{limpert_830W} (and more 
  recently 2~kW \cite{limpert2kW}) was demonstrated on a table top experiment. 
  Besides, a fully connectorised and compact $Yb$ doped fibre amplifier system 
  providing 50~W has been operated over days at ATF/KEK~\cite{KEK4mirror}
  in gamma ray production experiments. This 
  system has been recently upgraded to 200~W at CELIA for the ThomX project. 
  This is what is needed for the PERLE photon beam facility.   
  Using a LBO crystal, the laser beam frequency can finally be doubled with 
  more than $50\%$ efficiency before entering the optical cavity to 
  provide a high average power beam at a wavelength close to $515$~nm. 
  Eventually one can also parallelize two fiber amplifiers to compensate for 
  the second harmonic generation limited efficiency \cite{kienel2014energy,guichard2015high,kienel2015multidimensional}. 
  
  To reach a stored average power of more than 300~kW, the cavity finesse must be $\approx 30000$ leading to $\Delta\nu/\nu\approx 2\cdot 10^{-12}$. A strong feedback between laser and cavity is clearly required to keep the system on resonance. However, it should be mentioned that such a high average power has never been demonstrated for a wavelength of 515~nm. Apart from higher absorption in SiO$_2$, one of the dielectric dioxide used for high reflective coating, one does not expect tremendous differences for the cavity finesse foreseen here, experimental tests could be done at LAL and CELIA. The laser beam and cavity parameters are summarized in Tab. \ref{laser_cavity}. 
  
  For other laser beam wavelengths  one could also use gain media doped with the other rare earth elements Er (1.5~$\mu$m) or Tm (1.9~$\mu$m) \cite{revue_rare_earth}. Performances would be reduced with regard to Yb but still useful. Using quarter wave stack cavity mirror coatings one could also consider filling a single cavity with $\lambda$ and $\lambda/3$ (e.g. doubled Yb: 515~nm and Er: 1545~nm) to provide a gamma frequency together with its third harmonic.

\subsection{Cavity design}
\begin{figure}[htbp]
\centering
\makebox[\columnwidth]{
\subfigure[]{
\includegraphics[width=.5\linewidth]{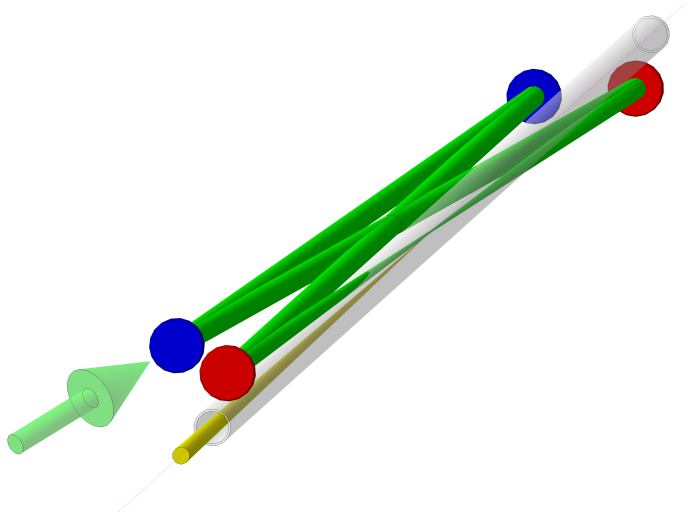}
\label{sfig:Zernike_ELDIM_distrib_15_06}
}\quad
\subfigure[]{
\includegraphics[width=.5\linewidth]{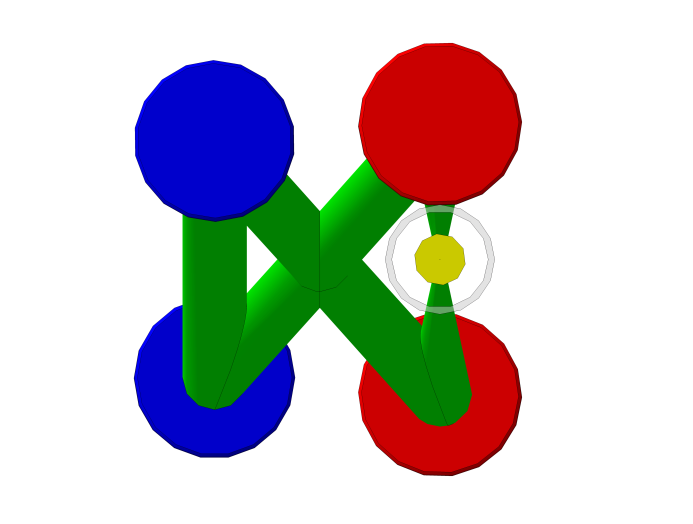}
\label{sfig:Zernike_ELDIM_coeffs_15_06}
}}
\caption{Schematic view of a possible four-mirror cavity implementation. (a) Isometric view; (b) face projection view. Red discs: concave mirrors; blue disks: plane mirrors. The cavity mode is represented as a green tubes (radius corresponding to $\approx 6\sigma$ of the intensity Gaussian profile) and cones, the beam pipe as a gray tube and the gamma ray beam as a yellow cone.}
\label{fig:3d_cavity}
\end{figure}

\begin{figure}[htbp]

\centering
\makebox[\columnwidth]{
\subfigure[]{
\includegraphics[width=.5\linewidth]{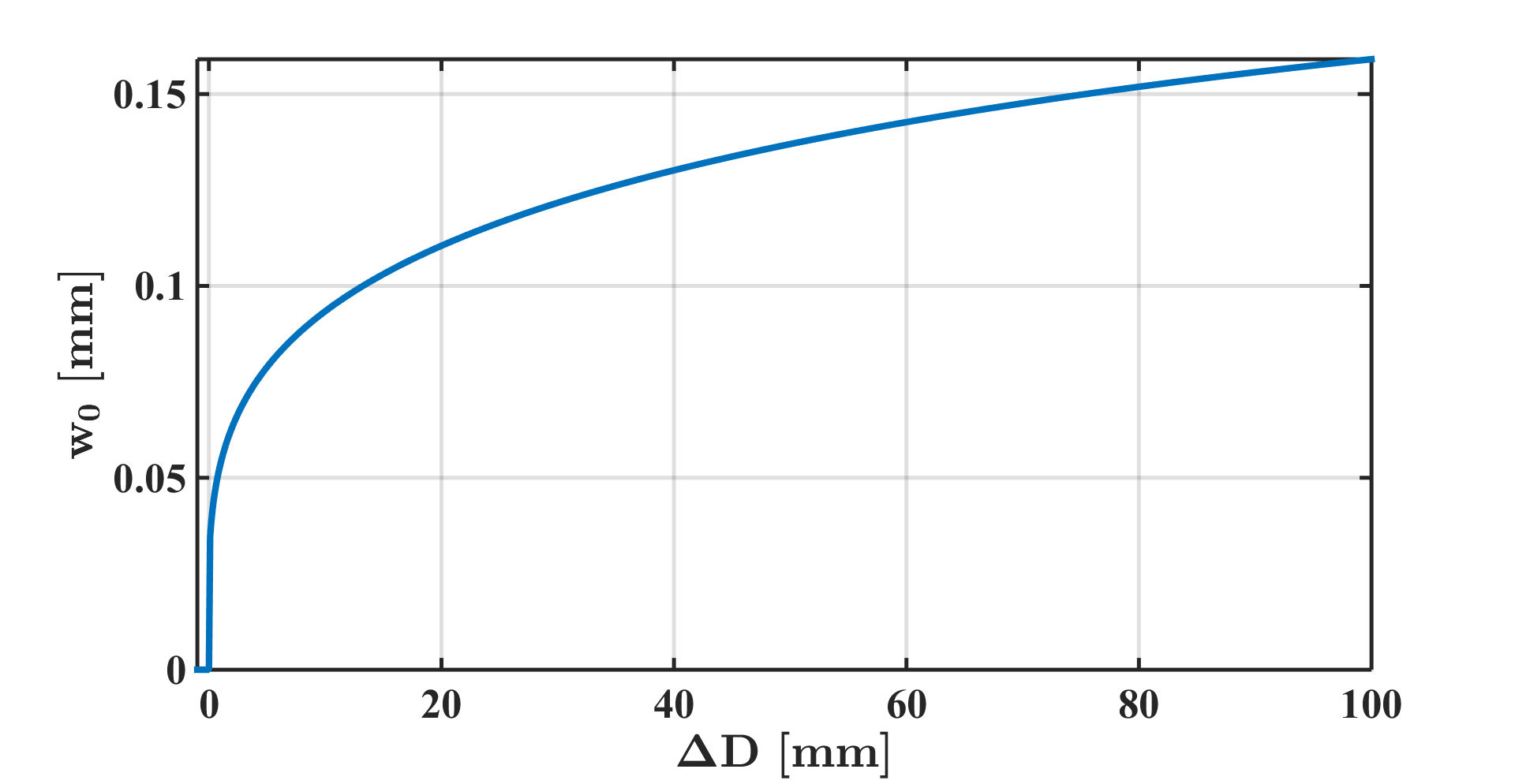}
\label{sfig:Zernike_ELDIM_distrib_15_06}
}\quad
\subfigure[]{
\includegraphics[width=.5\linewidth]{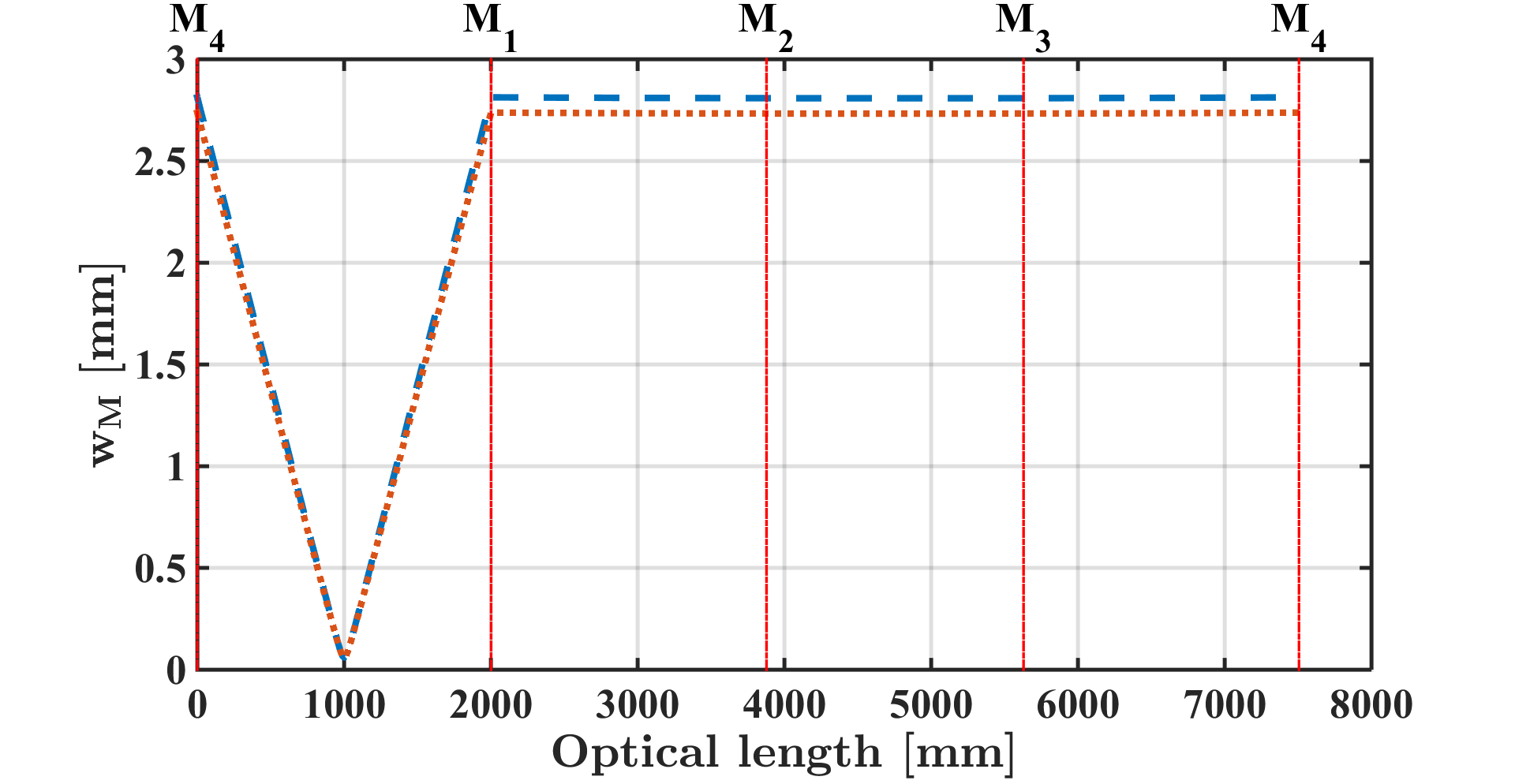}
\label{sfig:Zernike_ELDIM_coeffs_15_06}
}}
\caption{(a) Minimum mode cavity waist as a function of the distance between the two concave mirrors ($\Delta D=D-D_0$).
(b) Main mode radii as a function of the optical path inside the cavity. Dashed curve: maximum beam radius; dotted curve: minimum beam radius. Positions of the four mirrors are also indicated by vertical lines.}
  		  			\label{fig:w}
 		  		\end{figure}

There is  freedom in choosing the cavity geometry. Here a trade-off
is proposed  between a small laser-electron crossing angle, small enough laser beam spot size at the IP while ensuring reasonably large spot sizes on the mirror surfaces. 
To calculate the cavity mode one considers a planar four-mirror cavity (see Fig. \ref{fig:cavity} and Fig. \ref{fig:3d_cavity} for a possible implementation) with $L=7.5$~m seeded with 
a 40~MHz pulsed laser beam of wavelength 515~nm. Assuming a quasi symmetric geometry we set the distance between the concave mirrors $D$ close to $D_0= 2$~m and the distance $h=35$~mm to avoid beam vignetting effects induced by the 15~mm inner diameter beam pipe (see Fig. \ref{fig:3d_cavity}.b). The concave mirror radius of curvature is fixed to $D_0$ and the  mirror diameters to $1$\,inch. The laser beam waist $w_0$ is shown as a function of $\Delta D=D-D_0$ in Fig. \ref{fig:w}.a. Small waist values are thus obtained for the very mechanically stable confocal geometry ($D\gtrsim D_0$) \cite{dupraz_abcd} though very close to the modal instability region. 
 We choose $w_0=60~\mu$m ({\it i.e.} 30~$\mu $m Gaussian intensity spot size). As expected \cite{kogel_astigmatism}, the transverse mode profile is elliptical and the main radii are shown as a function of the optical path length in Fig. \ref{fig:w}.b. From this figure one sees that the mode is collimated between the two plane mirrors with a beam radius of approximately $ 2.7$~mm on the mirror surfaces. Such beam radius leads to negligible diffraction losses induced by the 1 inch mirror edges.
 We obtain a crossing angle between the laser beam and the electron bunch of 1.2$^\circ$. With $h/D=0.017$, the incident angle on the concave mirror is 0.53$^\circ$ leading to a small mode ellipticity of roughly $ 2.4 \%$ and negligible polarization instabilities \cite{nous-polar}. As for the mechanical mirror mounts, motion actuators and vacuum vessel, we propose to adopt the technical solutions tested successfully over years at ATF/KEK \cite{KEK4mirror},\cite{thomX}. It is noticeable that these elements were recommissioned without any difficulty after the 2011 earthquake, and the design can thus be considered as robust.

		\clearpage

\chapter{Monitoring and Operation}
 An energy-recovering linac (ERL) - though combining features of both linear and circular accelerators - is a non-equilbrium system that lacks a closed orbit and potentially does not possess global betatron or synchrotron stability. It is thus more closely equivalent to a time-of-flight spectrometer or injector than it is to a conventional accelerator, and so encounters a number of unique operational issues \cite{Douglas2009,Douglas2014}. Firstly, longitudinal motion is of paramount importance: timing and energy control set the system architecture, and thus RF phase and gradient control must be assured, as must the lattice momentum compaction - the dependence of the time of flight on energy. Secondly, as it is a non-equilibrium system (in contrast to, say, an electron storage ring), stability is a significant challenge. Thirdly, halo effects dominate operation, much as they do in injection systems, where losses can be performance limiting. Particular concerns include activation (as in injectors), damage (burn-through), and background for experimental users. Finally, as an inherently multi-pass system, an ERL must control multiple beams with different properties (e.g. energy or emittance) during transport through, and handing in, common beamline channels. Reliable machine operation thus requires a comprehensive strategy for machine commissioning, operations, monitoring machine health, system stabilisation, and machine protection.

\section{Operational Regimes}
ERL operation comprises a series of phases: commissioning, beam operations, and machine tuning/recovery. During each phase, system behaviour falls into various classes that can be differentiated by the time scales on which they are manifest: `DC' conditions - those associated with the machine set point intended to produce required beam conditions for users, `drift' effects - slow wandering of the set-point (due, for example, to thermal effects) degrading system output, and `fast' effects (at acoustical to RF time scales), resulting in beam instability. A fourth class - that of transient effects (for example, RF loading during beam on/off transitions and fast shut-down in the event of sudden beam loss for machine protection purposes) - can occur throughout all operational cycles.

\section{Machine Commissioning}
Machine commissioning has combined goals of validating system design architecture and defining a recoverable system operating point. For an ERL, this requires demonstration of the control of phenomena of concern - such as beam break-up (BBU) and the micro-bunching instability ($\mu$BI) - while generating settings for hardware components. 
Following pre-commissioning `hot' checkout of accelerator components and commissioning of hardware subsystems, beam operations commence with threading of low power beam so as to establish a beam orbit and correct it to specified tolerances. This requires orbit correction systems based on beam position monitors and steerers (typically every quarter-betatron wavelength); unique to a multipass ERL with common transport of multiple beams in a single beam line is the requirement that the system correct perturbations locally so that the multiple passes respond identically and the orbits not diverge unacceptably from turn to turn. Similarly, a baseline for longitudinal beam control must be established, by synchronising the beam to the RF using recirculator arcs as spectrometers for precision measurements of energy gain. Any path length adjustments needed to set RF phases and insure energy recovery per the design longitudinal match are thus determined.
 With a 6D phase space reference orbit thus defined, the beam and lattice behaviour is tuned and validated. Lattice performance is measured, tuned, and certified using differential orbit/lattice transfer function measurements; these, too, will require pass-to-pass discrimination amongst beams in common transport. Both transverse and longitudinal measurements (using phase transfer function diagnostics \cite{Evtushenko2011}) are necessary for a full analysis of lattice behaviour. Corrections must be applied to 'rematch the lattice' and bring both transverse (betatron motion/focusing) and longitudinal (timing/momentum compaction) motion into compliance with design (or to establish an alternative working point). 
Certification of lattice performance allows analysis, tuning, and validation of beam parameters, and matching of the beam to the lattice. This requires measurements of both betatron (emittance, beam envelope functions) and longitudinal (bunch length/energy spread/emittance, phase/energy correlation) properties. Disentangling the properties of multiple beams in common transport may prove challenging and require use of beyond-state-of-the-art techniques. If beam properties differ excessively from specification, 'matching' of the beam to the lattice is performed using appropriate correction algorithms. As with orbit correction, perturbations will likely require local correction so as to avoid excessive pass-to-pass divergence of beam properties. 
Given a validated working point, beam power scaling is performed, with currents increased from tune-up levels to full power CW. Transient control and beam stabilization (see below) will be initially investigated and demonstrated during commissioning; they remain a persistent activity through the operational lifetime of the machine, and are therefore discussed below.

\section{Machine Operation: Monitoring and Maintaining Machine Health}
Routine machine operations entail numerous monitoring and correction functions intended to provide beam stability for users and to control and preserve machine performance at a specific set point. These include timing and energy control, which is needed to provide synchronism, for example, at an interaction point, and to maintain the stability of delivered beam properties. This may require a high resolution timing system (if user timing is critical), and will require continuous measurement of energy and energy stability and control mechanisms for energy stability (see the following discussion of stabilisation). Similarly, user requirements may demand measurement and precise control of the orbit of the delivered beam. This can be provided by appropriate enhancements to - and utilisation of a subset of - the beam orbit correction system provided for orbit control during commissioning. Both transverse and longitudinal controls of this type are needed as the machine is used to explore beam dynamics, instability control, and beam quality preservation.
Machine performance is susceptible to degradation as system parameters change due to thermal effects and hardware parametric drift. Beam and lattice properties, control parameters, magnets, and RF variables are all susceptible to such effects; control algorithms providing appropriate monitoring of, and intervention/correction so as to restore RF gradients/phases, beam orbits, lattice focusing, and beam properties are required. These may be established as intermittent machine performance checks and retuning procedures, or, alternatively, be considered as `low speed feedback' systems in which critical beam and machine parameters are monitored and corrected. These provisions are also used for recovering machine configurations/working points after trips and system shutdowns. 
Halo control is critical to the operation of high power ERLs. Halo sources include field emission in SRF systems, cathode-driven sources (such as light scattered onto active areas and surface defects) that can change with ageing, beam/residual gas interactions, beam/wake interactions, and beam dynamical effects during beam formation and handling. All can lead to significant radiation background and potentially unacceptable levels of beam loss. Methods/hardware for monitoring and independent tuning of large amplitude components of multiple beams in common transport are therefore necessary to avoid activation and damage to system components. These can include collimation and/or nonlinear matching using, for example, higher order multipoles (sextupoles, octupoles, etc), and require the use of large dynamic range diagnostics \cite{Evtushenko2014}.
Transient control (maintaining machine and beam health through RF trips, other fast shutdowns, and/or inevitable hardware problems) is needed for all phases of machine operation and is discussed below.

\section{System Stabilisation}
ERLs are non-equilibrium systems subject to drift, jitter, and instability in any of numerous system variables on any of several time scales. They are typically under-constrained, with the number of noise-subjected control parameters much larger than the output observables of relevance to users. Specific strategies for system stabilisation are therefore needed. User requirements must be established from the outset of the system design process, and provision for hardware, software, and procedural control made so as to achieve adequate stability. Table ~\ref{StabilityIssues} outlines critical challenges. 
Globally, drift and jitter must be controlled - at the very least - for the key system parameters of energy and orbit. Beam energy will vary as a result of drift in RF phases; stabilisation by recovery of proper phasing will be necessary over the course of minutes or hours, and may be necessary on short time scales. This can be accomplished through the use of phase stabilisation and control and by providing energy verniers \cite{Krafft2000}. Energy control is coupled to synchronism and timing 
Orbit stability also varies over time and can be subject to jitter. Though orbit stabilisation techniques are well established, the presence of multiple beams in common transport places constraints on both the diagnostics on which the controls are based and on the feedback methods to be used so as to insure that beam- and pass-specific results are achieved.
Given the presence of both high beam brightness and high beam power, the possible need for instability control (BBU, wake effects, etc) must be considered, and the system design should provide opportunity for fast feedback if necessary. Similarly, stability of beam properties is not assured, and means of continuous monitoring/adjusting delivered beam quality (e.g. energy spread, bunch length, spot size/divergence, bunch, etc.) should be provided as necessary.

\begin{table}
\begin{center}
\begin{tabularx}{\textwidth}{l|lllllll}
 \hline 
\multicolumn{5}{c}{TIME SCALE/MAGNITUDE OF EFFECTS}\\* \hline 
Class of &DC&Slow &Fast &RF/dynamic \\*[-1ex]
Control &&(up to thermal)&(<1 kHz)&\\
\hline
Lattice&transfer map&transfer map&magnet jitter&& \\*[-1ex]	
&(set point)&(drift)&(power, vibration)\\
Beam orbit	&central orbit&orbit drift&orbit jitter&Beam stability  \\*[-1ex]
&&&&(e.g. BBU) \\
Beam &match to lattice&match drift&Instability&& \\*[-1ex]	
properties&(setpoint)&&&\\
Halo	&experimental &drift	&electron/ion &	electron/ion \\*[-1ex]
&background&&instability?&instability?\\
\hline
\end{tabularx}
\caption{System stability issues in energy recovery linacs}\label{StabilityIssues}
\end{center}
\end{table}

\section{Transient Control and Machine Protection}
ERLs are subject to numerous transient effects, two classes of which are of particular operational importance: the impact of RF transients (beam off/on transients, variable beam loading during current ramps, and RF trips), and machine protection fast shutdowns.
RF transients due to variations in beam loading \cite{Powers2007} are manageable with appropriate RF drive design. Care in choice of $Q_{ext}$ is of importance, as is planning for the type and operational range of the longitudinal match; implementation of incomplete energy can result in greater transient control requirements than encountered in systems with complete energy recovery. The RF drive system (control loops, feed-forward/back) must be configured to manage transients as experienced under different machine operating conditions and operating points; RF power and cavity tuning should be monitored during routine operation to insure that stability is maintained. 
Dramatic transients (particularly in beam loading) will occur during machine-protection-system (MPS) driven fast shutdowns. As ERL beam powers are very high, loss tolerances are tight and large losses must be prevented. Critical to machine safety, the MPS continually monitors the accelerator for beam loss and rapidly shuts off the beam if unsafe loss levels are observed \cite{Jordan2003}.  The machine control system monitors and records the interlock sequence precipitating the fast shutdown so as to characterise the source of the transient event and provide guidance on correction of the fault.
\clearpage

\chapter{Site Considerations}
The interest in PERLE, sketched in the present report, is threefold, regarding its
technology development potential, its physics and applied user programme and its importance for demonstrating and studying the technology choice of the LHeC.
At present there is no decision as to where PERLE may be placed. An initial study, of also general interest, considered the possibility of hosting PERLE at CERN. This is sketched below. It was subsequently studied to possibly build this facility  at LAL Orsay, may be at reduced beam energy for keeping its dimension fit to the available infrastructure and halls. This is also mentioned below. Recently, an idea has also been considered of building a low energy, lower current version of PERLE at Darmstadt in Germany. 

\section{Introduction}

As mentioned in the lattice section, the genuine footprint of the PERLE facility at its maximum energy of about $1$\,GeV occupies a rectangle of $42 \times 14$\,m$^2$. This area should be enclosed by shielding at a sufficient distance to allow passage and maintenance operations. We estimate the required passage and half thickness of the accelerator component to 2\,m. A concrete shielding of $50$\,cm thickness is assumed here to stop photons and neutrons produced by halo electrons. Detailed simulations of the radiation generated by the impinging electron will be necessary at a later stage. An increase of the shielding required could be alleviated by the use of denser materials like lead. Access conditions and the geographical location of the site may also influence the final choice of shielding. 
In addition to this central area,   space needs to be allocated for the auxiliary systems like:
\begin{itemize}
\item{Power converters for magnets, septa and kickers};
\item {RF power. Assuming IOTs or solid state amplifiers as close as possible to the SRF modules to minimize RF losses};
\item {Water cooling.The dimensioning of this system greatly depends on the operational modes};
\item {Cryogenics. The use of a dewars for storing liquid helium at 4.5 could avoid the cost of a liquefier. However it will limit flexibility of operation in non-recovery mode and needs to be studied further};
\item {Source};
\item {Dump. A design of the dump exists with a minimum length of 50 m (reference) but a more compact version could be used by limit the current or repetition rate when working on non recovery mode};
\end{itemize}

As a rough estimate one would like to double the area of the accelerator itself to accommodate all services. It is worth noting that some services like RF power generation or power supplies may be placed on a different level than the accelerator itself, while the source or the dump may not.
We do not consider here the use of the interior part of the ring as the escape routes would be compromised. It may however be used to house a low energy dump which itself needs to be shielded and which will have restricted access.

\section{CERN}
For an initial study, we have been considering existing buildings around the CERN site~\footnote{With the appearance of the Orsay option, detailed below, CERN as a site has not been 
considered further.}.  The building needs to be equipped with a crane, water and electricity services. The availability of cryogenic fluids would be an interesting option and provide considerable savings. The installation of electrical power and demineralised water seems to be less costly. The total area of the installation would be then of the order of $1500$\,m$^2$ with an incompressible area of approximately $45 \times 17$\,m$^2$ to host the accelerator footprint and shielding. There are not many buildings of this dimension at the CERN site and they are in general already in use for large facilities like the superconducting test facility in SM18 or the magnet  repair facility in building 180. A couple of sites have been identified which would suit the area requirements and present some advantage like the availability of cryogenics (b.973), power (b. 2275) or shielding (b. 2003). 
  
If one deemed to better  construct a new building one promising location is around the area 18, where a powerful cryogenic plant can serve the accelerator while the proximity to SM18 could ease the use of the electron beam for quench tests. This location would also be compatible with the possibility to use  PERLE as an injector to the LHeC. The detailed plans and costing of such a building would have to be studied for CERN. Naturally, a location of PERLE outside of CERN would pose other constraints and opportunities. 
\section{LAL Orsay}
\begin{figure}[H]
 \centering
  \includegraphics[width=0.85\textwidth]{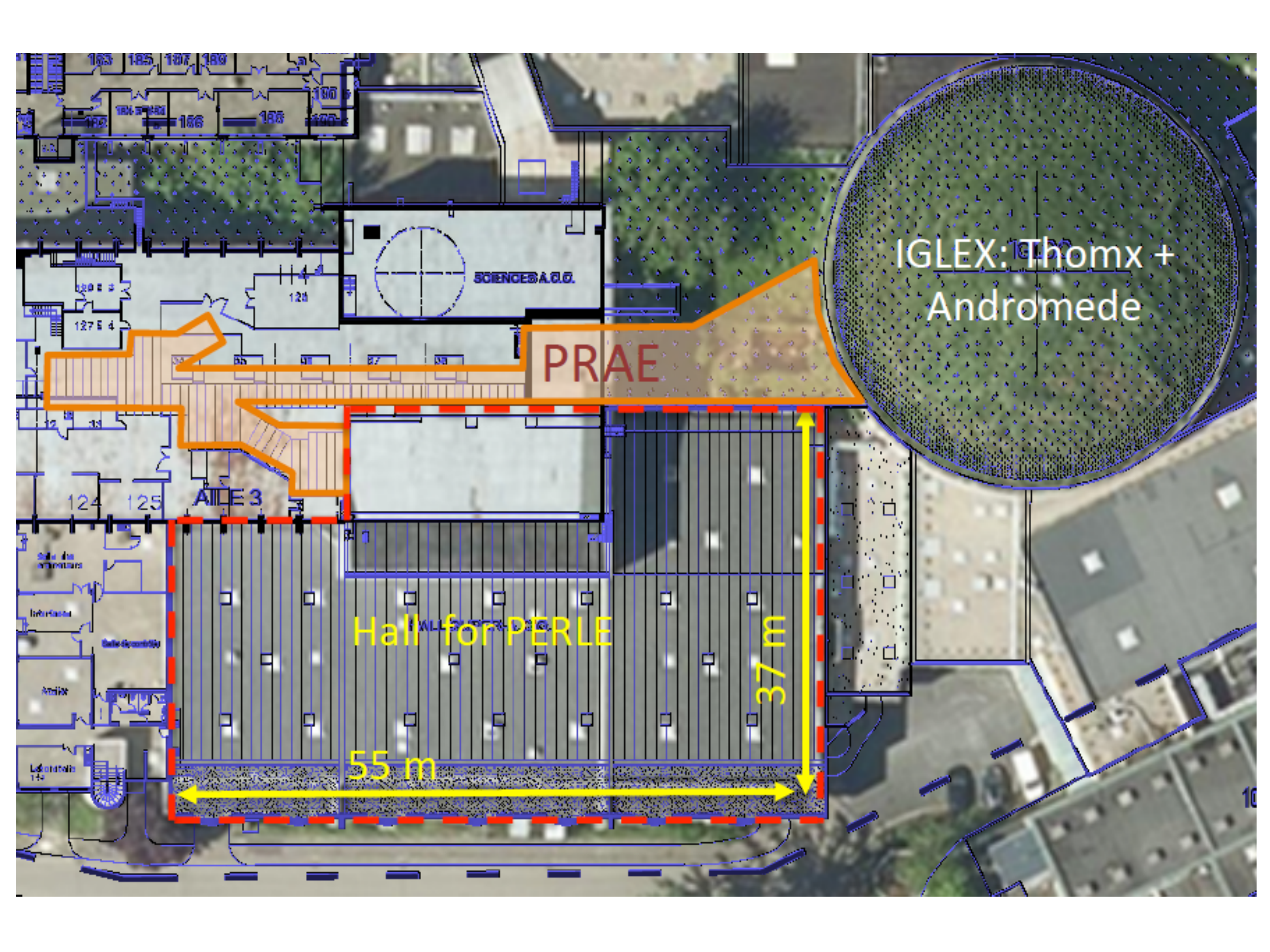}
\caption{Map of the site for PERLE at Orsay. The hall has in the upper part two
areas with particularly rigid ground, left top of $22.3 \times 11.5$\,m$^2$ and right 
top of $19.6 \times 12.9$\,m$^2$, which together may house PERLE with the larger part
available also but studies required for the status of the ground, see also text.}
\label{Fig:orsayhall}
\end{figure}

Prior to the publication of this report  it has been realised that  LAL Orsay would be very well prepared to house PERLE, preferentially  at up to $450$\,MeV energy which required an inner area of about $20 \times 7$\,m$^2$. 
 The building that would host this version of PERLE is a former experimental hall, the Super ACO hall, 
 which is equipped with cranes and electricity. The ground of the building is made of concrete slabs with variable ground resistance. Nevertheless, more than the half of the hall area has a sufficient resistance to allow the installation PERLE. 
 A complete study will be performed to confirm this fact.     
Being next to the tunnel of the old Orsay linac and close to the ``Igloo", where new accelerators are being installed currently, the building is partially shielded and water-cooling circuits could be shared with the other machines. The location is illustrated on Fig.\,\ref{Fig:orsayhall}.
The building gives the possibility to install the RF source and the power supplies at a different level than the accelerator. An existing  control room that overlooks the experimental hall could be used for PERLE. Since all the accelerators installed nearby are based on warm technology, a cryogenic plant has to be built. Altogether, this appears to be an available, suitable place. Hosting PERLE at Orsay would be of much interest for the development of  physics and technology at Orsay and internationally. At an initial meeting~\cite{kicko} at Orsay, in February 2017, a Collaboration has been launched, including LAL and IPN Orsay, BINP Novosibirsk, CERN, Daresbury, Jefferson Laboratory, Liverpool and possibly further partners, to pursue this.    

\clearpage

\markboth{Summary} {right head}\markright{Summary}
\chapter{Summary}

\vspace{-1cm}
Design concepts and applications have been presented 
of a novel, powerful energy recovery linac 
facility suitable to enable SCRF technology developments
and intense, low energy electron and photon physics experiments, termed PERLE. 
The two main goals of PERLE are to i) 
to enable technical developments as on SCRF and applications as well as future 
physics experiments in a novel, high current ERL facility environment;
ii) 
to develop and demonstrate the viability of the basic design assumptions
for a 60\,GeV electron multi-turn ERL linac as is proposed to be installed
tangential to the LHC, the HE-LHC  and/or a future FCC, for realising exploratory
electron-proton experiments at O($1000$) times the luminosity of HERA.
The PERLE target parameters and technology choices are largely derived from
the LHeC and in turn need to be compliant with the LHC.
This determines the frequency, chosen to be 802\,MHz, 
the number of turns to three and the electron beam current 
to be as large as about $15$\,mA.
   
PERLE is foreseen
to demonstrate and gain operational experience with low-frequency high-current SCRF cavities and cryomodules of a type suitable for scale up to a high-energy machine. Since the cavity design, HOM couplers, FPC's etc. will be all new or at least heavily modified, PERLE will serve as a technology test bed that will explore all the parameters needed for a larger machine. There is no other high current ERL test bed in the world that can do this. PERLE will  feature emittance preserving recirculation optics and this will also be an important demonstration that these can be constructed and operated in a flexible user-facility environment. The machine, when transformed from a test to a user facility, must run with high reliability to  provide test beams for experimenters or ultimately provide Compton or FEL radiation to light source users. This demonstration of stability and high reliability will be essential for any future large facility.   

As an example for technical impact, 
the present study has demonstrated the use of the electron beam to
perform quench tests on SC components and magnets. The facility may be used
for low energy test beam measurements and it may serve as a base to design
or build the injector of the LHeC.

The basic physics case is presented  for
new measurements of current outstanding importance. Relying
 on a luminosity of  O($10^{40}$)\,cm$^{-2}$s$^{-1}$,
 in elastic $ep$ scattering, most accurate investigations of 
electroweak loop effects and the proton radius as well as 
searches for new physics, such as dark photons, characterise the extremely attractive
physics potential of the  PERLE facility.

An exiting physics programme 
has been detailed from operating PERLE as a gamma ray facility with
a very high flux, at least two orders of magnitude above expected upgrades
of existing facilities, and superior spectral density. A path is shown
to discoveries using up to 30 MeV photons and for a variety of novel, unique
and precise measurements on photo-nuclear reactions, nuclear structure
as well as to important measurements for neutrino and nuclear astrophysics.
   
A thorough simulation study is presented  of the system architecture, the
transport optics and start-to-end beam dynamics. The paper presents
initial design concepts of the main components for PERLE, applicable also to
its possible lower energy version. These comprise descriptions of the
source and injector, the 802 MHz cavity, under design and 
construction by  us, of 
a cryomodule and HOM design considerations. 
Further, the inventory and novel designs are presented
of the arc magnets. A section is devoted to rather detailed 
considerations for the dumps and transfers. 

For CW electron bunches of larger than 10\,MHz repetition rate,
Fabry-Perot optical resonators are suitable to provide a
high quality photon beam and are presented in this paper as a
preferred reliable solution.

A final chapter is devoted to the monitoring and operations tasks
including the commissioning, system stabilisation and protection aspects.
Considerations have also been presented for the site and its infrastructure. These
naturally will be updated once a site is finally chosen which most likely will
be at the campus of the Linear Accelerator Laboratory at Orsay (Paris).
   
PERLE has the opportunity to be a clean-sheet globally optimised design for a new generation of high average power efficient ERL based machines,
a novel testing ground for far reaching experiments with
electron and photon beams of unique quality and, not least,
to become a  prime technical base for an electron beam upgrade of the LHC, i.e.  
a new generation of  deep inelastic scattering experiments entailing the precision
study of the Higgs boson and the exploration of new physics at TeV energies. 


%
\section*{Acknowledgement}
This work was pursued as part of the LHeC development and its
extension to the FCC-he. The authors are grateful to
the CERN directorate for its steady interest and support and the International Advisory Committee, led by Herwig Schopper, 
for encouragement and guidance. They thank the directorates of Thomas
Jefferson Laboratory, BINP Novosibirsk, IPNO and LAL Orsay, of ASTeC and the Cockcroft Institute at Daresbury, as well
as the Universities of Bordeaux, Darmstadt and Liverpool  
for supporting this study.  
 We especially thank our many collaborators participating in the development
of the accelerator, detector and the physics directed to a TeV energy
scale ep/eA collider for which the realisation of PERLE in suitably chosen steps will be essential.

   
\bibliography{bib.bib}  
\addcontentsline{toc}{chapter}{Bibliography}    
\bibliographystyle{ieeetr}

\end{document}